\documentclass[twocolumn]{aastex62}
\usepackage{amsmath,amstext}

\usepackage{pgffor}%for automated subfigures
\usepackage{subfigure}

\usepackage{xspace}
\newcommand{\Ned}{AT2018zr\xspace}
\newcommand{\Jon}{AT2018bsi\xspace}
\newcommand{\Sansa}{AT2018hco\xspace}
\newcommand{\Cersei}{AT2018lna\xspace}
\newcommand{\Arya}{AT2018lni\xspace}
\newcommand{\Jaime}{AT2019azh\xspace}
\newcommand{\Jorah}{AT2018iih\xspace}
\newcommand{\Petyr}{AT2019cho\xspace}
\newcommand{\Gendry}{AT2018hyz\xspace}
\newcommand{\Bran}{AT2019dsg\xspace}
\newcommand{\Brienne}{AT2019ehz\xspace}
\newcommand{\Varys}{AT2019bhf\xspace}
\newcommand{\Margaery}{AT2019meg\xspace}
\newcommand{\Bronn}{AT2019mha\xspace}
\newcommand{\Melisandre}{AT2019qiz\xspace}
\newcommand{\Catelyn}{AT2019eve\xspace}
\newcommand{\Robb}{AT2019lwu\xspace}

\graphicspath{{./}{fig/}{./fig/XMM/}}

\shorttitle{Winter is here}
\shortauthors{van Velzen et al.}
\begin{document}

\title{Seventeen Tidal Disruption Events from the First Half of ZTF Survey Observations:  \\Entering a New Era of Population Studies}

\correspondingauthor{Sjoert van Velzen}
\email{sjoert@astro.umd.edu}

\author[0000-0002-3859-8074]{Sjoert van Velzen}
\affiliation{Department of Astronomy, University of Maryland, College Park, MD 20742, USA}
\affiliation{Center for Cosmology and Particle Physics, New York University, NY 10003}

\author[0000-0003-3703-5154]{Suvi Gezari}
\affiliation{Department of Astronomy, University of Maryland, College Park, MD  20742, USA}
\affiliation{Joint Space-Science Institute, University of Maryland, College Park, MD 20742, USA}

\author[0000-0002-5698-8703]{Erica Hammerstein}
\affiliation{Department of Astronomy, University of Maryland, College Park, MD  20742, USA}

\author[0000-0002-6485-2259]{Nathaniel Roth}
\affiliation{Department of Astronomy, University of Maryland, College Park, MD  20742, USA}
\affiliation{Joint Space-Science Institute, University of Maryland, College Park, MD 20742, USA}

\author{Sara Frederick}
\affiliation{Department of Astronomy, University of Maryland, College Park, MD  20742, USA}

\author{Charlotte Ward}
\affiliation{Department of Astronomy, University of Maryland, College Park, MD  20742, USA}

\author{Tiara Hung} 
\affil{Department of Astronomy and Astrophysics, University of California, Santa Cruz, CA 95064,
USA}

\author[0000-0003-1673-970X]{S. Bradley Cenko}
\affiliation{Astrophysics Science Division, NASA Goddard Space Flight Center, MC 661, Greenbelt, MD 20771, USA}
\affiliation{Joint Space-Science Institute, University of Maryland, College Park, MD 20742, USA}

\author{Robert Stein}
\affiliation{Deutsches Elektronensynchrotron, Platanenallee 6, D-15738, Zeuthen, Germany}

\author{Daniel~A.~Perley}
\affiliation{Astrophysics Research Institute, Liverpool John Moores University, 146 Brownlow Hill, Liverpool L3 5RF, UK}

\author[0000-0002-5748-4558]{Kirsty Taggart}
\affiliation{Astrophysics Research Institute, Liverpool John Moores University, 146 Brownlow Hill, Liverpool L3 5RF, UK}

\author{Jesper Sollerman}
\affiliation{The Oskar Klein Centre \& Department of Astronomy, Stockholm University, AlbaNova, SE-106 91 Stockholm, Sweden}

% start alphabetical (builders, observers)

\author{Igor Andreoni}
\affiliation{Division of Physics, Mathematics, and Astronomy, California Institute of Technology, Pasadena, CA 91125, USA}

\author[0000-0001-8018-5348]{Eric C. Bellm}
\affiliation{DIRAC Institute, Department of Astronomy, University of Washington, 3910 15th Avenue NE, Seattle, WA 98195, USA} 

\author{Valery Brinnel}
\affiliation{Institute of Physics, Humboldt-Universit\"at zu Berlin, Newtonstr. 15, 12489 Berlin, Germany}

\author{Kishalay De}
\affiliation{Division of Physics, Mathematics, and Astronomy, California Institute of Technology, Pasadena, CA 91125, USA}

\author{Richard Dekany}
\affiliation{Caltech Optical Observatories, California Institute of Technology, Pasadena, CA 91125, USA}

\author{Michael Feeney}
\affiliation{Caltech Optical Observatories, California Institute of Technology, Pasadena, CA 91125, USA}

\author{Ryan J. Foley}
\affil{Department of Astronomy and Astrophysics, University of California, Santa Cruz, CA 95064,
USA}

\author{Christoffer Fremling}
\affiliation{Division of Physics, Mathematics, and Astronomy, California Institute of Technology, Pasadena, CA 91125, USA}

\author{Matteo Giomi}
\affiliation{Institute of Physics, Humboldt-Universit\"at zu Berlin, Newtonstr. 15, D-12489 Berlin, Germany}

\author[0000-0001-8205-2506]{V. Zach Golkhou}
\affiliation{DIRAC Institute, Department of Astronomy, University of Washington, 3910 15th Avenue NE, Seattle, WA 98195, USA} 
\affiliation{The eScience Institute, University of Washington, Seattle, WA 98195, USA}
%\altaffiliation{Moore-Sloan, WRF Innovation in Data Science, and DIRAC Fellow}

\author[0000-0002-9017-3567]{Anna.~Y.~Q.~Ho}
\affiliation{Division of Physics, Mathematics, and Astronomy, California Institute of Technology, Pasadena, CA 91125, USA}

\author{Mansi M. Kasliwal}
\affiliation{Division of Physics, Mathematics, and Astronomy, California Institute of Technology, Pasadena, CA 91125, USA}

\author{Charles~D.~Kilpatrick} 
\affil{Department of Astronomy and Astrophysics, University of California, Santa Cruz, CA 95064,
USA}

\author[0000-0001-5390-8563]{Shrinivas R. Kulkarni}
\affiliation{Division of Physics, Mathematics, and Astronomy, California Institute of Technology, Pasadena, CA 91125, USA}

\author[0000-0002-6540-1484]{Thomas Kupfer}
\affiliation{Kavli Institute for Theoretical Physics, University of California, Santa Barbara, CA 93106, USA}

\author[0000-0003-2451-5482]{Russ R. Laher}
\affiliation{IPAC, California Institute of Technology, 1200 E. California Blvd, Pasadena, CA 91125, USA}

\author[0000-0003-2242-0244]{Ashish Mahabal}
\affiliation{Division of Physics, Mathematics, and Astronomy, California Institute of Technology, Pasadena, CA 91125, USA}

\affiliation{Center for Data Driven Discovery, California Institute of Technology, Pasadena, CA 91125, USA}

\author[0000-0002-8532-9395]{Frank J. Masci}
\affiliation{IPAC, California Institute of Technology, 1200 E. California
             Blvd, Pasadena, CA 91125, USA}

\author{Jakob Nordin}
\affiliation{Institute of Physics, Humboldt-Universit\"at zu Berlin, Newtonstr. 15, 12489 Berlin, Germany}

\author{Reed Riddle}
\affiliation{Caltech Optical Observatories, California Institute of Technology, Pasadena, CA 91125, USA}

\author[0000-0001-7648-4142]{Ben Rusholme}
\affiliation{IPAC, California Institute of Technology, 1200 E. California
             Blvd, Pasadena, CA 91125, USA}

\author{Yashvi Sharma}
\affiliation{Division of Physics, Mathematics, and Astronomy, California Institute of Technology, Pasadena, CA 91125, USA}

\author{Jakob van Santen}
\affiliation{Deutsches Elektronensynchrotron, Platanenallee 6, D-15738, Zeuthen, Germany}

\author[0000-0003-4401-0430]{David L. Shupe}
\affiliation{IPAC, California Institute of Technology, 1200 E. California
             Blvd, Pasadena, CA 91125, USA}

\author[0000-0001-6753-1488]{Maayane T. Soumagnac}
\affiliation{Lawrence Berkeley National Laboratory, 1 Cyclotron Road, Berkeley, CA 94720, USA}
\affiliation{Department of Particle Physics and Astrophysics, Weizmann Institute of Science, Rehovot 76100, Israel}



\begin{abstract}
While tidal disruption events (TDEs) have long been heralded as laboratories for the study of quiescent black holes, the small number of known TDEs and uncertainties in their emission mechanism have hindered progress towards this promise. Here present 17 new TDEs that have been detected recently by the Zwicky Transient Facility along with {\it Swift} UV and X-ray follow-up observations. Our homogeneous analysis of the optical/UV light curves, including 22 previously known TDEs from the literature, reveals a clean separation of light curve properties with spectroscopic class. 
The TDEs with Bowen fluorescence features in their optical spectra have smaller blackbody radii, as well as  longer rise times and higher disruption rates compared to the rest of the sample. The Bowen fluorescence mechanism requires a high density which can be reached at smaller radii, which in turn yields longer diffusion time scales. Thus, the difference in rise times suggests the pre-peak TDE light curves are governed not by the fallback timescale, but instead by the diffusion of photons through the tidal debris. The small subset of TDEs that show only helium emission lines in their spectra have the longest rise times, the highest luminosities and the lowest rates.
We also report, for the first time, the detection of soft X-ray flares from a TDE on $\sim$~day timescales. Based on the fact the flares peak at a luminosity similar to the optical/UV blackbody luminosity, we attribute them to brief glimpses through a reprocessing layer that otherwise obscures the inner accretion flow.
~
~
~
~
~
~
~
~
~
\end{abstract}

\section{Introduction}
The occasional ($\sim 10^{-4}$ yr$^{-1}$) luminous flare of radiation from a galaxy nucleus due to the tidal disruption of a star by an otherwise dormant central massive black hole originated as a theoretical concept \citep{Lidskii_Ozernoi79, Rees88}, but thanks to the rapid increase in wide-field survey capabilities across the electromagnetic spectrum, is now a well established class of transients.  While the first candidates were detected as soft X-ray outbursts in previously quiescent galaxy nuclei by the {\it ROSAT} All-Sky Survey \citep{Donley02},
these tidal disruption events (TDEs), have more recently emerged as a unique class of nuclear transients in optical surveys with common photometric properties:  persistent blue colors, a relatively long rise time compared to most supernovae (SNe), and a smooth, power-law decline from peak \citep{vanVelzen10,Hung17,vanVelzen18_NedStark}.  The spectroscopic features of TDEs are characterized by a hot, blue thermal continuum, and very broad ($5-15,000$ km s$^{-1}$; \citealt{Arcavi14,Hung17}) emission lines, which are distinct from nearly all SNe (when observed post peak) and AGN. The inferred volumetric rate of photometric and spectroscopic TDEs class falls off steeply above the ``Hills mass", for which a star can be disrupted before being disappearing behind the black hole event horizon \citep{Hills75}, further strengthening the association of this class of transients as bonafide stellar disruptions \citep{vanVelzen18}.  
																														   
However, while discoveries of TDEs are becoming increasingly more common in wide-field optical surveys such as iPTF \citep{Blagorodnova17, Hung17, Blagorodnova19}, ZTF \citep{vanVelzen18_NedStark}, ASAS-SN \citep{Holoien14,Holoien16,Holoien16b,Wevers19,Holoien19}, and Pan-STARRS \citep{Gezari12, Chornock14, Holoien18, Nicholl19}, the nature of what is powering their relatively uniform optical light curves is uncertain.  Unlike the soft X-ray component detected in some optically selected TDEs, which is consistent with thermal emission from the inner radii of an accretion disk \citep{Komossa15,Miller_14li, Gezari17, vanVelzen18_NedStark, Wevers19}, the inferred blackbody radius of the UV/optical thermal component is a factor of 10-100 larger than expected for the size of the nascent debris disk expected to form from the circularization of the stellar debris streams.  This implies the existence of an unknown, larger structure, potentially produced as a result of an outflow or wind \citep{Miller15,Metzger17, Dai18}, or the intersecting debris streams themselves \citep{Piran15,Jiang16,Bonnerot17}. Indeed, there are now several examples of outflow signatures from optical \citep{Hung19}, UV \citep{Cenko16, Brown18, Blagorodnova18} and X-ray \citep{Miller_14li, Kara17} spectroscopy, plus potentially also in the radio \citep{Alexander16, Alexander17}---however see \citet{vanVelzen16b,PashamvanVelzen17} for a different explanation of the radio emission from optical TDEs. 

There has been a recent expansion of the spectroscopic sub-classes for TDEs. From the first optical TDE spectra \citep{vanVelzen10,Gezari12}, one of which surprisingly showed only broad He II lines and no hydrogen emission \citep{Gezari12}, to the He-rich to H-rich sequence proposed by \citet{Arcavi14}, to including classes with Bowen fluorescence emission line features OIII and NIII \citep{Blagorodnova18, Leloudas19}, low-ionization Fe~II lines \citep{Wevers19}, as well as a TDE that showed the gradual disappearance of broad H lines, while broad He II$\lambda 4686$ line remained strong \citep{Nicholl19}.  The UV spectra of TDEs are also unique, characterized by strong N~III]~$\lambda$1750 emission but weak Mg~II~$\lambda\lambda$2896, 2803 and C~III]~$\lambda1909$ \citep{Cenko16}.  The nature of this spectral diversity has been attributed to the chemical composition of the star \citep{Gezari12, Kochanek16}, ionization state of the debris \citep{Guillochon14}, radiative transfer effects in an optically thick envelope \citep{Roth15}, and reprocessing of X-ray emission through dense, optically thick gas \citep{Leloudas19, Wevers19}.  In this paper, we present the largest spectroscopic TDE sample to date. We discovered correlations between the spectroscopic sub-class of the TDE and the host galaxy and flare properties. These correlation provide new insights into the origin of the spectral diversity in TDEs.  
%In \S \ref{sec:results} we present statistically significant correlations between TDE spectral class and the blackbody radius and rise time.

We were able to discover these correlations thanks to a homogeneous treatment of well-sampled optical/UV light curves of 32 spectroscopic TDEs. This factor $\approx 2$ increase in sample size of known TDEs can, for a large part, be attributed to the start of the Zwicky Transient Facility \citep[ZTF;][]{Bellm19} in March 2018. We searched the ZTF data for new TDEs using a combination of photometric selection and spectroscopic and multi-wavelength follow-up \citep{vanVelzen18_NedStark}. While ZTF is not always the first survey to report these events to the Transient Name Server (TNS) and thus claim discovery credit (see Table~\ref{tab:discovery}), for most sources ZTF provides the deepest difference imaging light curve that are publicly available \citep{Patterson19,Masci19}. %For 17 of the 32 spectroscopic TDEs in our sample, the optical light curves used in this paper were provided by ZTF.

Besides the origin of optical emission, a second important (and unexpected) observation of optically-selected TDEs is their X-ray faintness.  The most common explanation is that the soft X-rays from accretion in the inner disk are absorbed and reprocessed into optical photons \citep[e.g.,][]{Guillochon14,Auchettl16,Dai18}. In this scenario the X-rays can only break out after the obscuring gas has expanded enough to become transparent to X-rays \citep{Metzger16, Lu18}.  However, intrinsically faint soft X-ray TDEs have also been proposed as a result of delayed accretion due to the timescale required for circularization of the debris into an appreciable accretion disk \citep{Piran15,Krolik16, Gezari17}. Discriminating between these models, and thus determining if the optical emission is powered by accretion or the stream kinetic energy, is possible by looking at the relative timing and response of the optical flare to the soft X-ray emission from TDEs \citep{Pasham17}. Significant soft X-rays variability has recently been observed, including a late-time brightening, that is anti-correlated with the smooth decline of the optical component \citep{Gezari17,vanVelzen18_NedStark,Wevers19}.  In this paper, we present four more optically-selected TDEs with soft X-ray detections, including both flaring and late-time X-ray brightening, which provide new constraints on the emission mechanisms.

In \S \ref{sec:cand} we present the selection of TDE candidates from the ZTF stream and spectroscopic follow-up, as well as our naming scheme for three spectroscopic classes. In \S \ref{sec:host} we investigate the host galaxies of our TDEs, obtaining  estimates of their mass and star formation histories, followed by \S \ref{sec:multiw} which contains the details of our multi-wavelength follow-up observations. In \S \ref{sec:lc} we present our light-curve model that is applied to  39 spectroscopic+photometric TDEs.  In \S \ref{sec:results} we present correlations between features extracted from our light-curve model, plus a discovery of differences in the photometric features between the TDEs of each spectroscopic class. %We discuss these finding in \S \ref{sec:discussion} and close with a list conclusions. 

We adopt a flat cosmology with $\Omega_\Lambda=0.7$ and $H_0=70~{\rm km}\,{\rm s}^{-1}{\rm Mpc}^{-1}$. All magnitudes are reported in the AB system \citep{oke74}.

\begin{table*}
\caption{Names and discovery name (bold)}\label{tab:discovery}
\begin{tabular}{l l l l l l}
\hline
IAU Name & ZTF Name & GOT Name & Other/Discovery Name & First TDE Classification Report \\
\hline
AT2018zr   & ZTF18aabtxvd           & Ned        & \textbf{PS18kh}  & ATel\#11444 \citep{2018ATel11444....1T} \\
AT2018bsi  & \textbf{ZTF18aahqkbt}  & Jon        &                  & ATel\#12035 \citep{2018ATel12035....1G} \\
AT2018hco  & ZTF18abxftqm           & Sansa      & \textbf{ATLAS18way} & ATel\#12263 \citep{2018ATel12263....1V} \\
AT2018iih  & ZTF18acaqdaa           & Jorah      & \textbf{ATLAS18yzs}, Gaia18dpo  & This paper \\
AT2018hyz  & ZTF18acpdvos           & Gendry     & \textbf{ASASSN-18zj}, ATLAS18bafs & ATel\#12198 \citep{2018ATel12198....1D}\\
AT2018lni  & \textbf{ZTF18actaqdw}  & Arya       & & This paper \\
AT2018lna  & \textbf{ZTF19aabbnzo}  & Cersei     & & ATel\#12509 \citep{2019ATel12509....1V}\\ 
AT2019cho  & \textbf{ZTF19aakiwze}  & Petyr      & & This paper \\
AT2019bhf  & \textbf{ZTF19aakswrb}  & Varys      & & This paper \\
AT2019azh  & ZTF17aaazdba           & Jaime      & \textbf{ASASSN-19dj}, Gaia19bvo & ATel\#12568 \citep{2019ATel12568....1V}\tablenotemark{a}  \\
AT2019dsg  & \textbf{ZTF19aapreis}  & Bran       & ATLAS19kl & ATel\#12752 \citep{2019ATel12752....1N} \\
AT2019ehz  & ZTF19aarioci           & Brienne    & \textbf{Gaia19bpt} & ATel\#12789 \citep{2019ATel12789....1G}\\
AT2019eve  & \textbf{ZTF19aatylnl}  & Catelyn    & Gaia19bti, ATLAS19kfv & This paper \\
AT2019mha  & ZTF19abhejal           & Bronn      & \textbf{ATLAS19qqu} & This paper  \\
AT2019meg  & \textbf{ZTF19abhhjcc}  & Margaery   & Gaia19dhd  & AN-2019-88 \citep{2019TNSAN..88....1V}\tablenotemark{b}\\
AT2019lwu  & \textbf{ZTF19abidbya}  & Robb       & ATLAS19rnz, PS19ega & This paper \\
AT2019qiz  & {\bf ZTF19abzrhgq}     & Melisandre & ATLAS19vfr, Gaia19eks, PS19gdd & ATel\#13131 \citep{2019ATel13131....1S} \\
\hline
\end{tabular}
\\
Note: Names in boldface indicate the discovery name, i.e. the first survey to report photometry of the transient detection to the TNS. $^{a}$First spectrum obtained by \citet{2019ATel12529....1H} on 2019 Feb 21 but classification not yet conclusive. $^{b}$First spectrum published by \citet{2019TNSAN..59....1N} on 2019 Aug 1 but classification not yet conclusive. 
\end{table*}

\begin{figure}
\gridline{\fig{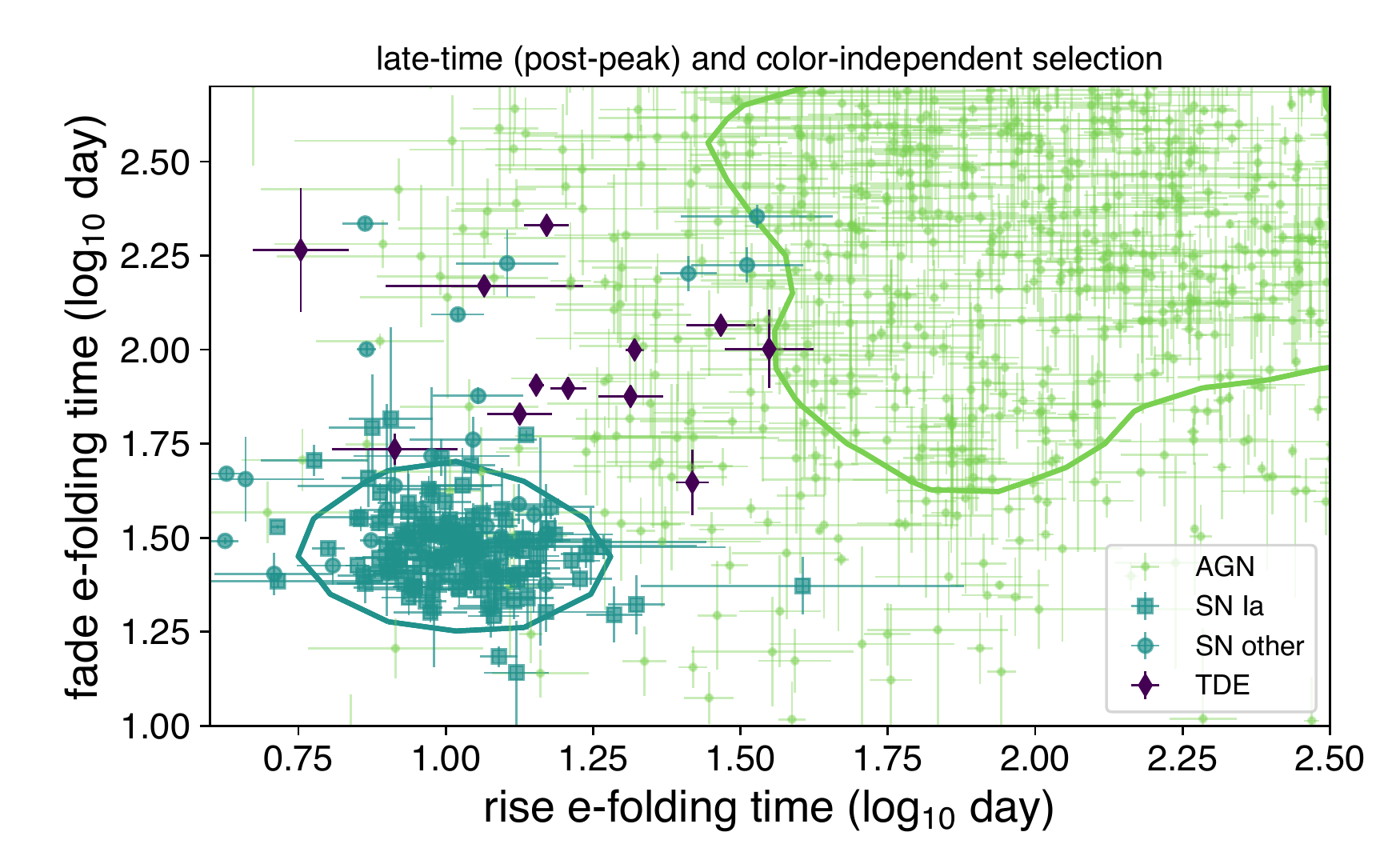}{0.48 \textwidth}{}\\[-20 pt]}
\gridline{\fig{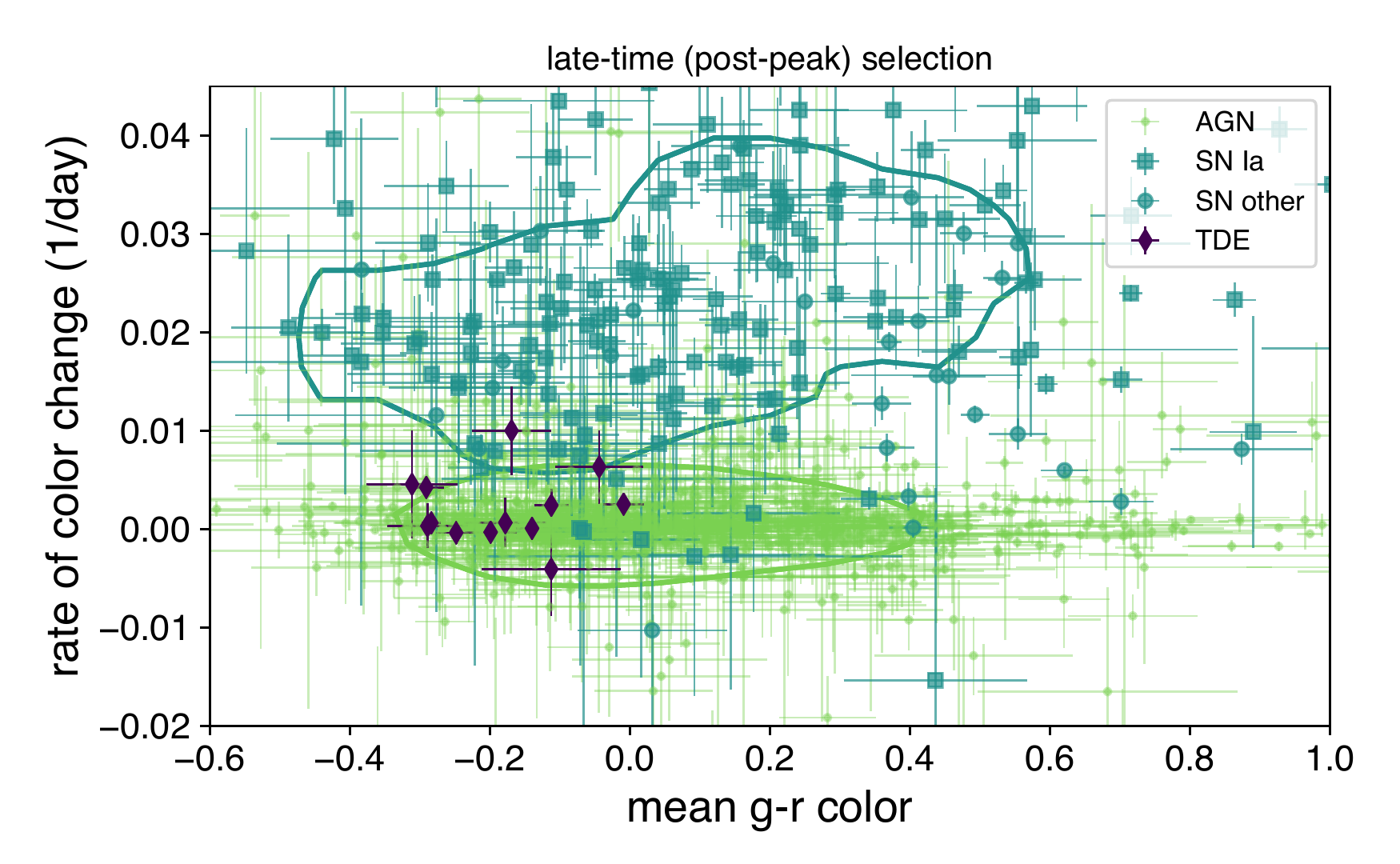}{0.48 \textwidth}{}\\[-20 pt]}
\gridline{\fig{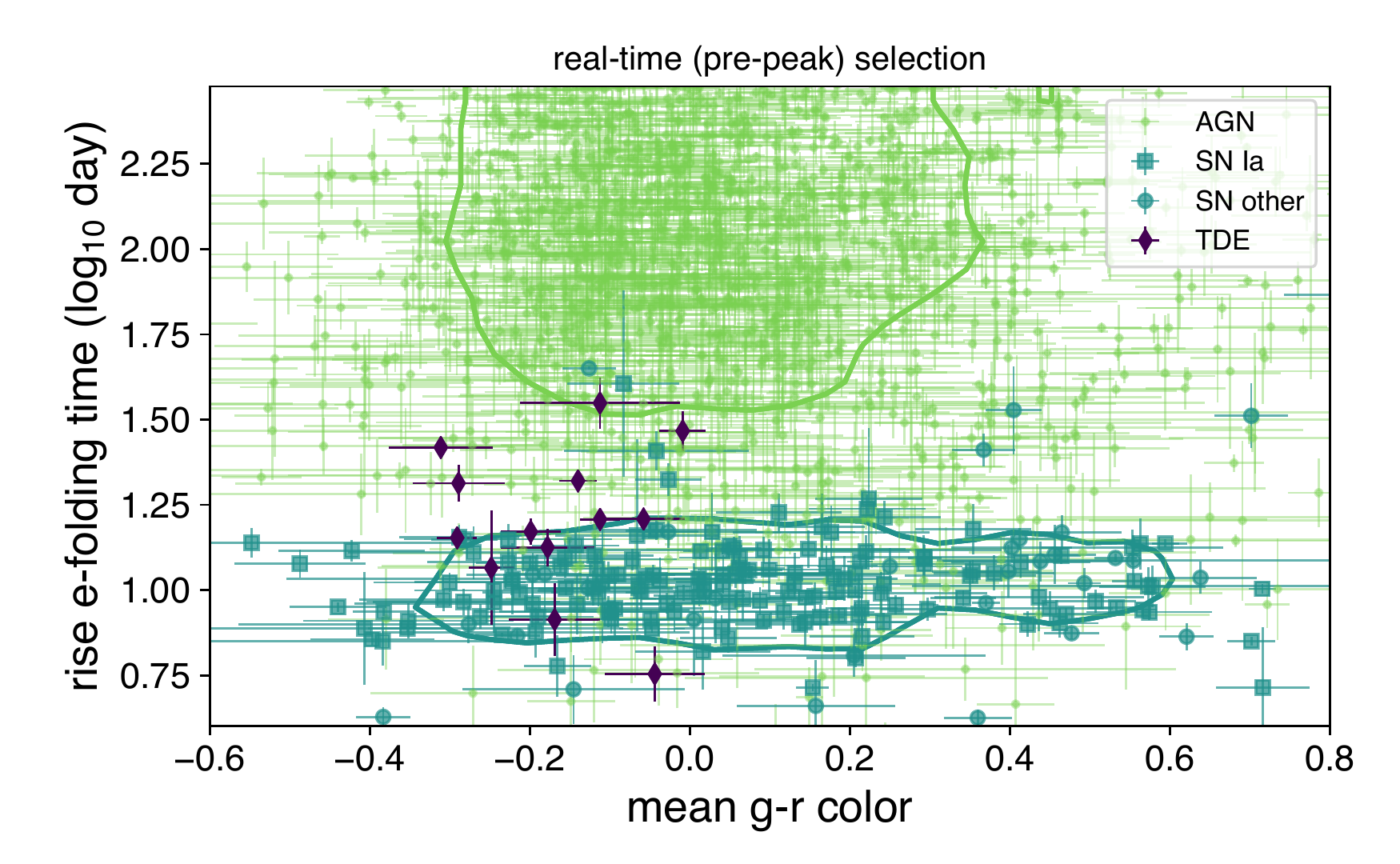}{0.48 \textwidth}{}\\[-20 pt]}
\caption{Yield of nuclear transients after 1.5 years of ZTF observations. Contours enclose two thirds of all spectroscopically classified nuclear supernovae (SNe) in our sample and two thirds of the AGN. The latter are classified based on archival data or prior variability. In the top panel we see that that TDEs have both longer rise times and a longer fading timescale compared to the majority of SNe. The middle panel demonstrates that color evolution provides further separation of TDEs from SNe. Here we display the mean $g-r$ color and the color change ($\Delta(g-r)/t$), both measured using all detections of the light curve. Tidal disruption flares show an almost constant optical color, while in post-peak observations most SNe show cooling (i.e., an increase of the color). For photometric selection of TDEs detected before maximum light, their blue color and slow rise time of can be used (bottom panel), although this metric yields a larger background of SNe. }\label{fig:photosel}
\end{figure}

\section{Candidate selection and classification}\label{sec:cand}

\subsection{Zwicky Transient Facility}
Our search for new TDEs is done exclusively using ZTF data. The strength of ZTF \citep{Graham19} is a combination of depth ($m\approx 20.5$ per visit) and area (47~deg$^2$ field of view). Most of our sources originate from the public MSIP survey, which aims \citep{Bellm19a} to visit the entire visible Northern sky every 3 nights in both the $g$ and $r$ filters. The use of two filters is an essential ingredient to our TDE selection pipeline, since it allows for efficient photometric filtering (Fig.~\ref{fig:photosel}) to narrow down the number of targets for spectroscopic follow-up observations.

\subsection{ZTF Alert Filtering}
We use the information from the data stream \citep{Patterson19} of ZTF alerts, which contains the difference imaging photometry and astrometry of transients and variable sources \citep{Masci19}. 

Except for rejecting galaxies that can be classified as broad-line AGN, we place no requirement on the host galaxy type. For AGN identification we use the Million quasar catalog \citep[][v5.2.]{Flesch15}. In addition, we construct a light curve from the neoWISE \citep{Mainzer11} photometry and reject any galaxies with significant variability ($\chi^2/{\rm dof}>10$) or a mean W1-W2 color that exceeds the AGN threshold of \citet{Stern12}. Our filter is executed by Ampel \citep{Nordin19}, which includes fast catalog matching by catsHTM \citep{Soumagnac18a}, and we use the GROWTH marshal \citep{Kasliwal19} to coordinate our follow-up observations and spectroscopic classifications. %We use catsHTM \citep{Soumagnac18} for efficient and fast catalog cross matching. 

Compared to our TDE search in ZTF commissioning data \citep{vanVelzen18_NedStark}, we use a more liberal cut on the star-galaxy score \citep{Tachibana18} of $<0.8$. This increases the galaxy sample at the cost of a much higher background due to bright variable stars (these often have a score equal to 0.5 due to issues with the PS1 photometry for bright and variable objects). We therefore veto the star-galaxy score if the source has a detected parallax in Gaia DR2 \citep{Gaia-Collaboration16,Gaia-Collaboration18} or if the ratio of the Gaia $G$-band flux to the PS1 PSF {$g$,$r$,$i$} flux (converted to the $G$-band) is consistent with a point source. Since we require a match to a known source in the ZTF reference image, we can use a relatively liberal cut on the real-bogus score \citep{Mahabal19} of 0.3. 

As demonstrated in Fig.~\ref{fig:photosel}, TDEs can be discriminated from SNe and AGN based on their rise/fade timescale, $g$-$r$ color, and lack of color evolution. We rank photometric TDE candidates for spectroscopic follow-up based on their distance from the locus of SNe and AGN these  photometric properties.  In general, we rejected transients that are significantly off-center (mean offset $>0.4"$), or have significant $g-r$ color evolution ($d(g-r)/dt>0.015$~day$^{-1}$), or show only a modest flux increase when comparing the difference flux to the PSF flux in the ZTF reference image ($m_{\rm diff}-m_{\rm ref}>1.5$). We also rejected all objects that can be classified as SNe or broad-line AGN in our spectroscopic follow-up observations. The details of our photometric selection, including estimates for the completeness and selection effects which are required to compute rates, will be presented in a forthcoming publication.

\begin{table*}
\begin{center}
\caption{Spectroscopic Observations and TDE classification}\label{tab:spectra}
\begin{tabular}{l l c l l l l}
\hline
IAU Name & Date & Phase & Telescope/Inst. & TDE class & Redshift & ID\\
\hline
AT2018zr & 2018 Mar 28 &  25 &  WHT/ISIS\tablenotemark{a}     & TDE-H      & 0.075   & 1 \\
AT2018bsi & 2018 May 13  & 34 & DCT/DeVeny                    & TDE-Bowen  & 0.051   & 2 \\
AT2018hco & 2018 Nov 10  & 29 & Keck/LRIS                     &  TDE-H     & 0.088   & 3 \\
AT2018iih & 2019 Mar 10  & 102 & DCT/DeVeny                   & TDE-He     & 0.212   & 4 \\
AT2018hyz & 2018 Nov 12  & 6 & FTN/Floyds-N\tablenotemark{b}  & TDE-H      & 0.0458  & 5 \\
AT2018lni & 2019 Mar 01 & 81 & DCT/DeVeny                     & TDE-Bowen  & 0.138   & 6 \\
AT2018lna & 2019 Jan 26  & 0 & Palomar/DBSP                   & TDE-Bowen  & 0.091   & 7 \\
AT2019cho & 2019 May 02  & 58 & DCT/DeVeny                    & TDE-Bowen  & 0.193   & 8 \\
AT2019bhf & 2019 May 29 & 90 & DCT/DeVeny                     & TDE-H      & 0.1206  & 9 \\
AT2019azh & 2019 May 01  & 46 & Keck/LRIS                     & TDE-Bowen  & 0.022   & 10\\
AT2019dsg & 2019 May 13 & 13 & NTT/EFOSC2\tablenotemark{d}    & TDE-Bowen  &0.0512   & 11\\
AT2019ehz & 2019 Jun 14 & 35 & Lick/Kast                      & TDE-H      & 0.074   & 12\\
AT2019eve & 2019 Jun 29 & 50 & DCT/DeVeny                     & TDE-H      & 0.064   & 13\\
AT2019mha & 2019 Aug 27  & 17 & Palomar/DBSP                  & TDE-H      & 0.148   & 14\\
AT2019meg & 2019 Aug 10  & 8 & Palomar/DBSP                   & TDE-H      & 0.152   & 15\\
AT2019lwu & 2019 Aug 27 & 31 & DCT/DeVeny                     & TDE-H      & 0.117   & 16\\
AT2019qiz & 2019 Nov 05 & 29 & DCT/DeVeny                     & TDE-Bowen  & 0.0151  & 17\\
\hline
\end{tabular}
\\[-10pt]
\end{center}
$^{a}$Spectrum published in \citep{Hung19}. 
$^{b}$Publically available spectrum on TNS posted by \citet{2018ATel12198....1D}. 
$^{d}$Publically available spectrum on TNS posted by \citet{2019ATel12752....1N}. 
\end{table*}

\begin{figure*}
\gridline{	\fig{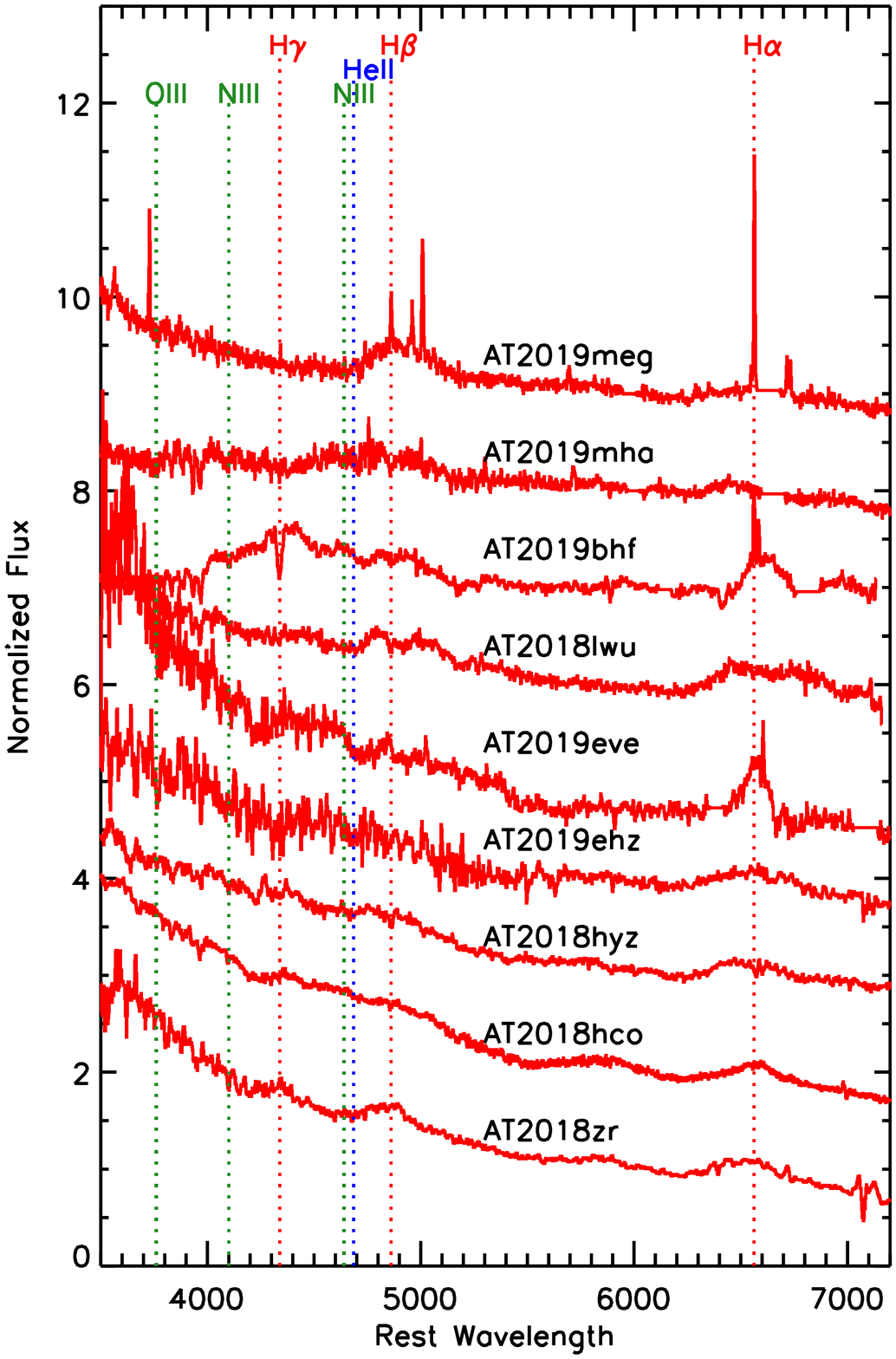}{0.48 \textwidth}{}
			\fig{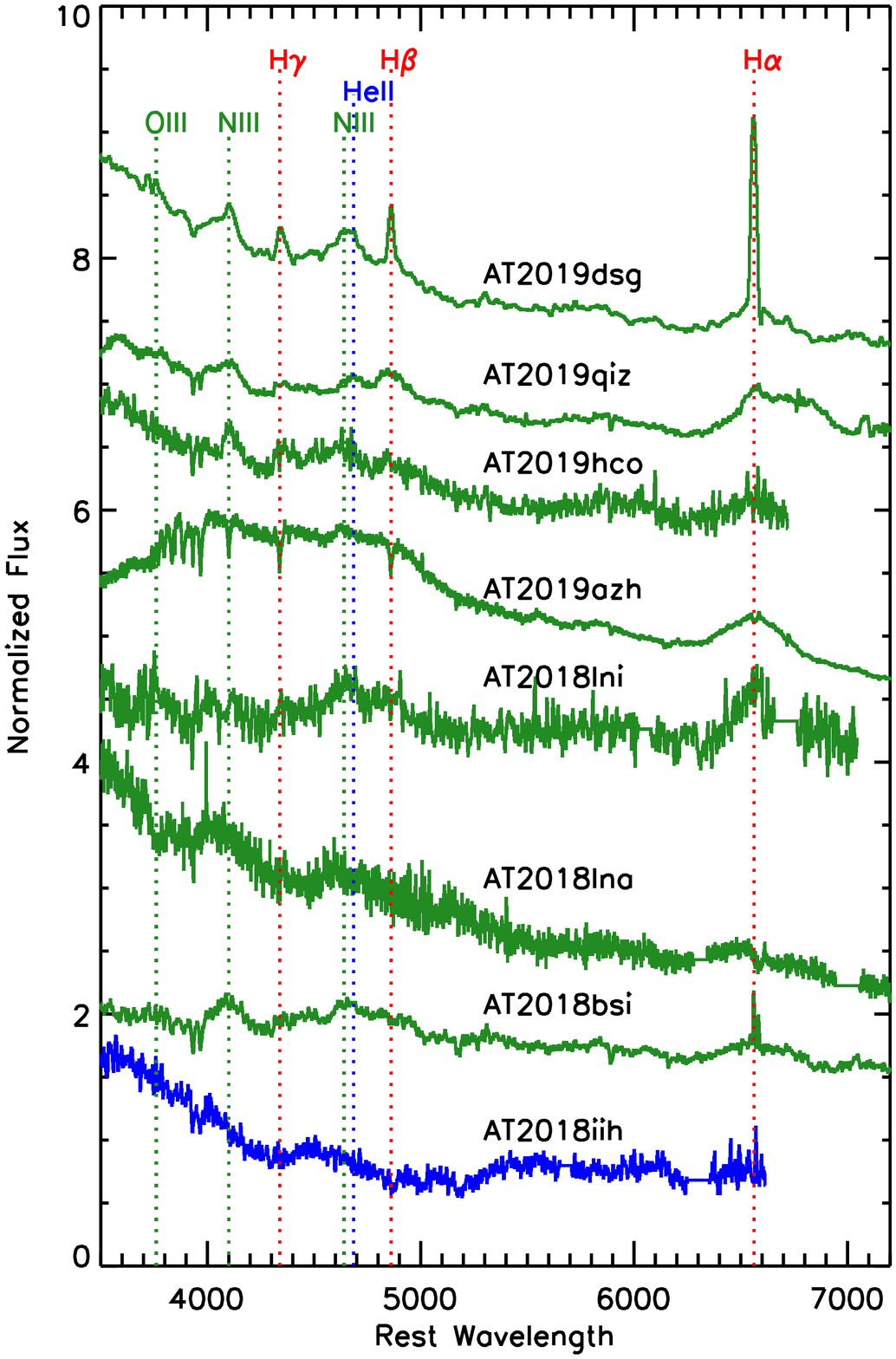}{0.48 \textwidth}{} \\[-20pt]}
\caption{Spectroscopic classifications of our ZTF TDE sample from medium resolution spectroscopy. {\it Left:} TDEs with Balmer line features only (TDE-H, in red). {\it Right:} TDEs with Balmer and He II, and N III and O III Bowen fluorescence features (TDE-Bowen, in green), He II only (TDE-He, in blue).  Spectra have not been host galaxy subtracted.}\label{fig:spec}
\end{figure*}

\subsection{Discovery and classification history}
In Table~\ref{tab:discovery} we list the IAU name, the ZTF name, our internal nickname\footnote{Given the 7-character length of the ZTF names required by the large volume of ZTF transient alerts, for ease of communication, we chose an internal naming scheme for our TDE candidates based on characters from the HBO TV show, Game of Thrones (GOT)}, the name given by other optical transient surveys, and reference to the first public spectroscopic classification of this transient as a TDE. The table is sorted by the date of the first ZTF detection and credit for discovery of the transient, based on the first report to the Transient Name Server (TNS), is indicated using bold face.

Our TDE discovery pipeline does not use the TNS as input, we read and filter the ZTF alerts directly from their source \citep{Patterson19}. The TNS reporting of ZTF alerts is mainly provided by AMPEL \citep{Nordin19} and by the Redshift Completeness Factor project \citep{Fremling19}, plus more recently by a filter implemented in the ALeRCE broker. For 10 of the 17 sources in our sample ZTF was the first survey to report a detection to TNS. As listed in Table~\ref{tab:discovery}, ATLAS provided 4 discoveries,  ASAS-SN 2 discoveries, and Gaia and PS1 each claim one more discovery.

%Some cases are ambiguous. For example, a spectroscopic observation for the transient \Jaime based on the ZTF data was obtained by Heikkila et al.~(2019) {\it before} any photometric detection was reported to TNS, but they reported their spectral follow-up in ATel\#12529 after the ASAS-SN survey reported their photometric detection in ATel\#12526.   

\begin{table}
\caption{Known TDEs included in population analysis}\label{tab:known}
\begin{tabular}{l c l l}
\hline
Discovery name/IAU name  & Ref. & Spectral type 	& ~~~$z$ \\
\hline
GALEX-D1-9               & 1  & No spectrum    & 0.326 \\
GALEX-D3-13              & 2  & No spectrum    & 0.3698 \\
GALEX-D23H-1             & 2  & No spectrum    & 0.1855 \\
SDSS-TDE1                & 3  & No spectrum    & 0.136 \\
SDSS-TDE2                & 3  & TDE-H         & 0.256 \\
PS1-10jh                 & 4  & TDE-He        & 0.1696 \\
PS1-11af                 & 5  & Featureless    & 0.4046 \\
PS17dhz/AT2017eqx        & 6  & TDE-Bowen       & 0.1089 \\
PTF-09ge                 & 7  & TDE-He        & 0.064 \\
PTF-09axc                & 7  & TDE-H         & 0.1146 \\
PTF-09djl                & 7  & TDE-H         & 0.184 \\
ASASSN-14ae              & 8  & TDE-H         & 0.0436 \\
ASASSN-14li              & 9  & TDE-Bowen       & 0.0205 \\
ASASSN-15oi              & 10 & TDE-He       & 0.0484 \\
ASASSN-15lh              & 11 & Unknown        & 0.2326 \\
ASASSN-18pg/AT2018dyb    & 12 & TDE-Bowen       & 0.018 \\
ASASSN-18ul/AT2018fyk    & 13 & TDE-Bowen       & 0.059 \\
ASASSN-19bt/AT2019ahk    & 14 & TDE-H         & 0.0262 \\
iPTF-15af                & 15 & TDE-Bowen       & 0.0789 \\
iPTF-16axa               & 16 & TDE-Bowen       & 0.108 \\
iPTF-16fnl               & 17 & TDE-Bowen      & 0.0163 \\
OGLE16aaa                & 18 & Unknown   	   & 0.1655 \\
\hline
\end{tabular}
1. \citet{Gezari06}; 2. \citet{Gezari08}; 3. \citet{vanVelzen10}; 4. \citet{Gezari12}, 5. \citet{Chornock14}, 6. \citet{Nicholl19}, 7. \citet{Arcavi14}, 8. \citet{Holoien14}; 9. \citet{Holoien16}; 10. \citet{Holoien16b}; 11. \citet{Dong16}; 12. \citet{Leloudas19}; 13. \citet{Wevers19a}; 14. \citet{Holoien19}; 15. \citet{Blagorodnova19}; 16. \citet{Hung17}; 17. \citet{Blagorodnova17}; 18. \citet{Wyrzykowski17}. 
\end{table}

\subsection{Spectroscopic classification}\label{sec:specclass}
%For xx SEDM??. 
In order to classify the TDEs into spectroscopic sub-classes, we use the ``best" spectrum, high signal to noise and prominent line features, for each of our TDEs from our various follow-up programs with: the 4.3m Discovery Channel Telescope De Veny Spectrograph (DCT/DeVeny, PI: Gezari), the 200in Palomar Telescope Double Spectrograph (P200/DBSP, PI: Kulkarni), the 10m Keck Low Resolution Imaging Spectrograph (Keck/LRIS, PI: Kulkarni), and the 3m Lick Kast Double Spectrograph (Lick/Kast, PI: Foley).  Spectra were reduced with \texttt{PyRAF} using standard long-slit spectroscopy data reduction procedures.  For those spectra not corrected for telluric absorption, we show the spectra in Figure \ref{fig:spec} with those wavelength regions masked out.  In three cases, we use publically available spectra from the Transient Name Server (TNS).  In Table \ref{tab:spectra} we indicate the IAU name, date, phase in days since peak, and telescope and instrument of the spectrum we use for determining the spectroscopic sub-classification shown in Figure \ref{fig:spec}, the TDE class, and the redshift.

We find that our ZTF TDE sample can be divided into three spectroscopic classes:  
\begin{itemize}
    \item[\it i.]  {\it TDE-H}: broad H$\alpha$ and H$\beta$ emission lines. 
    
    \item[\it ii.] {\it TDE-Bowen}: broad H$\alpha$ and H$\beta$ emission lines, a broad complex of emission lines around He~II$~\lambda4686$ and N~III~$\lambda4640$ and emission at $\lambda4100$ identified as N~III~$\lambda4100$ instead of H$\delta$, and in some cases also O~III~$\lambda3760$.  
    
    \item[\it iii.] {\it TDE-He}: no broad Balmer emission lines, a broad emission line near He~II~$\lambda4686$ only. 
\end{itemize}

In our flux-limited sample of TDEs with ZTF observations the relative ratios of the classes are H:Bowen:He = 9:7:1. In \S \ref{sec:discussion} we will elaborate on how the rarity of the TDE-He class might be an important clue to understand what conditions are needed to provide the spectroscopic properties of TDEs.  Two of the TDEs (\Margaery~ and \Bran) also have strong narrow emission lines from star-formation in their host galaxies.  Note that these classifications are based on a single spectral epoch.  There is at least one case in which a TDE showed the late-time disappearance of H$\alpha$ emission \citep{Nicholl19}, which according to our classification scheme, would result in a change of spectral class from TDE-Bowen to TDE-He.  A detailed analysis of the spectroscopic evolution of the ZTF TDEs and their line features with time will be presented in a future paper (Hung et al.~in prep.).

\begin{figure}
\includegraphics[width=0.40 \textwidth]{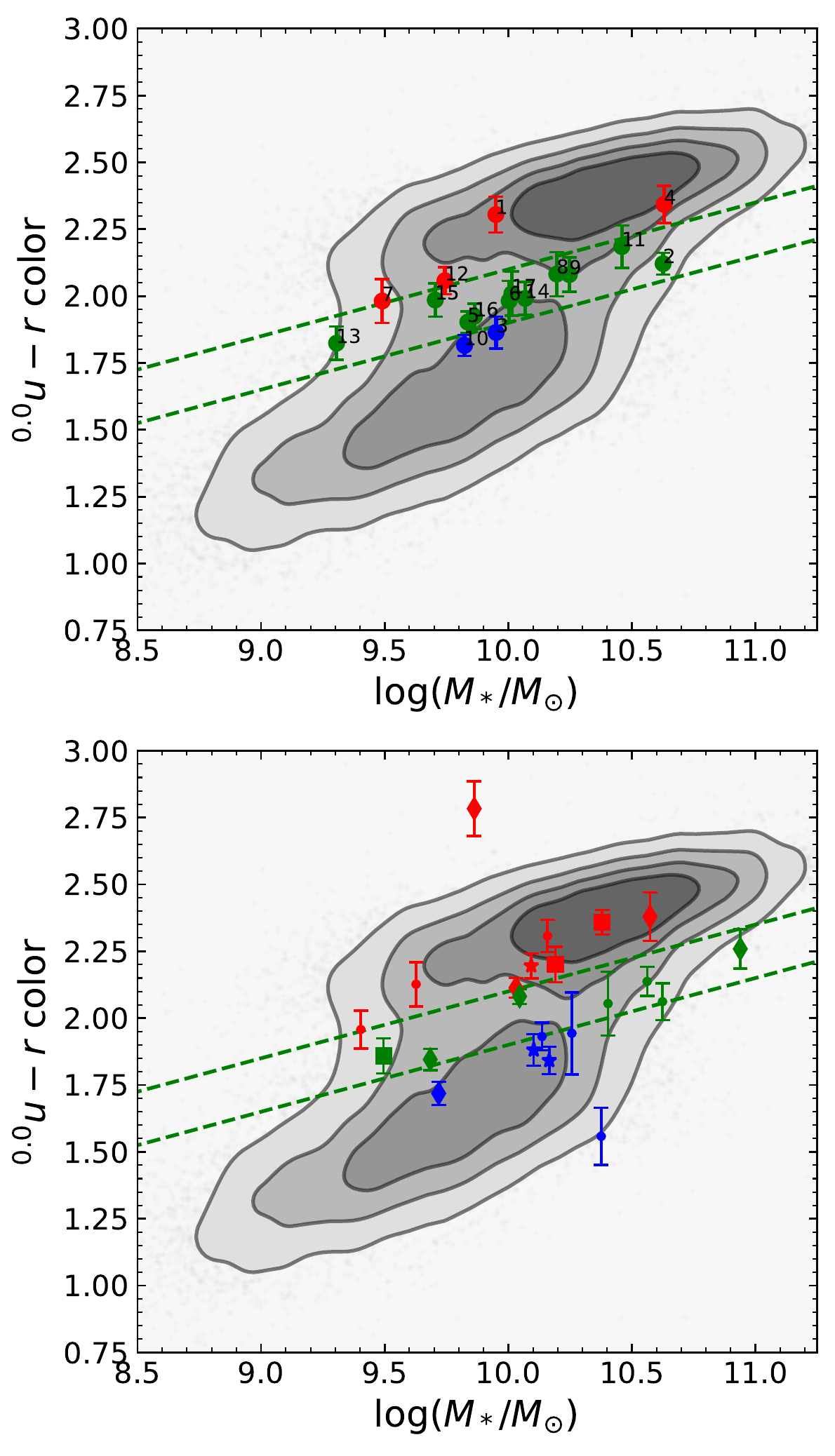}
\caption{The extinction-corrected rest-frame $u-r$ color and the total stellar mass, both  obtained from the best-fit population synthesis model. In the top panel we show the 17 TDE host galaxies (labeled by their corresponding number in the ID column of Table~\ref{tab:spectra}). In the bottom we show known TDEs from the literature (Table \ref{tab:known}), with diamonds indicating ASAS-SN sources, squares for iPTF sources, stars for PTF sources, and dots for the remaining surveys. Colors of symbols correspond to the spectral classifcations given in Tables \ref{tab:spectra} and \ref{tab:known}, with TDE-H in red, TDE-Bowen in green, TDE-He in blue, and either unclassified or featureless in black. The contours enclose a volume-limited comparison sample of galaxies, matched to the depth of ZTF, decreasing in step of 0.5$\sigma$ to 2$\sigma$ for the outer contour. We see that host galaxies in the green valley are significantly over-represented in both samples. 
}\label{fig:urhost}
\end{figure}

\section{Host Galaxy Properties}\label{sec:host}
The stellar mass of the TDE host galaxies is estimated from the pre-flare photometry of the host. For most of our sources, no pre-flare spectroscopic observations of the host are available and the redshift is obtained from the spectrum of the TDE. We use SDSS model magnitudes \citep{stoughton02} or Pan-STARRS Kron magnitudes \citep{Chambers16} for sources outside the SDSS footprint. We also include GALEX NUV and FUV photometry \citep{Martin05}, both detections or upper limits. 

To obtain a posterior distribution for parameters of the flexible stellar population synthesis \citep[FSPS][]{Conroy09} module we use the Prospector \citep{Johnson17} software to run a Markov Chain Monte Carlo (MCMC) sampler \citep{Foreman-Mackey13}. We adopted the same model choices that were used by \citet{Mendel14}, who applied the FSPS module to the SDSS galaxy sample. The 5 free parameters are the stellar mass, the \citet{Calzetti00} dust model optical depth, the age of the stellar population, the metalicity ($Z$), and the e-folding time of the star formation history ($\tau_{\rm sfh}$).  We use flat priors over the same parameter range as \citet{Mendel14}. Sufficient sampling of the posterior is ensured by using only the second half of 1000 steps, taken by 100 walkers. 
%We follow the same method as \citet{vanVelzen18_FUV}.  

\begin{figure*}
    \includegraphics[width=\textwidth]{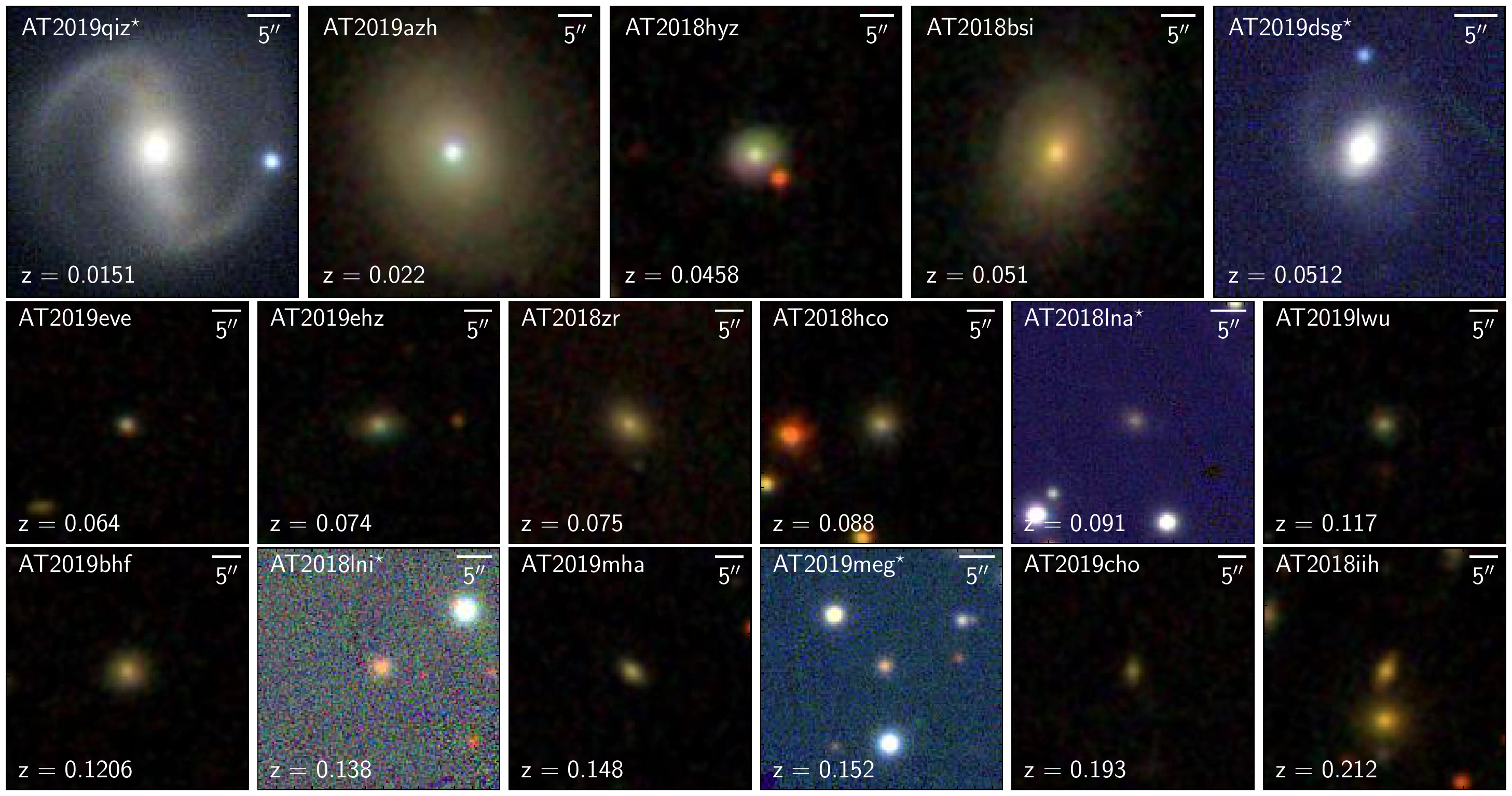}
    \caption{SDSS \textit{gri} images of the TDE host galaxies in order of increasing redshift. Galaxies with a star next to the name are not in the SDSS footprint and therefore have Pan-STARRS \textit{gri} images. All images are $34\arcsec\times34\arcsec$.}
    \label{fig:hostcutout}
\end{figure*}

Figure \ref{fig:urhost} shows the extinction-corrected, synthetic rest-frame $u-r$ color vs. total stellar mass for the TDE host galaxies from the stellar population synthesis fits to the pre-flare spectral energy distributions (SEDs) described above, together with a sample of approximately 17000 comparison galaxies from SDSS with observed $u-r$ colors. Figure \ref{fig:urhost} also shows the sample of known TDEs listed in Table \ref{tab:known} with the same comparison sample of SDSS galaxies. This comparison sample is based on the \citet{Mendel14} value added catalog of bulge, disk, and total stellar mass estimates.
%and the \citet{simard11} value added catalog of bulge+disk decompositions. 
% The \citet{simard11} decompositions are available for galaxies with an $r$-band Petrosian magnitude to be between 14 and 18. 
This catalog contains spectroscopically classified  galaxies \citep{strauss02}, with mass estimates based on FSPS (i.e., the same software we used for the TDE host galaxies). In Figure \ref{fig:hostcutout} we show cut-outs of the host galaxy color $gri$ images from SDSS (or Pan-STARRS1 when SDSS is not available) for the 17 ZTF TDE galaxy hosts, in order of increasing redshift.  The morphology of the host galaxies appears to be dominated by an elliptical component, typically for early-type galaxies; some of the lowest redshift TDE hosts ($z = 0.015 - 0.05$) show a compact core and an extended spiral and/or disk structure.

Our highest redshift TDE with ZTF observations is at $z=0.21$, which implies a nuclear transient search with ZTF is sensitive to a volume-complete sample of galaxies with $M_r\lesssim-18$. To match this absolute magnitude limit, we restrict the comparison catalog from SDSS by applying a redshift of $z < 0.04$. We have indicated the location of the green valley in Figure \ref{fig:urhost}, as originally defined in \citet{schawinsk14}. However, as our sample has a different redshift cut, we have redefined the upper bound of the green valley based on our galaxy distribution:
\begin{equation}\label{eq:greenval}
    ^{0.0}u-r(M_{\rm gal}) = -0.40 +0.25\times M_{\rm gal}
\end{equation}
but kept the width of the green valley fixed to that of \citet{schawinsk14}, of 0.2 mag, to define the lower bound.

The ZTF sample of TDE host galaxies is dominated by green valley galaxies (Fig.~\ref{fig:urhost}), with $\approx 65$\% of the TDE hosts falling within the limits of the green valley region compared to 13\% of the SDSS comparison sample. \citet{Law-Smith17} used the definition of the green valley based on total star formation rate and found that their sample of TDE host galaxies may be transitioning from star-forming to quiescent, a time during which quenching of star formation causes galaxies to cross into the green valley \citep{schawinsk14}. The green valley is also known to host quiescent, Balmer-strong galaxies (including post-starburst or E+A galaxies), which previous studies have shown to be overrepresented in TDE host galaxy populations \citep{Arcavi14,French16, Law-Smith17, Graur17}. Late-time spectra of the ZTF host galaxies can be used to obtain better star-formation rate estimates, but these are not yet available for the entire sample (in some case the flare still dominates the optical emission). %can be used investigate the star formation histories of the hosts in order to further understand the host galaxy preferences of TDEs.

%\clearpage

\section{Follow-up observations}\label{sec:multiw}

%\subsection{Radio: VLA}

% \subsection{Infrared: Spitzer} 

\subsection{Optical: SEDM and LT}
For a few TDEs we acquired multi-band images with P60/SEDM \citep{Blagorodnova18,Rigault19} and/or the optical imager (IO:O) on the Liverpool Telescope \citep[LT;][]{ssr+04}. For LT data, image reductions were provided by the IO:O pipeline. For both LT and SEDM, image subtraction was performed versus PS1 ($g’$,$r’$,$i’$,$’z$-bands) or SDSS ($u’$-band) reference imaging, following the techniques of \citet{fst+2016}. PSF photometry was performed relative to PS1/SDSS photometric standards. 

\subsection{UV: Swift/UVOT}\label{sec:uvot}

All of our 17 ZTF TDEs have \textit{Neil Gehrels Swift Observatory} \citep{Gehrels04} follow-up observations in the UV and X-ray from UVOT \citep{Roming05} and XRT \citep{Burrows05}, respectively, with a typical cadence of $3-5$ days, and in most cases triggered within 2 weeks of the peak.  Some of the fainter TDEs have only a few epochs of {\it Swift} observations, but they are sufficient to measure the average temperature of the UV/optical component, given that the optical colors of these TDEs are relatively constant with time.

The {\it Swift} photometry was measured using the \texttt{uvotsource} package with an aperture of 5 arcsec, in the AB system, and corrected for the enclosed energy within the aperture (for \Jaime \Jon, and \Bran we use a larger aperture to make sure we capture all the flux of the host galaxy). We estimate the host galaxy flux in the UVOT bandpass from the posterior distribution of the population synthesis models. The uncertainty on this baseline level is propagated into our measurement of the TDE flux. Our host-subtracted UVOT aperture photometry, as well as the ZTF, SEDM, and LT photometry, will be made available at the publisher website. 

\subsection{X-ray: Swift/XRT}

The 0.3-10 keV X-ray light curves for the four TDEs with XRT detections were produced using the UK Swift Data center online XRT data products tool, which uses the HEASOFT v6.22 software \citep{Arnaud96}.  We used a fixed aperture at the ZTF coordinate of the transient, and converted to flux using the best fit blackbody model to the stacked XRT spectrum.  The XRT stacked spectra were processed by the XRT Products Page \citep{Evans09}, with Galactic extinction fixed to values from the HI4PI survey \citep{2016A&A...594A.116H}: $N_{\rm H}/10^{20}\,$cm$^{-2}$ = 2.59, 6.46, 1.42, and 4.16, for \Gendry, \Bran, \Brienne, and \Jaime, respectively. The resulting temperatures are $kT$/keV = $0.132 \pm  0.026$, $0.071 \pm  0.003$, $0.101 \pm  0.004$, and  $0.053 \pm 0.001$, again for \Gendry, \Bran, \Brienne, and \Jaime, respectively (uncertainties correspond to the 90\% confidence levels). 
%which an ``unabsorbed flux" in the 0.3-10 keV band of 3.10$10^{-11}$, 6.87$10^{-11}$, 3.18$10^{-11}$, and 6.27$10^{-11}$ erg s$^{-1}$ cm$^{-2}$ for \Gendry, \Bran, \Brienne, and \Jaime, respectively. 
 These soft blackbody temperatures are similar to the previously known X-ray detected optically-selected TDEs: ASASSN-14li ($kT = 0.050$~keV; \citealt{Miller_14li}), ASASSN-15oi ($kT = 0.045$~keV; \citealt{Gezari17}), AT2018zr/PS18kh ($kT =0.10$~keV; \citealt{vanVelzen18_NedStark}), AT2018fyk/ASASSN-18ul ($kT = 0.12$~keV; \citealt{Wevers19}).

\begin{deluxetable}{l l l}

\tablewidth{0pt}
\tablecolumns{7}
\tablecaption{Priors for MCMC light-curve analysis.}\label{tab:priors}
\tablehead{Parameter &  Description &Prior}
\startdata
$\log L_{\rm peak}$ & Peak luminosity  &  [$L_{\rm max}/2$, $2 L_{\rm max}$]\tablenotemark{a}  \\
$t_{\rm peak}$ & Time of peak &  [-20, 20]~day\tablenotemark{a}  \\
$\log T_{0}$ & Mean temperature  &  [4,5] Kelvin \\
$\log \sigma$ & Gaussian rise time  &   [0,1.5] day  \\
$\log \tau$ & Exponential decay time  &   [0,3] day \\
$ p$ & Power-law index  &   [-5,0]  \\
$ \log t_{0}$ & Power-law normalization  &   [0,3] day \\
$ dT/dt$ & Temperature change  &   [$-200$,200] K\,day$^{-1}$ \\
$ \ln(f)$ & White noise factor  &   [$-5$,$-1.8$]  \\
\enddata
\tablenotetext{a}{$L(t=0) \equiv L_{\rm max}$ is the observed maximum luminosity. When we fit for the blackbody luminosity (Eq.~\ref{eq:pl}), we compute $L_{\rm max}$ using the mean temperate measured for the first 100~days since peak (i.e., $T_0$ as obtained from fitting Eq.~\ref{eq:exp}).  }
\end{deluxetable}

\section{Light curve analysis}\label{sec:lc}
We extract the properties of the light curve by fitting a model to the data from ZTF, {\it Swift}/UVOT, and if available, LT and SEDM. We also include known TDEs from the literature, selecting all source used in the luminosity function analysis of \cite{vanVelzen18} as well as spectroscopically confirmed TDEs that have been published since. We use the published photometry and our own analysis of the public {\it Swift}/UVOT data (cf. \S \ref{sec:uvot}). We add all TDEs listed in Table~\ref{tab:known} to obtain a total of 39 sources.  

We consider two models to describe the TDE light curve. First, for the first 100 days after maximum light we use a Gaussian rise and exponential decay:
\begin{align}
L_\nu(t) &= L_{\nu_0\, \rm peak}~\frac{B_\nu(T_0)}{B_{\nu_0}(T_0)}  \nonumber \\
    &\times \begin{cases}  e^{-(t-t_{\rm peak})^2/2\sigma^2} & t\leq t_{\rm peak} \\ 
    e^{-(t-t_{\rm peak})/\tau} & t>t_{\rm peak}\\
    \end{cases}\label{eq:exp}
\end{align} 
Here $L_{\nu_0\,\rm peak}$ is the peak luminosity, measured at the reference frequency $\nu_0$ (in the rest-frame of the source). To predict the luminosity in other bands we assume the spectrum follows a blackbody, $B_\nu(T_0)$ with a constant temperature $T_0$. We pick the $g$-band ($6.3\times 10^{14}$~Hz) as our reference frequency. We adopt $T_0$ as our default temperature measurement and we use this temperature to estimate the maximum bolometric luminosity ($L_{\rm bb}$) and blackbody radius ($R$). 

One advantage of Eq.~\ref{eq:exp} is simplicity: measuring the rise/decay timescale independently with multi-band observations would not be possible with fewer free parameters. However,  for observations longer than 100~days, all TDEs show deviations from an exponential decay. A power-law is required to properly describe the light curve which introduces an extra free parameter. In our second light-curve model, we therefore use a power-law decay and also allow for evolution of the blackbody temperature.  
\begin{align}
\nonumber
L(t,\nu) &= L_{\rm peak}~\frac{\pi B_\nu(T(t))}{\sigma_{\rm SB} T^4(t)} \\
    &\times \begin{cases}  e^{-(t-t_{\rm peak})^2/2\sigma^2} & t\leq t_{\rm peak} \\ 
     [(t-t_{\rm peak})/t_0]^p & t>t_{\rm peak}\\ 
    \end{cases}    
\label{eq:pl}
\end{align}
Here $\sigma_{\rm SB}$ is the Stephan-Boltzmann constant. While this model allows for temperature evolution, we cannot measure this at the same cadence as the observations because (i) the {\it Swift} and ZTF observations were not obtained simultaneous and (ii) the uncertainty on the temperature estimated from a single (near-simultaneous) epoch are often very large. We therefore first try a simple linear relation for the post-peak temperature evolution:
\begin{equation}\label{eq:Tevo}
T(t) = (T_0) + dT/dt \times (t-t_{\rm peak}) \quad.
\end{equation} 
Here $dT/dt$ is a free parameter with units Kelvin/day and we enforce ${\rm min}(T)>10^4$~K and ${\rm max}(T)<10^5$~K. This simple model works well for most TDEs, since they only show very modest temperature evolution (about 50 K/day, which corresponds to a 20\% increase over 100 days, see Fig.~\ref{fig:RT})
However for some sources more rapid temperature changes have been observed \citep[e.g.,][]{Holoien18}. 

To allow for more flexibility in our description of temperature evolution we use linear interpolation of the temperature on a grid of fixed points in time. The points on this grid are the free parameters of our fit. The grid starts at the time of peak and the spacing is $\pm 30$ days. At each grid point we use a log-normal Gaussian prior with a dispersion of 0.1~dex centered on the best-fit mean temperature from our simplest light-curve model (Eq.~\ref{eq:exp}). We adopt this non-parametric approach as our default model to estimate the parameters of the power-law ($p$, and $t_0$) as well as the rise timescale. 

We apply our single power-law decay model (Eq.~\ref{eq:pl}) only the first year of data (measured after maximum light) because at later times many TDE light curves show significantly flattening that is not consistent with the early-time power-law decay \citep{vanVelzen18_FUV}, likely due to the contribution of an accretion disk to the optical/UV light. 

To estimate the parameters of our two models we use the \texttt{emcee} sampler \citep{Foreman-Mackey13}. Following \citet{vanVelzen18_FUV}, we use a Gaussian likelihood function that includes a ``white-noise'' term, $\ln(f)$, which allows for additional variance in the data that is not captured by the reported measurement uncertainty. We use 100 walkers and 2000 steps, discarding the first 1500 steps to ensure convergence. We use a flat prior for all parameters (except the grid points that anchor the temperature evolution): the boundaries of the parameters are listed in Table~\ref{tab:priors}. 

%We make an exception for sources with less than three $u-$band or UV observations, because in this case the temperature evolution is very poorly constrained. For these sources we force the temperature to be constant using delta-function prior. 
An exception is made for sources with no detections prior to maximum light, i.e., TDEs discovered post peak. For these we force $t_{\rm peak}=0$ when measuring $L_{\rm peak}$ using Eq.~\ref{eq:exp}. However we always use the default priors (Table~\ref{tab:priors}) when estimating the best-fit parameters of the power-law decay (Eq.~\ref{eq:pl}), since this allows the uncertainty on the true time of peak to enter the posterior distributions of power-law parameters. Finally we also make an exception for the 3 TDEs from PTF (Table~\ref{tab:known}). These are the only sources with light curves that have no UV coverage in the first year and we therefore keep their blackbody temperature fixed at the value measured from the optical spectrum by \citet{Arcavi14}. 

To estimate the blackbody radius and blackbody luminosity at peak, we sample the posterior distribution of $T_{0}$ and $L_{\rm peak}$ as obtained from the model of Eq.~\ref{eq:exp}. For all parameters, the reported uncertainty follows from a credible interval of [0.16, 0.84], i.e., $\pm 1\sigma$ for Gaussian statistics.

The ZTF light curve of two sources in our sample was included in the reference image, hence the IPAC difference-imaging light curves are compromised and excluded from the light-curve analysis. In one case (\Jon), both $r$-band and $g$-band light curves are affected, but we were able to use alerts based on an earlier, TDE-free, $r$-band reference frame that had been created during the ZTF commission period. For the second source (\Gendry) only the $g$-band light curve is affected. For both sources, sufficient {\it Swift}/UVOT photometry is available to obtain a good estimate (uncertainty $<0.1$~dex) of the light curve features. 

\begin{figure}
\includegraphics[trim=5mm 2mm 8mm 1mm, width=0.48 \textwidth]{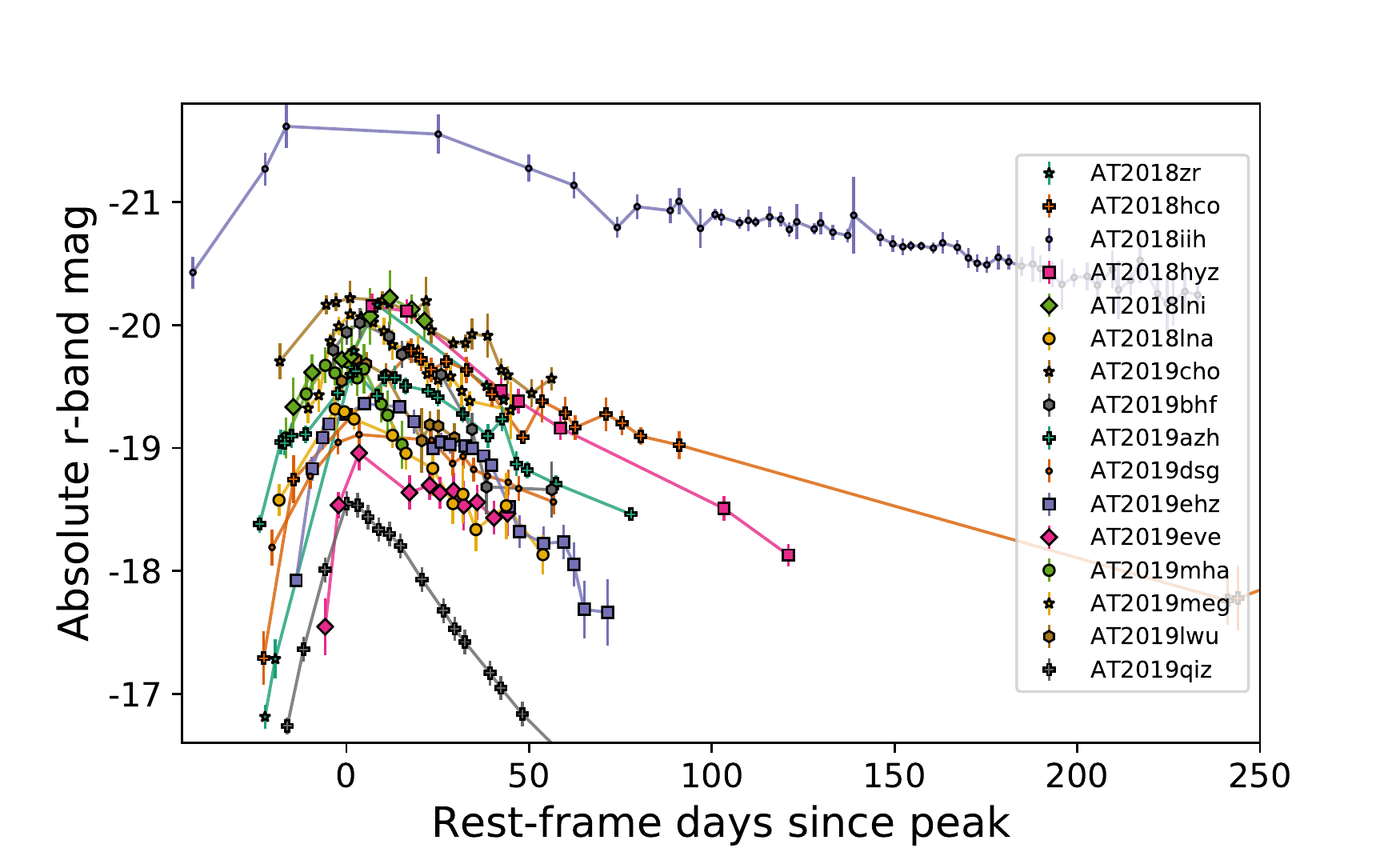}

\includegraphics[trim=5mm 2mm 8mm 1mm, width=0.45 \textwidth]{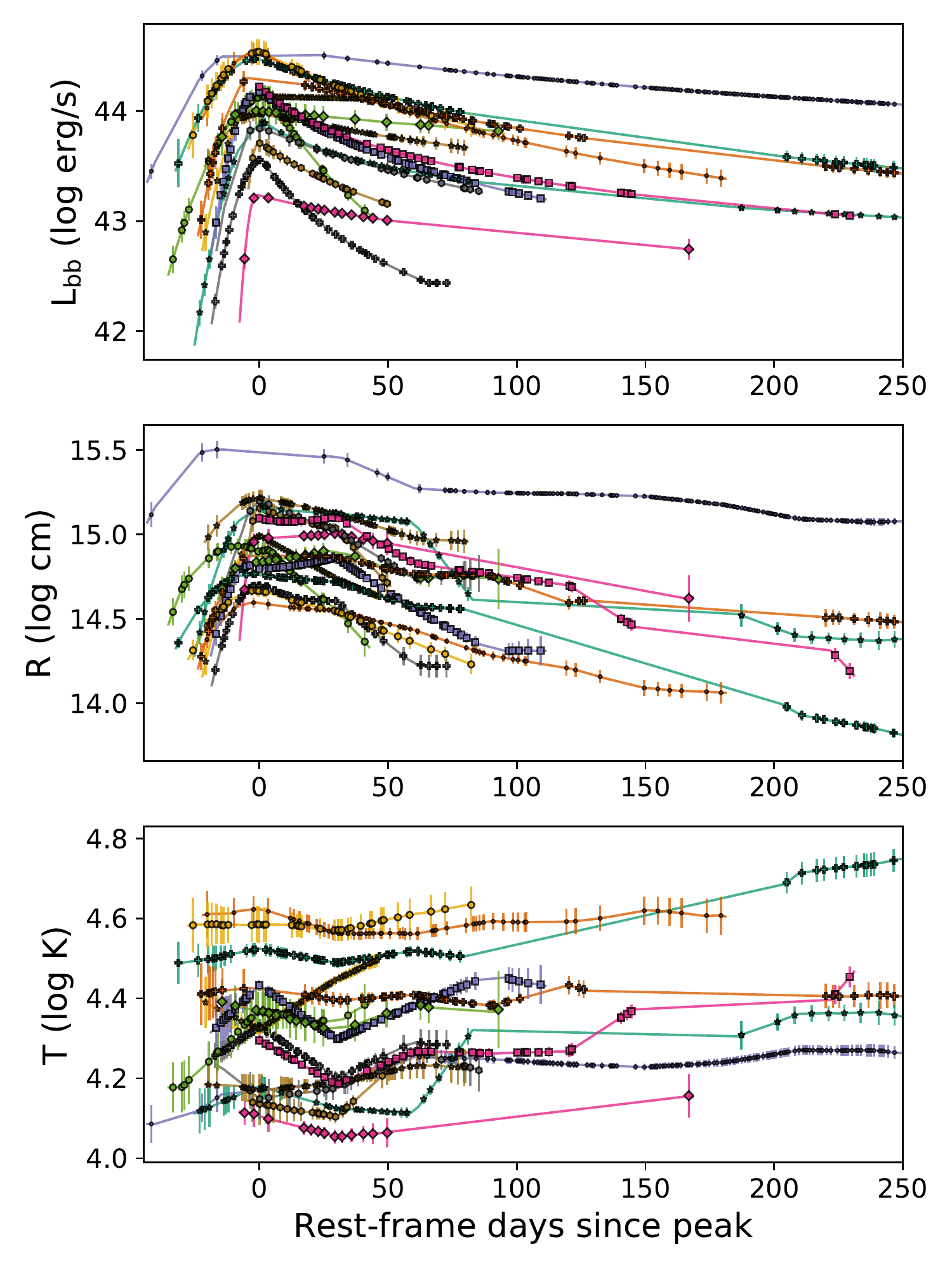}
\caption{The $r$-band absolute magnitude, blackbody luminosity, blackbody radius, and rest-frame blackbody temperature for the 17 TDEs in our sample with ZTF and Swift/UVOT observations. We see that all TDEs show a decrease of the blackbody radius after maximum light and most sources show an small but significant increase of the blackbody temperature.  }\label{fig:LRT}
\end{figure}

We list the results for the 17~TDEs with ZTF data in Table~\ref{tab:lcfit}. We show the rest-frame absolute $r$-band magnitude, and derived blackbody luminosity, radius, and temperature with time in Figure \ref{fig:LRT}.  This study increases the number of TDEs with well characterized pre-peak light curves by a factor of 3, and illustrates the remarkable homogeneity in the shape of their luminosity and radius evolution with time. In particular, we point out that most sources show a (modest) increase of the temperature with time. 

We find that the typical value of the power-law index of the bolometric luminosity is close to the canonical $p=-5/3=-1.67$, albeit with large scatter. For the ZTF sources we find a median power-law index of $\bar{p}=-1.66$ with a root-mean-square (rms) dispersion of 0.75. For the entire sample of 39 TDEs we find $\bar{p}=-1.64$; restricting the sample to the 32 spectroscopic TDEs yields a similar value of $\bar{p}=-1.62$ with an rms of 0.63. To conclude, the mean power-law index is consistent with $p=-5/3$, but we also find some significant deviations. 

\subsection{Optical to X-ray Ratio}\label{ref:opt_xray}
\begin{figure}
\includegraphics[trim=3mm 2mm 8mm 8mm, clip, width=0.48 \textwidth]{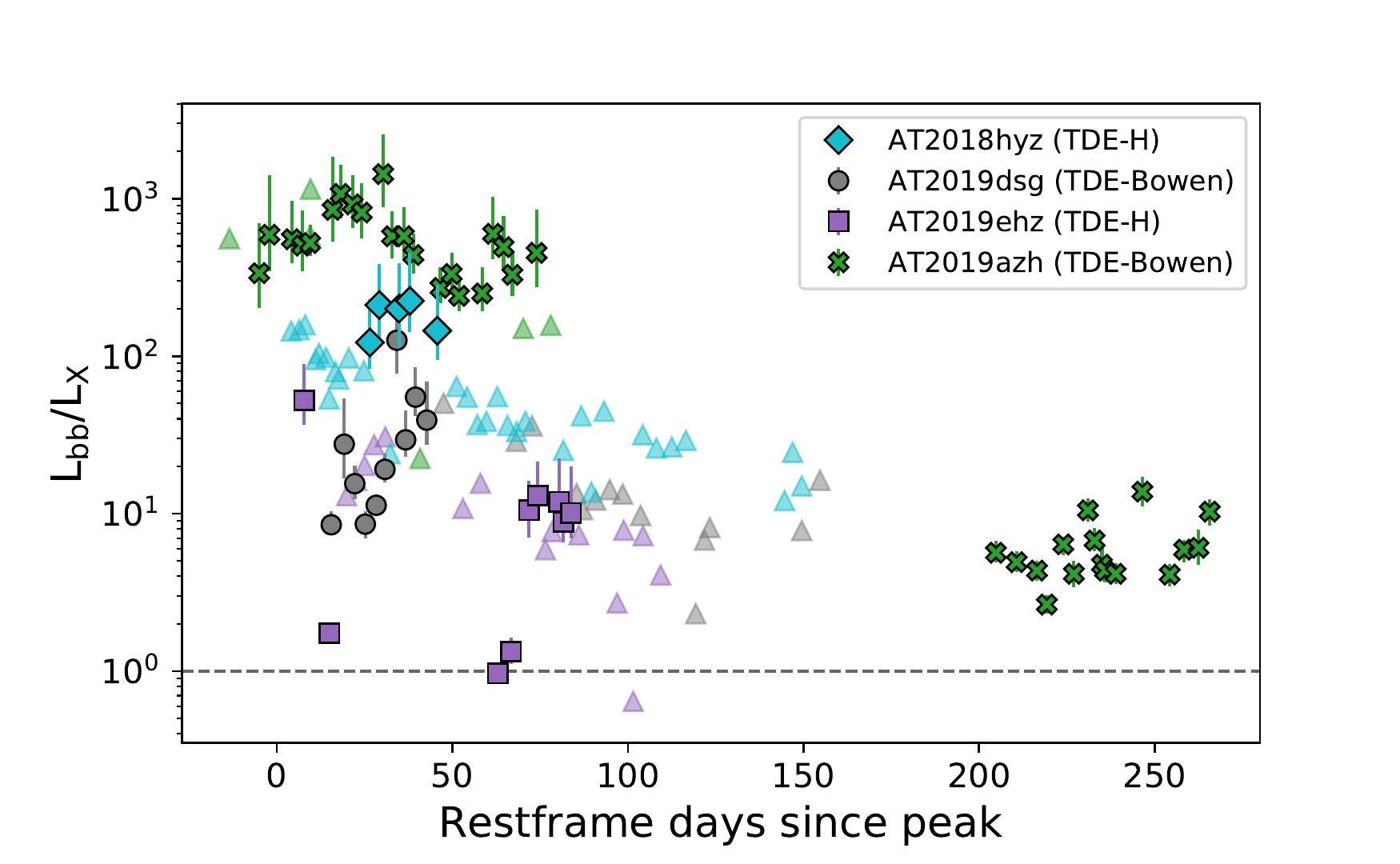}
\caption{Ratio of the blackbody luminosity derived from the optical/UV light curves and the X-ray luminosity (0.3--10 keV, based on {\it Swift}/XRT observations). We call attention to the X-ray flares of \Brienne, which reach an X-ray-to-optical ratio close to unity. Triangles indicate 3$\sigma$ lower limits.}\label{fig:LbbLX}
\end{figure}

In Figure \ref{fig:LbbLX} we show a comparison of the 0.3-10 keV X-ray luminosity measured by {\it Swift}/XRT to the luminosity of the UV/optical component derived in \S \ref{sec:lc}, for the four new ZTF TDEs with {\it Swift}/XRT detections.  Unlike the UV/optical luminosity, which has a smooth evolution over time and is well described with a single power-law decline post peak, the soft X-ray component shows variability on several timescales. Both large amplitude flaring on the timescale of just a few days (\Brienne) and an dramatic increase in luminosity over a timescale of a few months (\Jaime) have been observed.  

There are only three other TDEs with well sampled soft X-ray light curves from {\it Swift}: ASASSN-14li \citep{Holoien14}, ASASSN-15oi \citep{Gezari17, Holoien18}, and AT2018fyk/ASASSN-18ul \citep{Wevers19}. While ASASSN-14li showed a soft X-ray flare that followed the general power-law decline of the UV/optical component, with a characteristic ratio of $L_{\rm opt}/L_{\rm X} \sim 1$ for over a year, the other two TDEs show quite dramatic variability, with a variability and a systematic brightening in the soft X-rays at late times.  The first TDE in our ZTF sample, AT2018zr/PS18kh, was detected with a soft X-ray component ($kT \sim 100$ eV) in XMM-Newton observations \citep{vanVelzen18_NedStark}, but with a weak level relative to the optical, with $L_{\rm opt}/L_{\rm X} \sim 100$.

\begin{figure*}
\gridline{
\fig{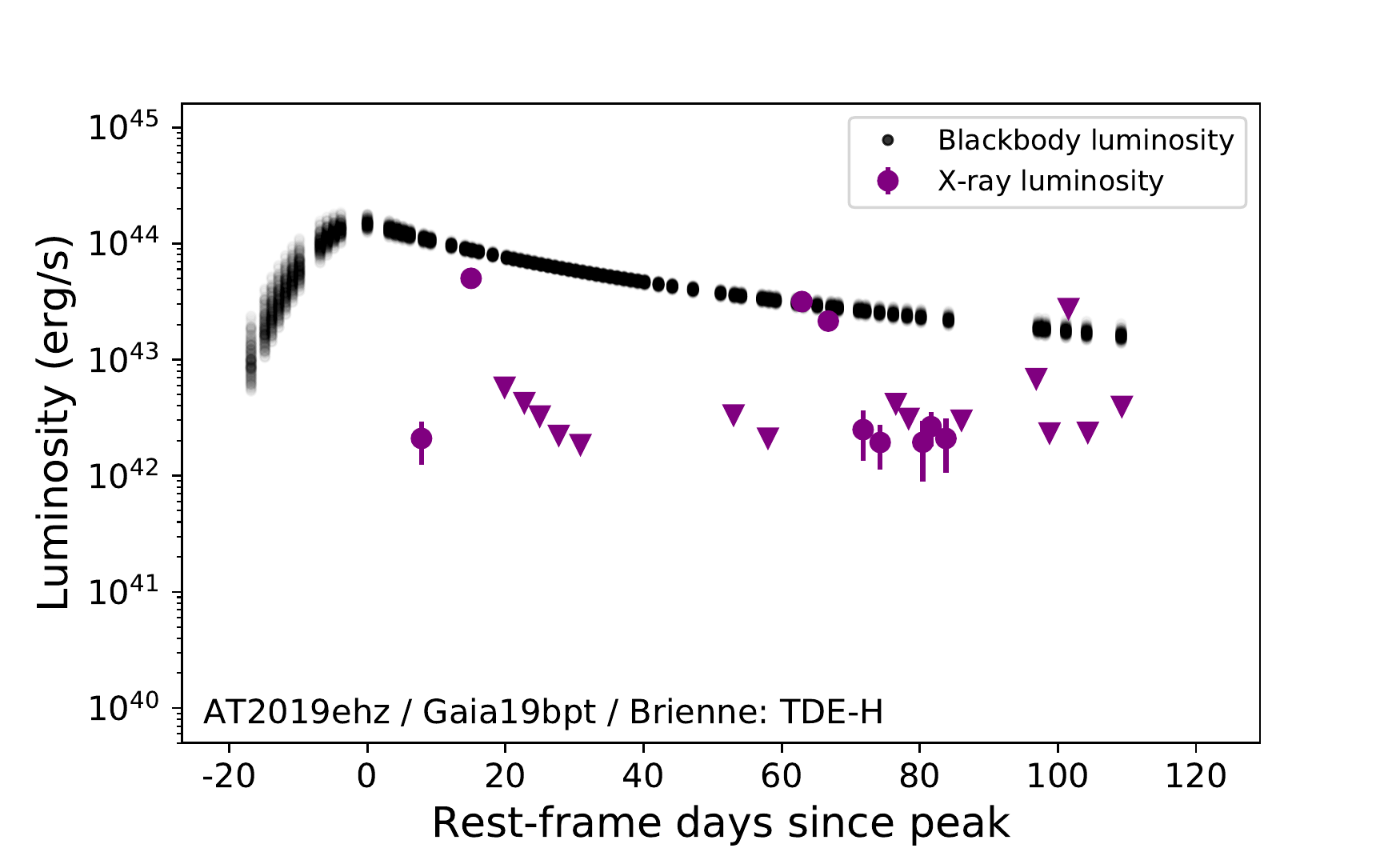}{0.48 \textwidth}{}
\fig{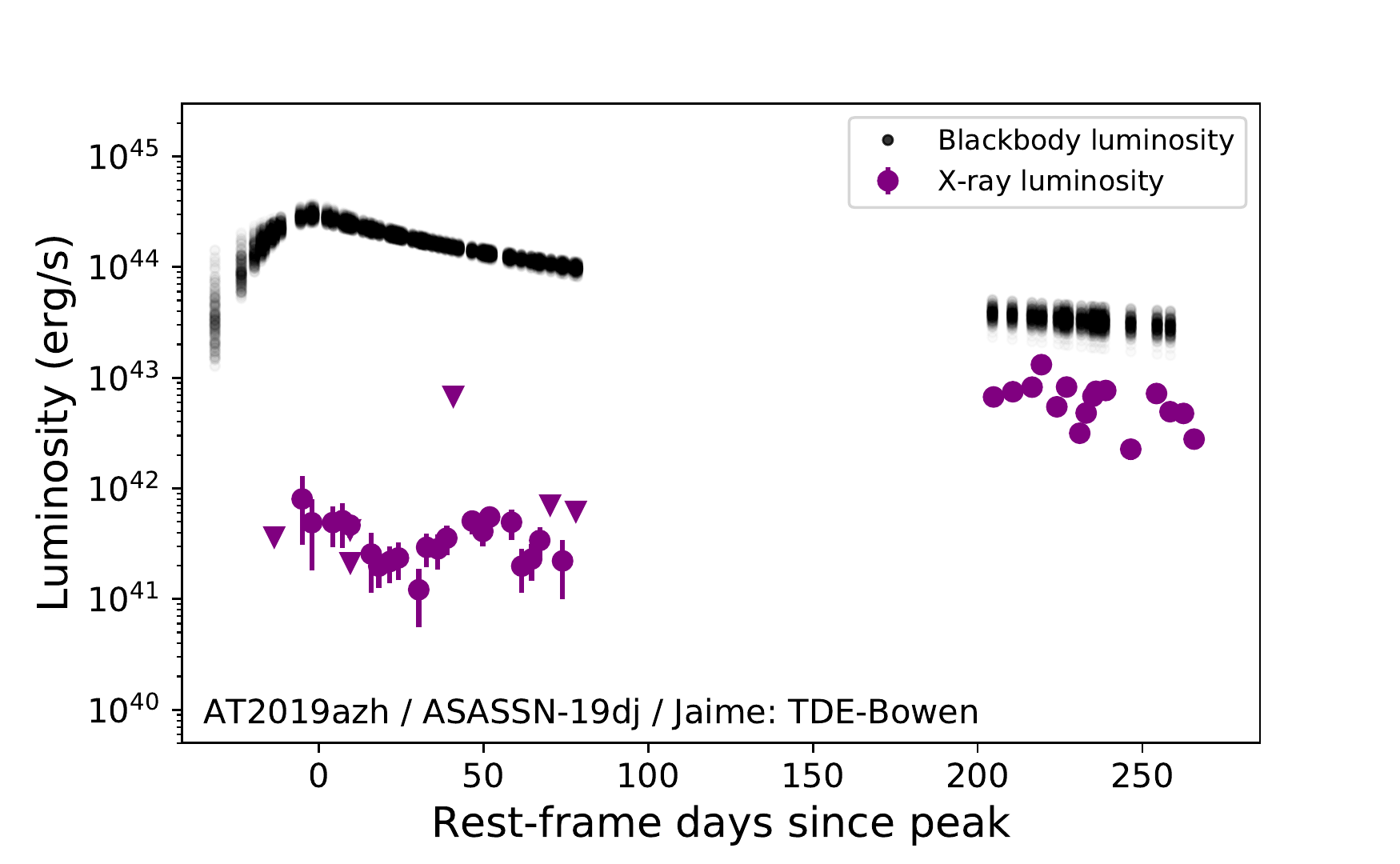}{0.48 \textwidth}{} \\[-30pt] }
\gridline{
\fig{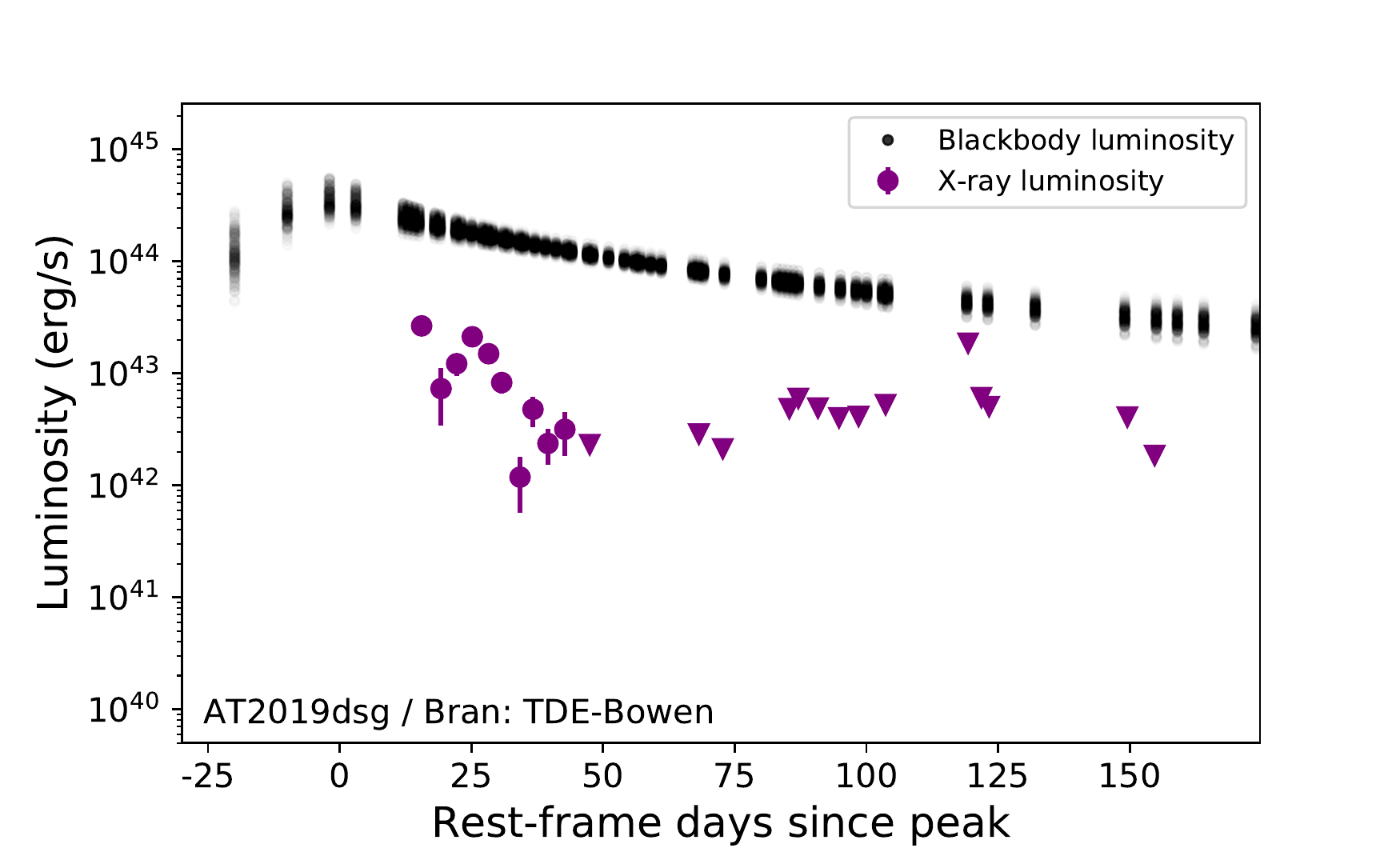}{0.48 \textwidth}{}
\fig{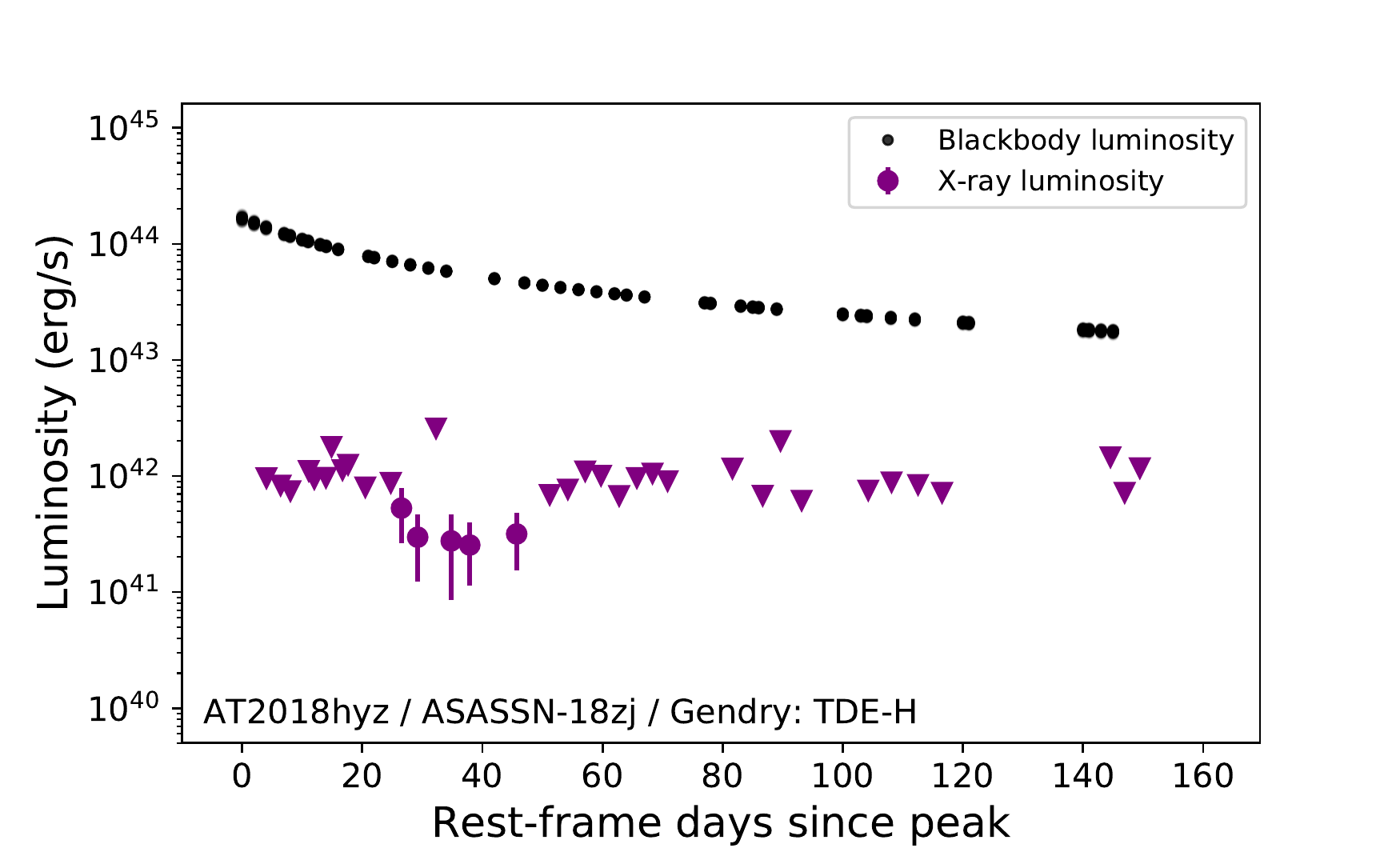}{0.48 \textwidth}{}
 \\[-10pt]}
\caption{Optical/UV blackbody luminosity (sampled from the posterior distribution) and the observed X-ray 0.3--10 keV luminosity for four TDEs with detections in {\it Swift}/XRT. We see rapid and luminous X-ray flares for \Brienne, a steady increase in the X-ray luminosity for \Jaime, and relatively weak early-time detections for remaining two sources (\Bran and \Gendry). Triangles indicate 3$\sigma$ upper limits.}\label{fig:LbbLX_all}
\end{figure*}

% \begin{table}
% \caption{Light curve peak luminosity and temperature}\label{tab:temp}
% \begin{tabular}{l c c c c}
% \hline
%  name                 & $L_g$         & $L_{\rm bb}$          & $T$               & $dT/dt$  \\ 
%                      & $\log$ erg/s & $\log$ erg/s & $\log$ K     & $10^{2}$ K/day  \\
% \hline
% \input{tables/ZTF_temp_and_lum.tex}
% \hline
% \end{tabular}
% \end{table}

% \begin{figure}
% %\gridline{\fig{R-dT.pdf}{0.48\textwidth}{}} 
% \gridline{\fig{T-dT.pdf}{0.48\textwidth}{}}
% \caption{Temperature change}\label{fig:dT}
% \end{figure}

\section{Results}\label{sec:results}
At this point we have extracted the following characteristics from our sample of TDEs: host galaxy properties (e.g., mass, color, age), three different spectral types of the flare (TDE-H, TDE-He, or TDE-Bowen), and  $\approx$4 independent light curve features  (blackbody temperature, blackbody radius, rise timescale, and fade timescale). In this section we investigate which of these properties are correlated.

\subsection{Comparing spectroscopic TDE classes}\label{sec:ks}
We first use a Kolmogorov-Smirnov (KS) test to assess wether our three TDE spectral classes show different distributions of light curve or host properties. For the light curve properties, we use only those measured with an uncertainty smaller than 0.3~dex (this requirement only affects the rise/decay timescale distributions). 
%We adopt $p<0.05$ as our definition of a significantly different population. 
The results are summarized in Table~\ref{tab:KS} and examples of cumulative distributions are shown in Fig.~\ref{fig:cumu}. 

Comparing the two biggest spectral classes, TDE-H and TDE-Bowen, each containing 14 TDEs, we find a striking difference in the distribution of blackbody radius. The typical radius of the TDE-H population is a factor two larger than the  TDE-Bowen TDEs. The hypothesis that these two classes are drawn from the same distribution of blackbody radius can be rejected with $p<3\times 10^{-5}$. The TDE-H and TDE-Bowen classes also show a significantly different temperature distribution ($p=0.02$), the latter being hotter on average. Since on the Rayleigh-Jean tail the luminosity is given by $L_{\rm RL} \propto R^2 T$, we also find a significant difference ($p=0.005$) between the distributions of $g$-band luminosity: the TDE-H class has higher values of $L_g$. 

With only 4 events, the He-only TDEs are a much smaller sample, yet we still find evidence for differences in rise timescale when compared to the TDE-H class ($p=0.02$) and $g$-band luminosity when compared to the TDEs with H+He/Bowen lines ($p=0.03$). 

Finally, after noticing that the He-only TDEs appear to have more blue host galaxy $u-r$ colors, we also investigated differences between the TDE spectroscopic classes and their host properties as derived from our population synthesis model. The population parameters are: stellar mass, metalicity, age since the peak of star formation, e-folding time of the starformation rate ($\tau_{\rm sfh}$), and the dust optical depth. Since we only have 5 to 7 observables (the GALEX FUV and NUV flux or upper limits, plus 5 bands from SDSS or PS1) the stellar population parameters have large uncertainties and degeneracies. We therefore also consider a  Principal Components Analysis (PCA) of these parameters, which should capture the main correlations between the population synthesis parameters (e.g., the age-metacillity degeneracy). We find that the fourth principal component of the galaxy population parameters (PC4 hereafter) yields a significant separation of the TDE-H and TDE-Bowen TDE populations ($p=0.02$). The weights of PC4 are dominated by the dust parameter and $\tau_{\rm sfh}$.

\begin{figure}
\gridline{\fig{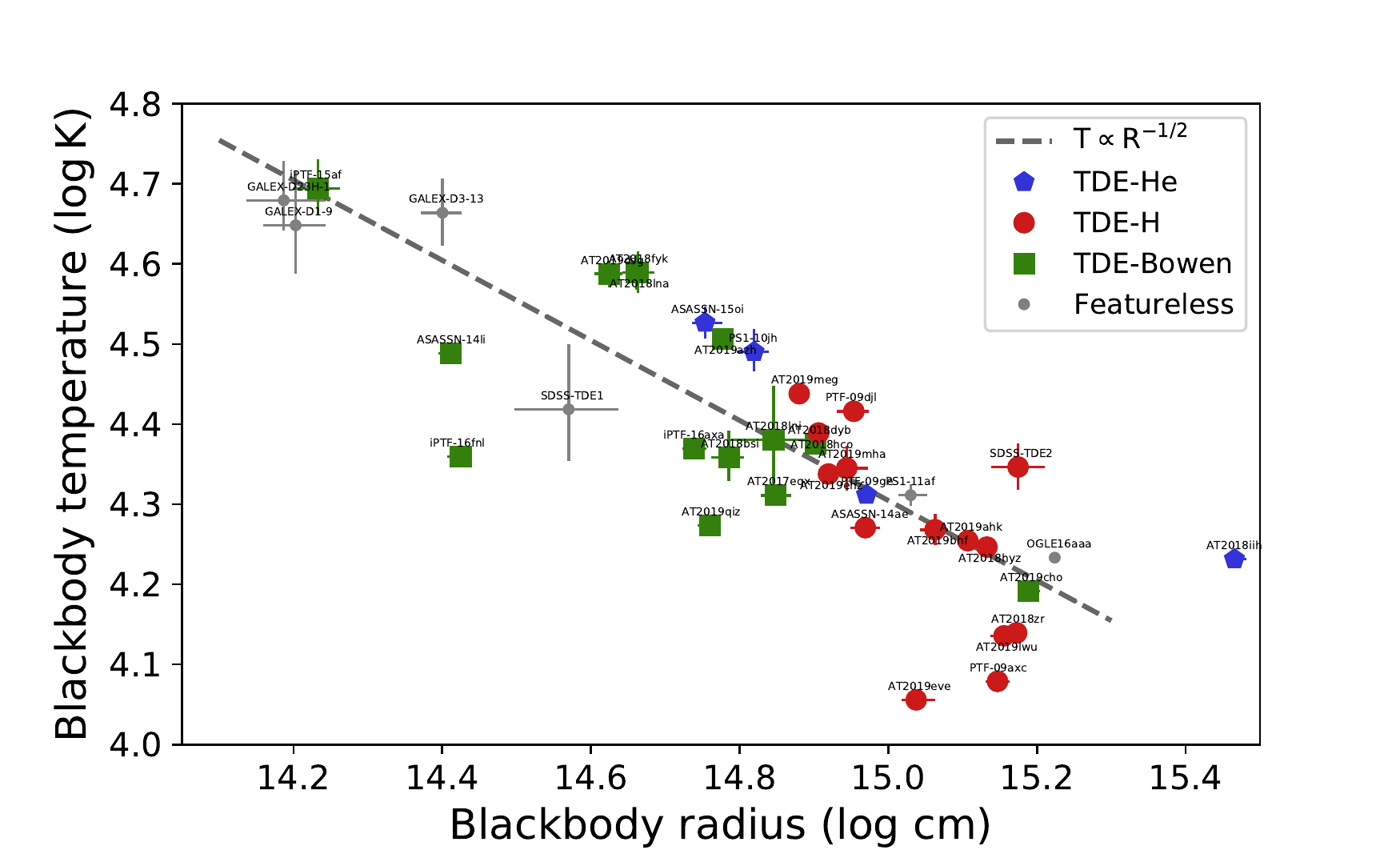}{0.48 \textwidth}{} \\[-30pt]}
\gridline{\fig{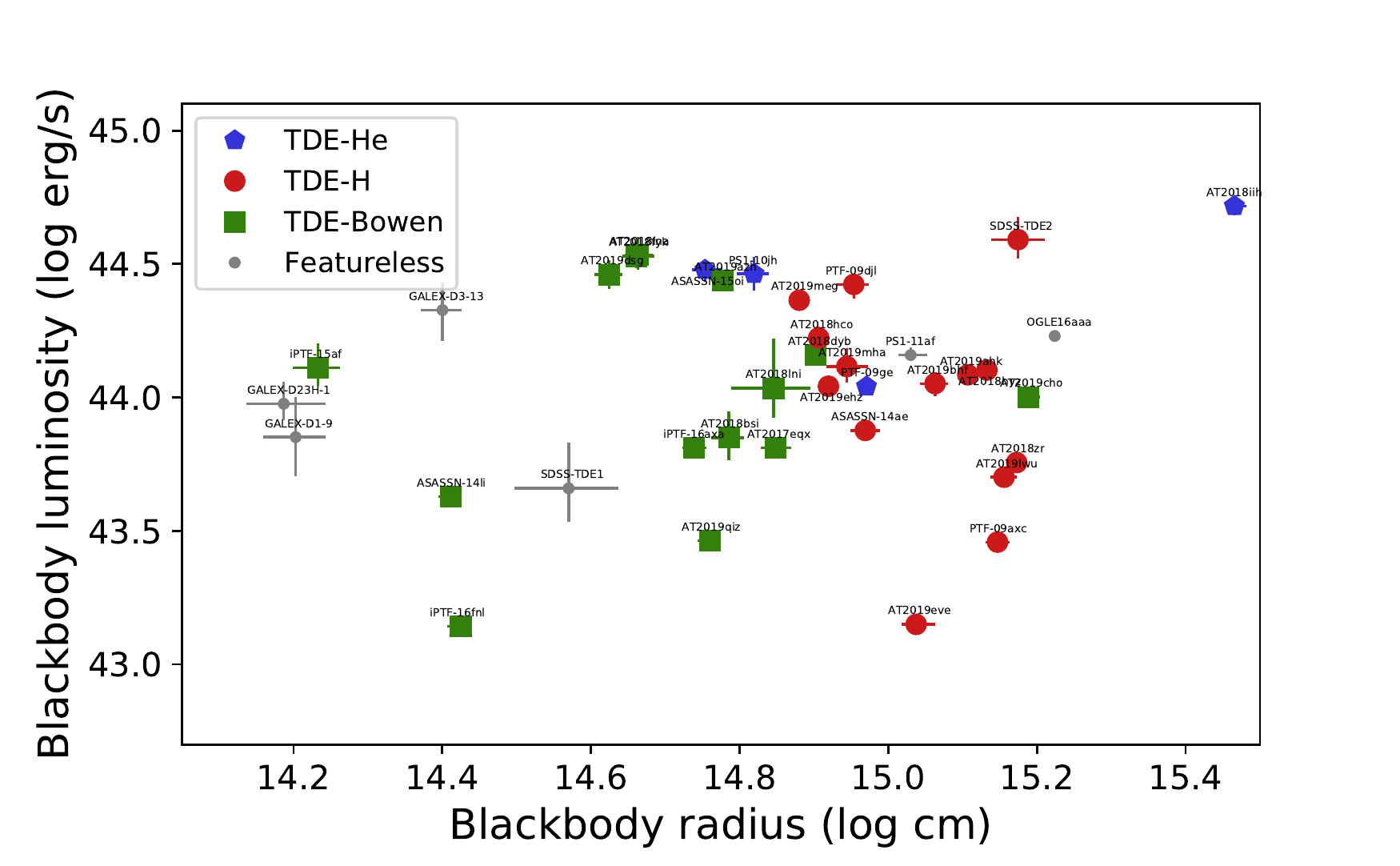}{0.48 \textwidth}{}\\[-15pt]}
%\gridline{\fig{Lbol-rise}{0.48 \textwidth}{}}
\caption{Black body temperature and blackbody luminosity versus the blackbody radius, all measured at the peak of the TDE light curve. In the top panel, the dashed line shows the relation expected for a single blackbody spectrum with a luminosity of $10^{44.1}$~erg\,s$^{-1}$. We see that the TDE-Bowen class has smaller radii and larger temperatures compared to the other two spectroscopic TDE classes. There is no difference in the blackbody luminosity of the TDE-Bowen and TDE-H class, but the TDE-He class appears to have a higher average luminosity.  }\label{fig:RT}
\end{figure}

\begin{figure}
\gridline{\fig{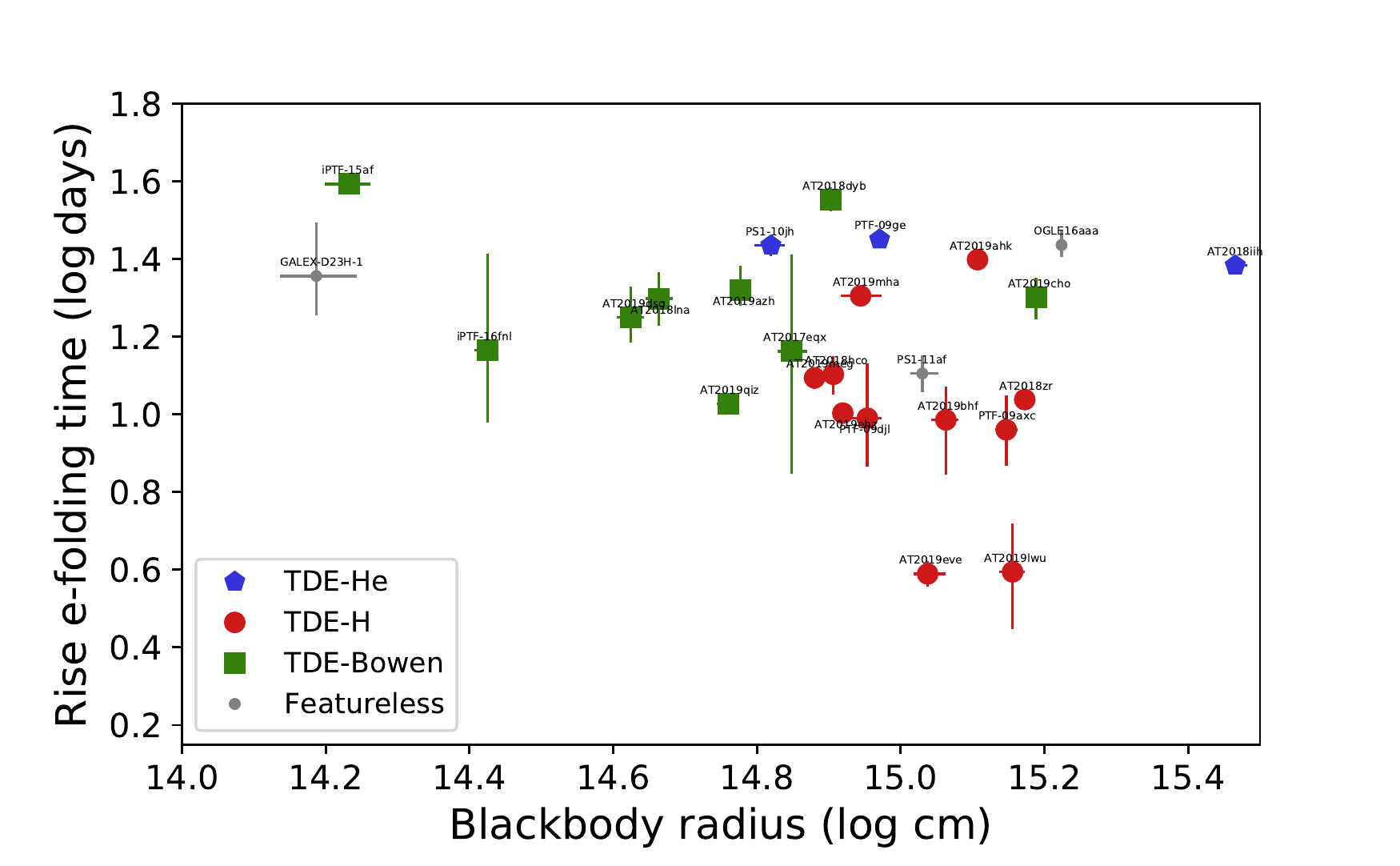}{0.48 \textwidth}{}  \\[-30pt]}
\gridline{\fig{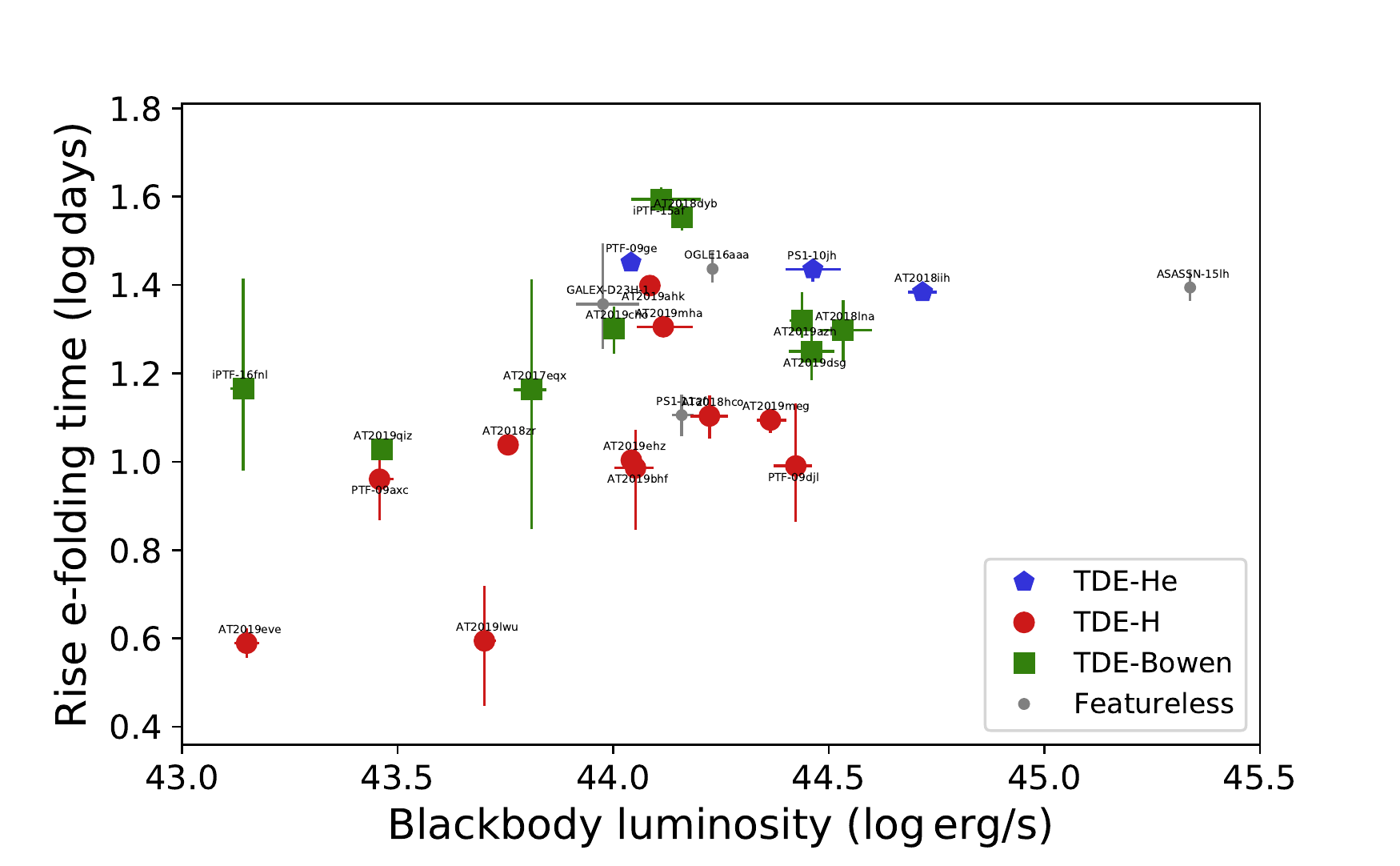}{0.48 \textwidth}{} \\[-30pt]}
\gridline{\fig{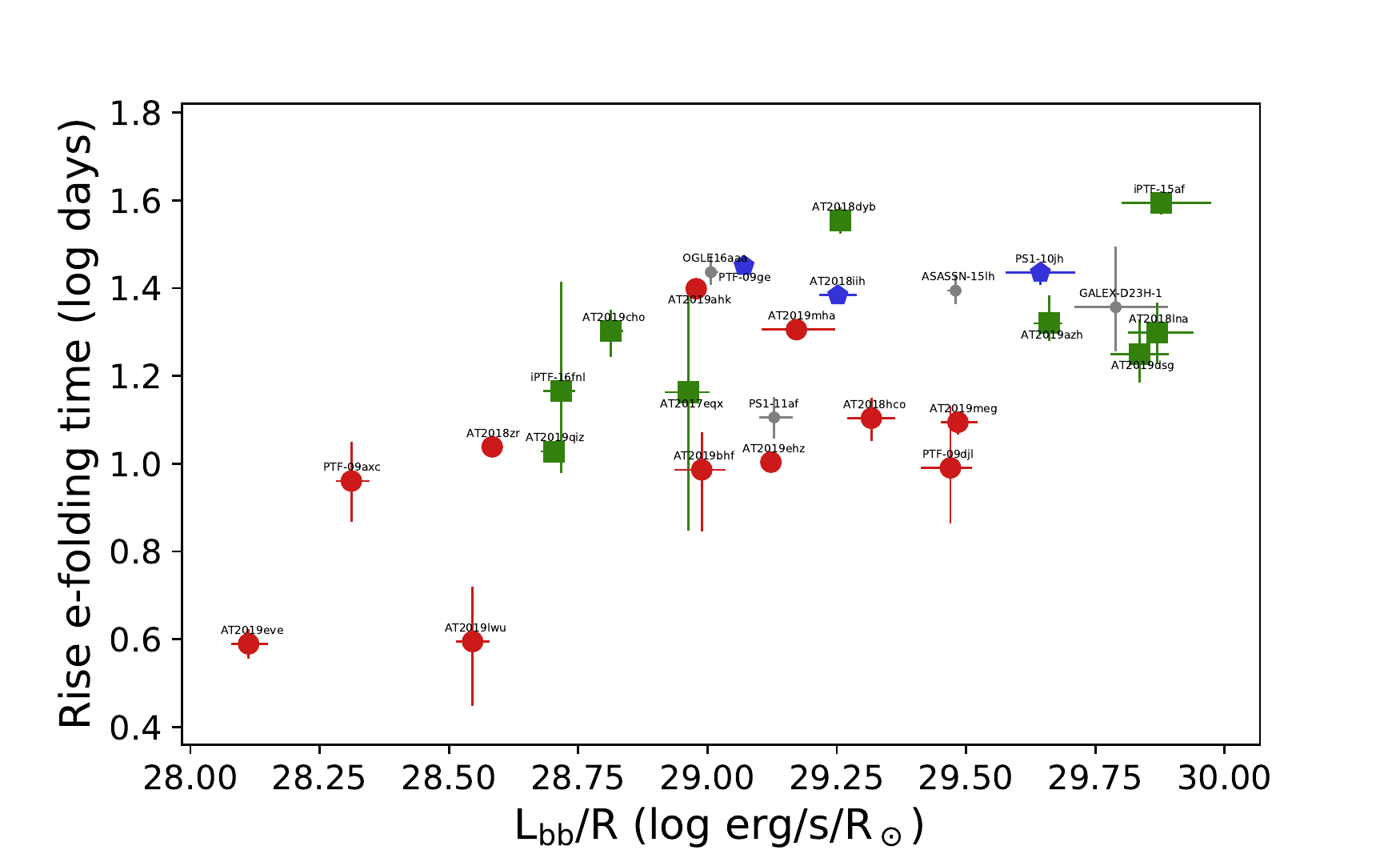}{0.48 \textwidth}{} \\[-15pt]}
\caption{Rise time versus blackbody radius ($R$), blackbody luminosity ($L_{\rm bb}$), and $L_{\rm bb}/R$. We see that sources with the smaller blackbody radii have the longest rise times and the TDEs with only helium emission lines in their optical spectra (TDE-He) have significantly longer rise times compared to the rest of the population. While the blackbody radius and luminosity are not correlated with each other (Fig.~\ref{fig:RT}), the rise time appears to be correlated with both of these two light curve properties (top and middle panel). The strongest correlation is found between the rise time and the ratio of the luminosity and radius (lower panel). These results can be explained if the rise time is proportional to the density inside the photosphere (\S ~\ref{sec:diff}).  }
\label{fig:rise}
\end{figure}

\subsection{The ``look elsewhere" effect for multiple KS tests}
The look elsewhere effect can cause one to overestimate the significance of a correlation because multiple trials have been made to search for correlations that pass the threshold for significance. If each of our $N$ photometric properties is counted as a trial, and each property is independent, the $p$-value for a single KS-test should be increased by $N(1-p)^{N-1}$. However, our photometric properties are not independent and we have multiple correlations that appear to be significant. We therefore need to use the data directly to estimate the importance of the look elsewhere effect.  

To account for multiple trials in our dataset of correlated parameters, we repeat the KS test after randomly reordering the spectroscopic TDE labels.   On this shuffled dataset we repeat the KS-test for each parameter, as well as all their principal components. From this set of ``trials", we pick the lowest $p$-value. We repeat this procedure, each time shuffling the spectroscopic labels, to obtain a distribution of KS $p$-values that account for the multiple parameters that we considered in our comparison. 

Applying this method to 8 parameters that we obtained from our light-curve models ($\tau$, $\sigma$, $p$, $t_0$, $L_{\rm bb}$, $L_{g}$, $T$, and $R$ of  Eqs.~\ref{eq:exp} \& \ref{eq:pl}) we find nearly identical values for the KS test when the $p$-value of the original (unshuffled) dataset is $p<0.02$. The only parameter that no longer yields a significant difference in the distribution ($p>0.05$) is the $g$-band luminosity for the He-only versus the TDE-Bowen class ($p=0.06$ after correcting for trials).

After correcting for multiple trials for the 5 galaxy population synthesis parameters (mass, dust, age, $Z$, and $\tau_{\rm sfh}$), we find that the significance of difference in PC4 between the TDE-H and TDE-Bowen class remains unchanged. However the significance of the difference in $u-r$ color between the TDE-He population and the TDE-H population decrease from $p=0.02$ to $p=0.05$. 

We note that a decrease of the significance is expected if the TDE-He share some properties with the two other spectroscopic classes (as suggested by the cumulative distributions, Fig.~\ref{fig:cumu}). Because the number of TDE-He's is small, the shuffled population that is used to assess the look-elsewhere effect can contain traces of real signal from TDE-H or TDE-Bowen population.   

We can conclude that the separations between the two largest spectroscopic TDE classes, TDE-H and TDE-Bowen, are unlikely to be explained by random fluctuation that got promoted due to an over-diligent search of the parameter space. Because most parameters are correlated the independent parameter space is quite small, i.e., the freedom to ``look elsewhere" is limited.   

% \begin{figure}
% \gridline{\fig{rise-decay.pdf}{0.48\textwidth}{}}
% %\gridline{\fig{rise-t053.pdf}{0.5\textwidth}{}}
% \caption{Rise time}
% \end{figure}

\begin{figure*}
\gridline{
\fig{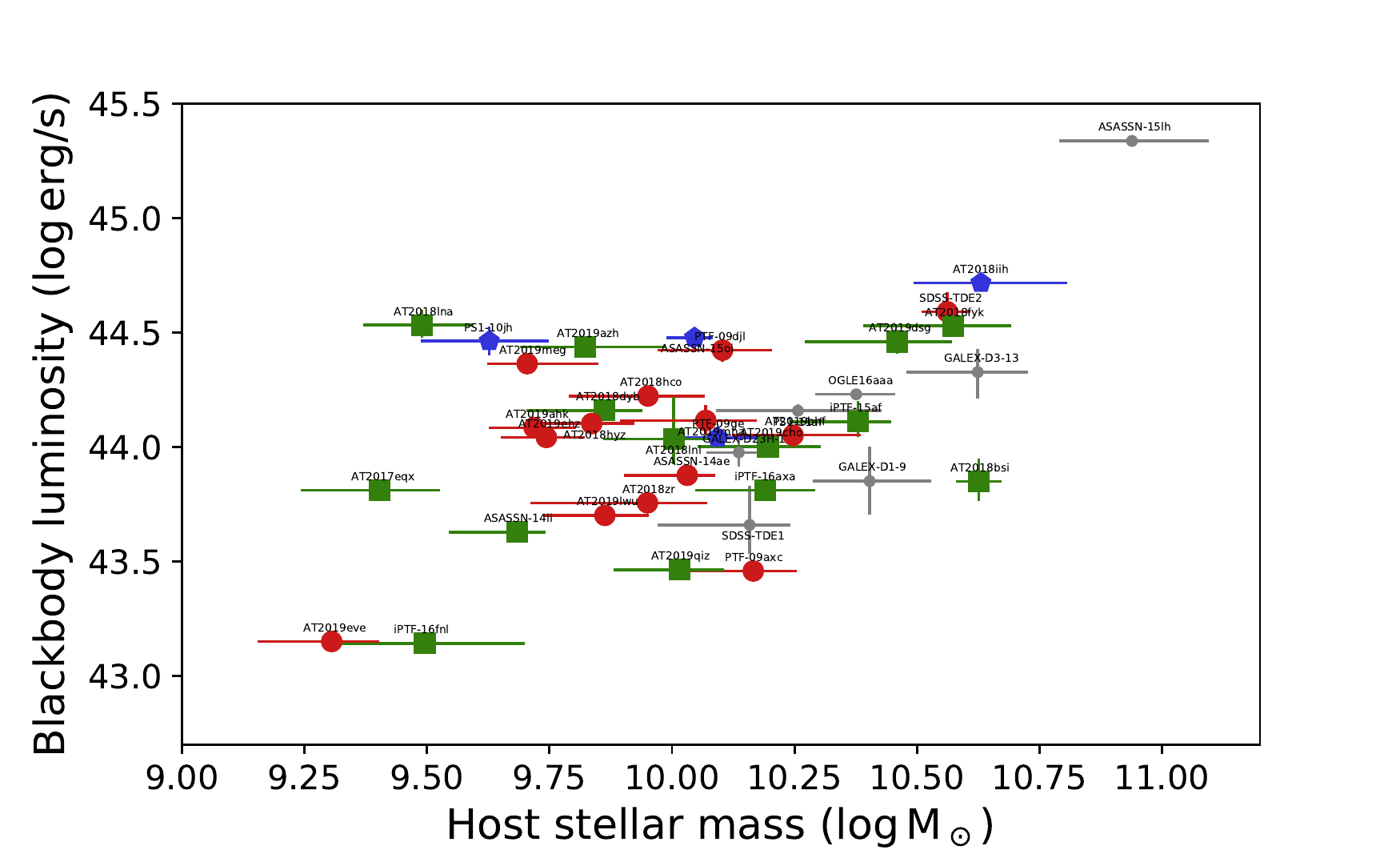}{0.33 \textwidth}{}
\fig{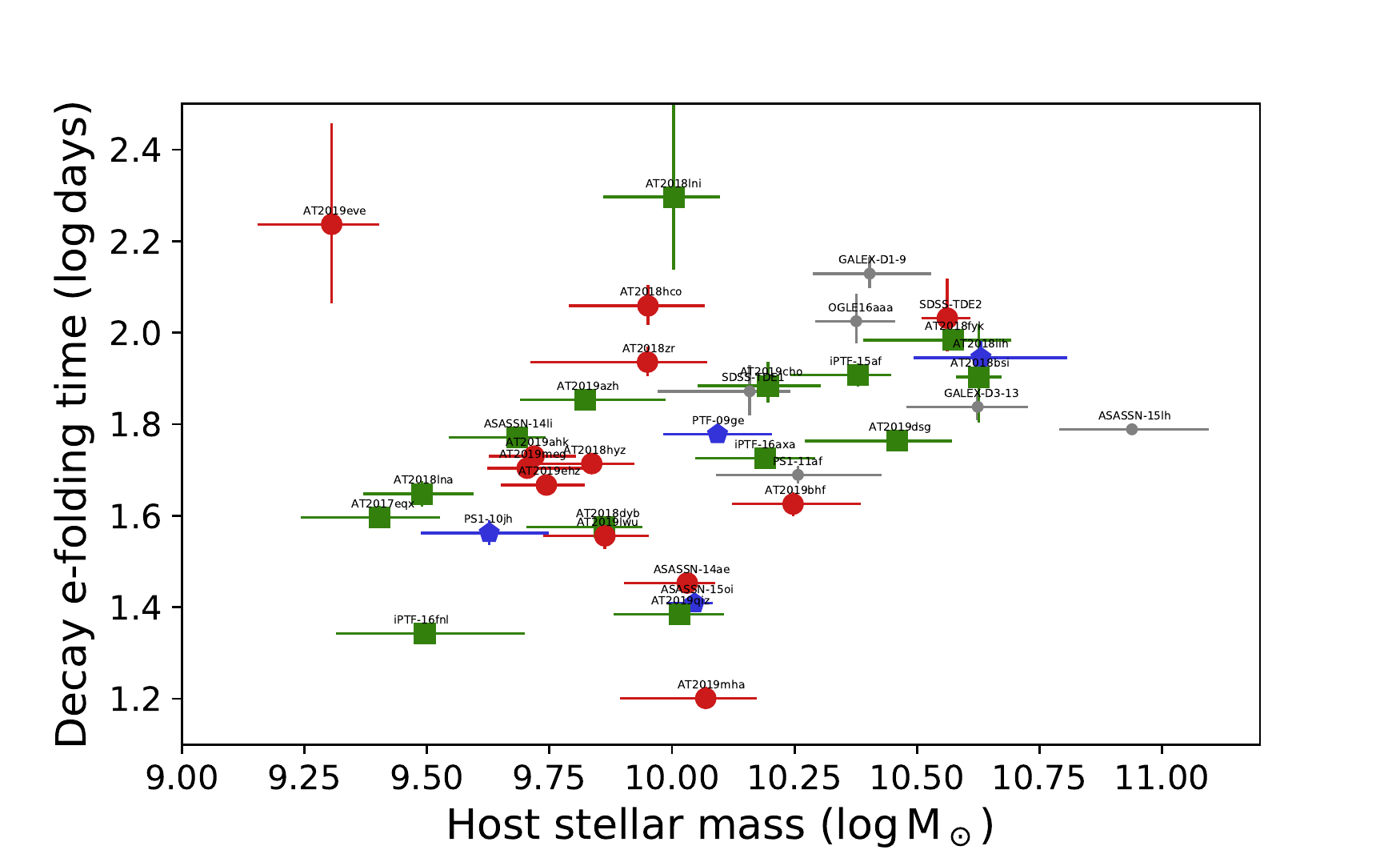}{0.33 \textwidth}{} 
\fig{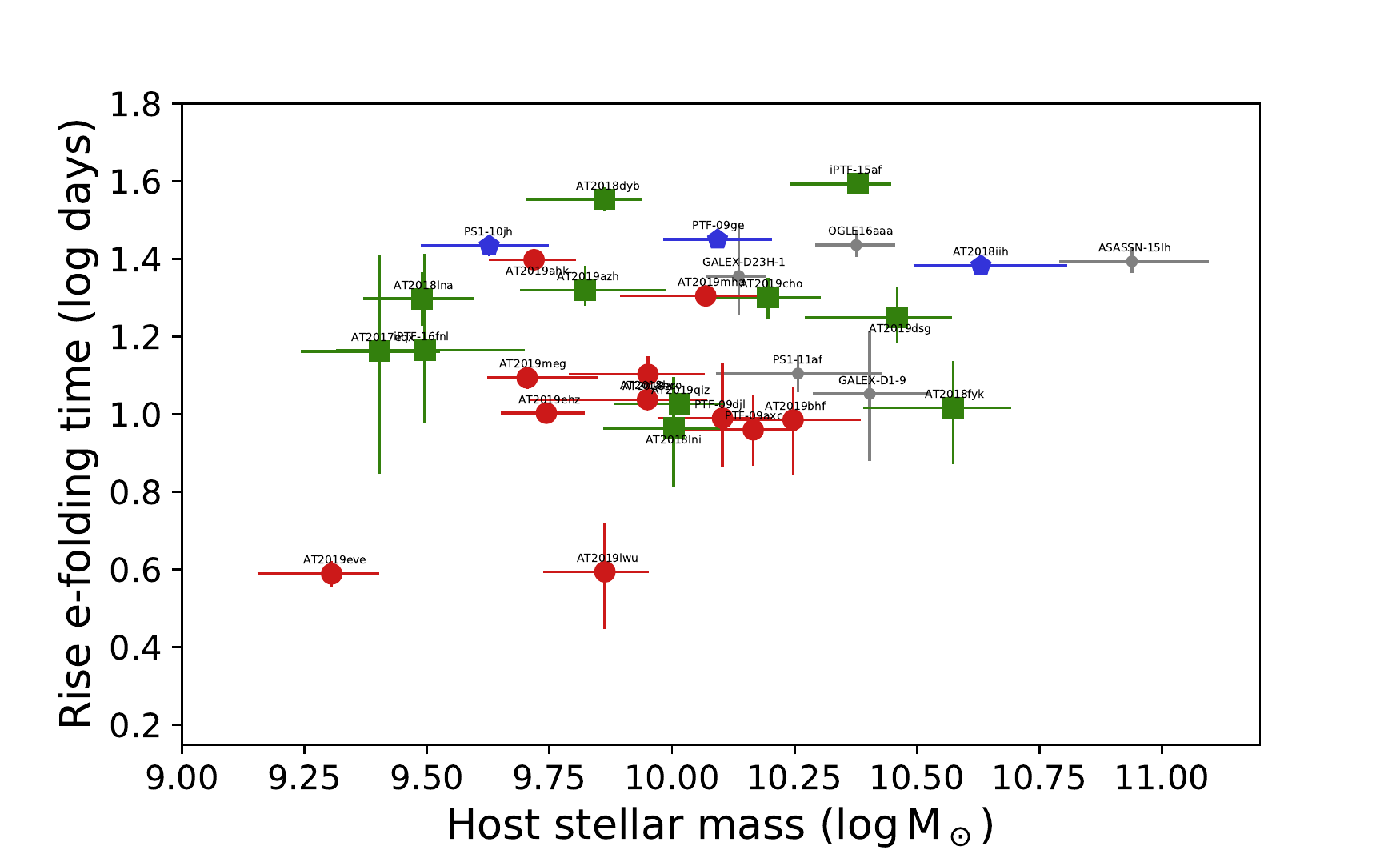}{0.33 \textwidth}{}
\\[-40pt]}

\gridline{
\fig{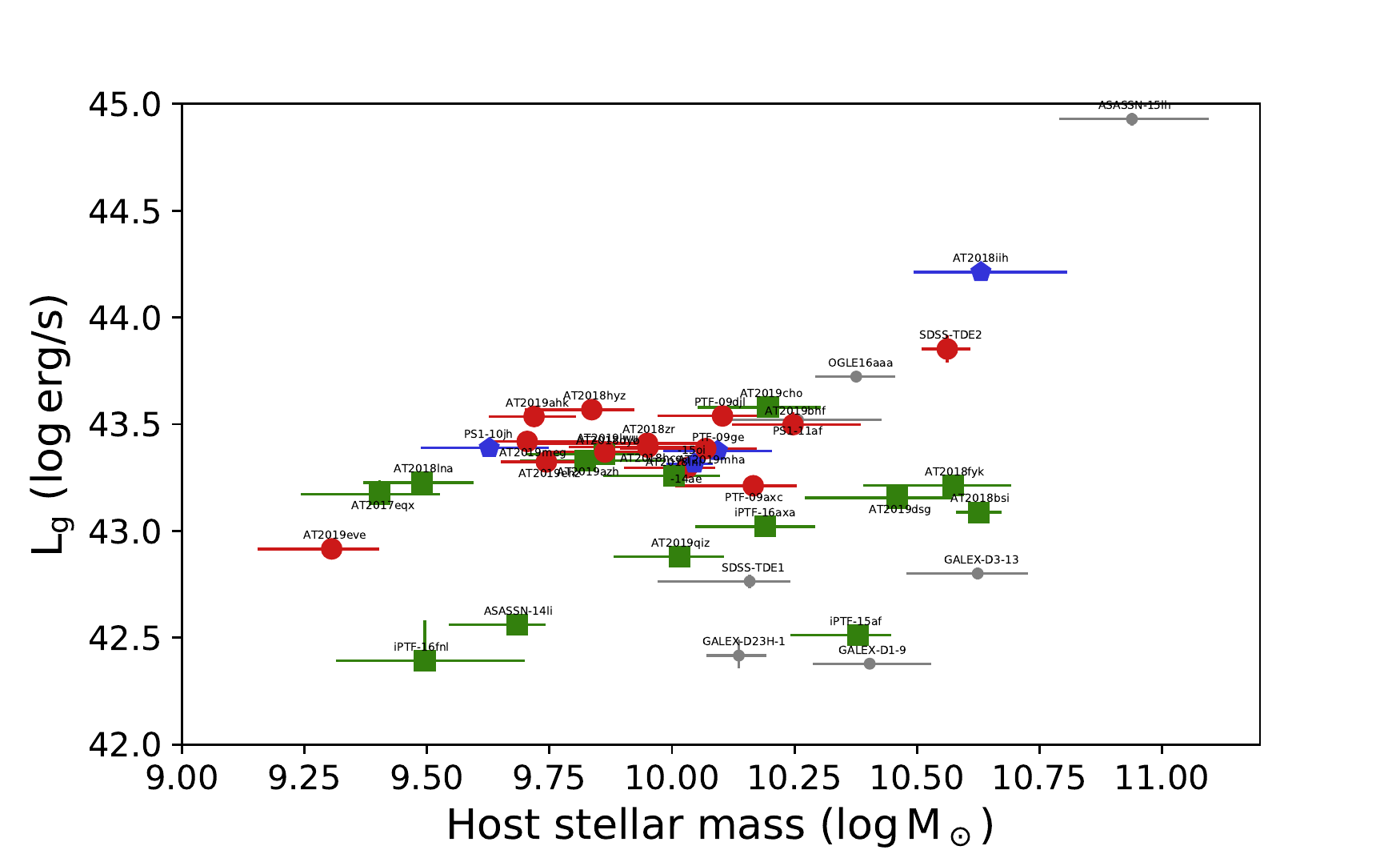}{0.33 \textwidth}{}
\fig{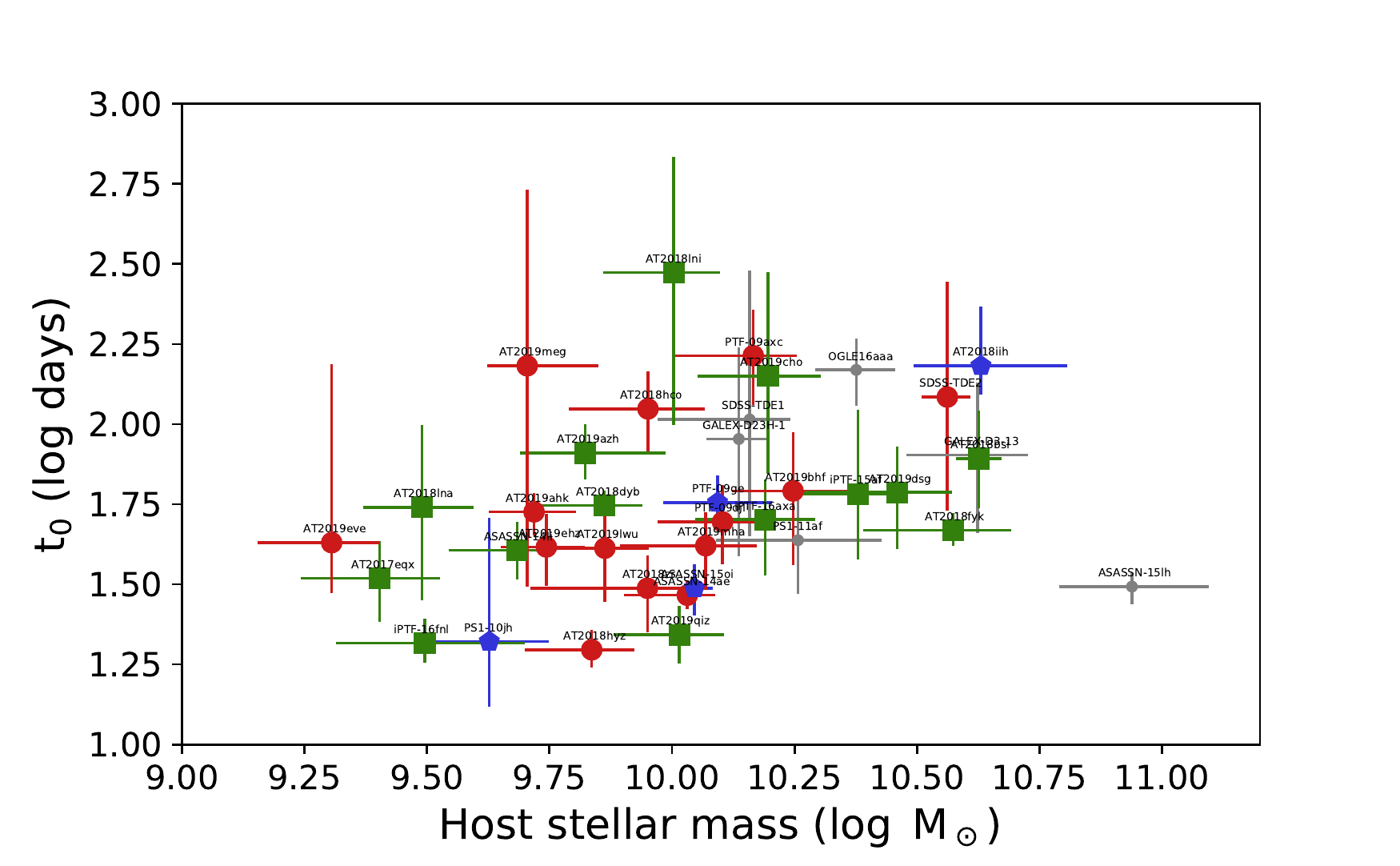}{0.33 \textwidth}{} 
\fig{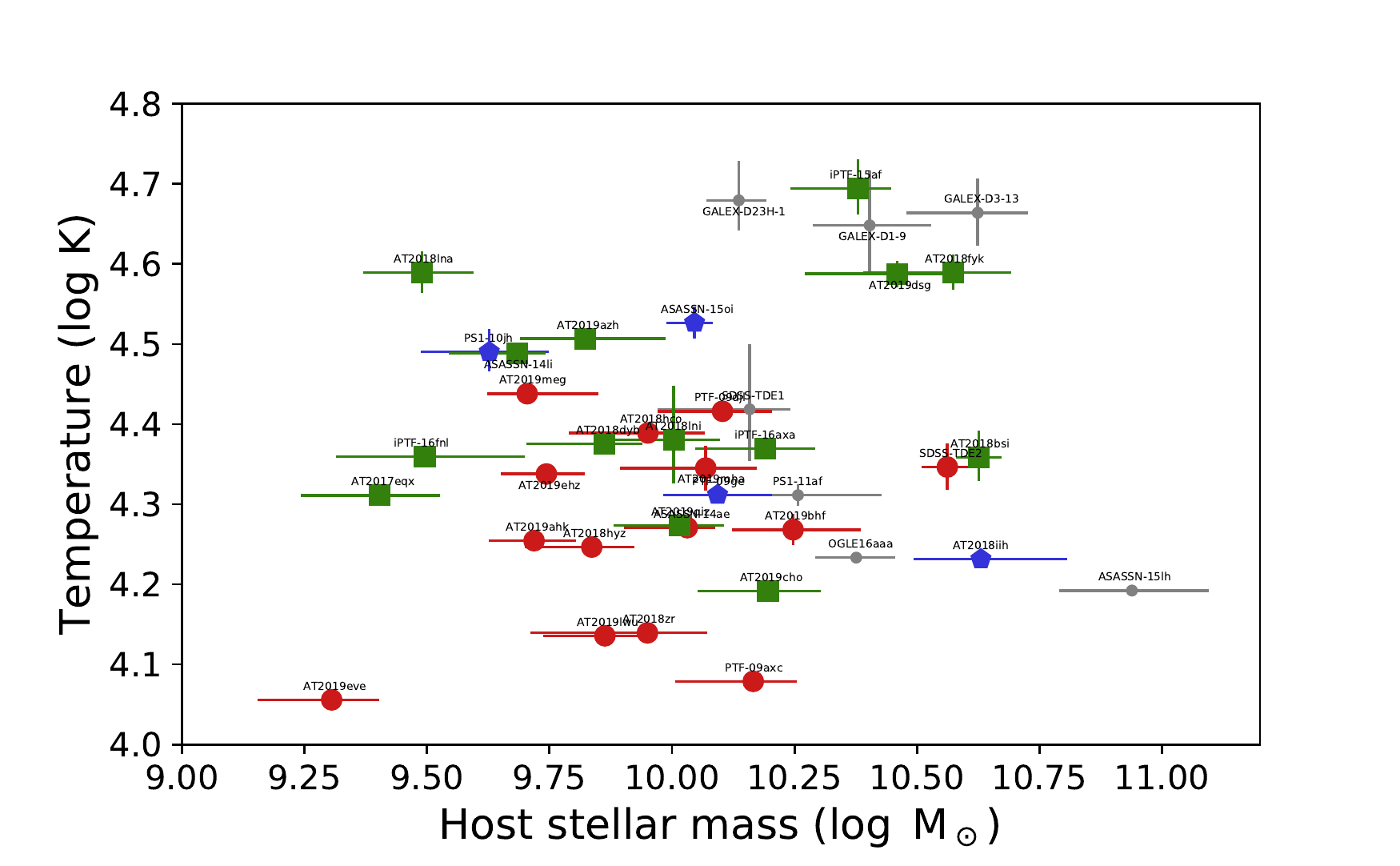}{0.33 \textwidth}{} 
\\[-40pt]}

\gridline{
\fig{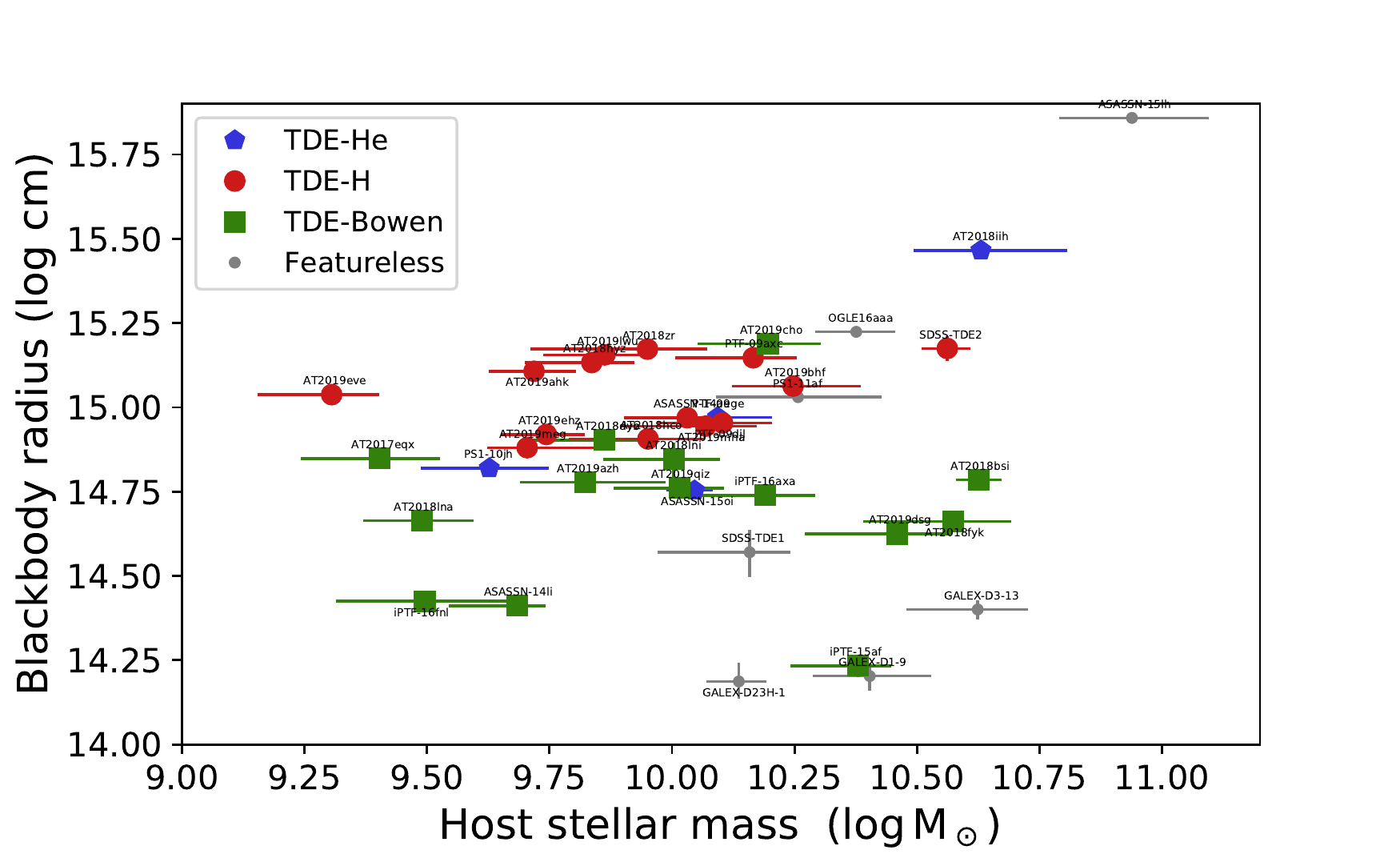}{0.33 \textwidth}{}
\fig{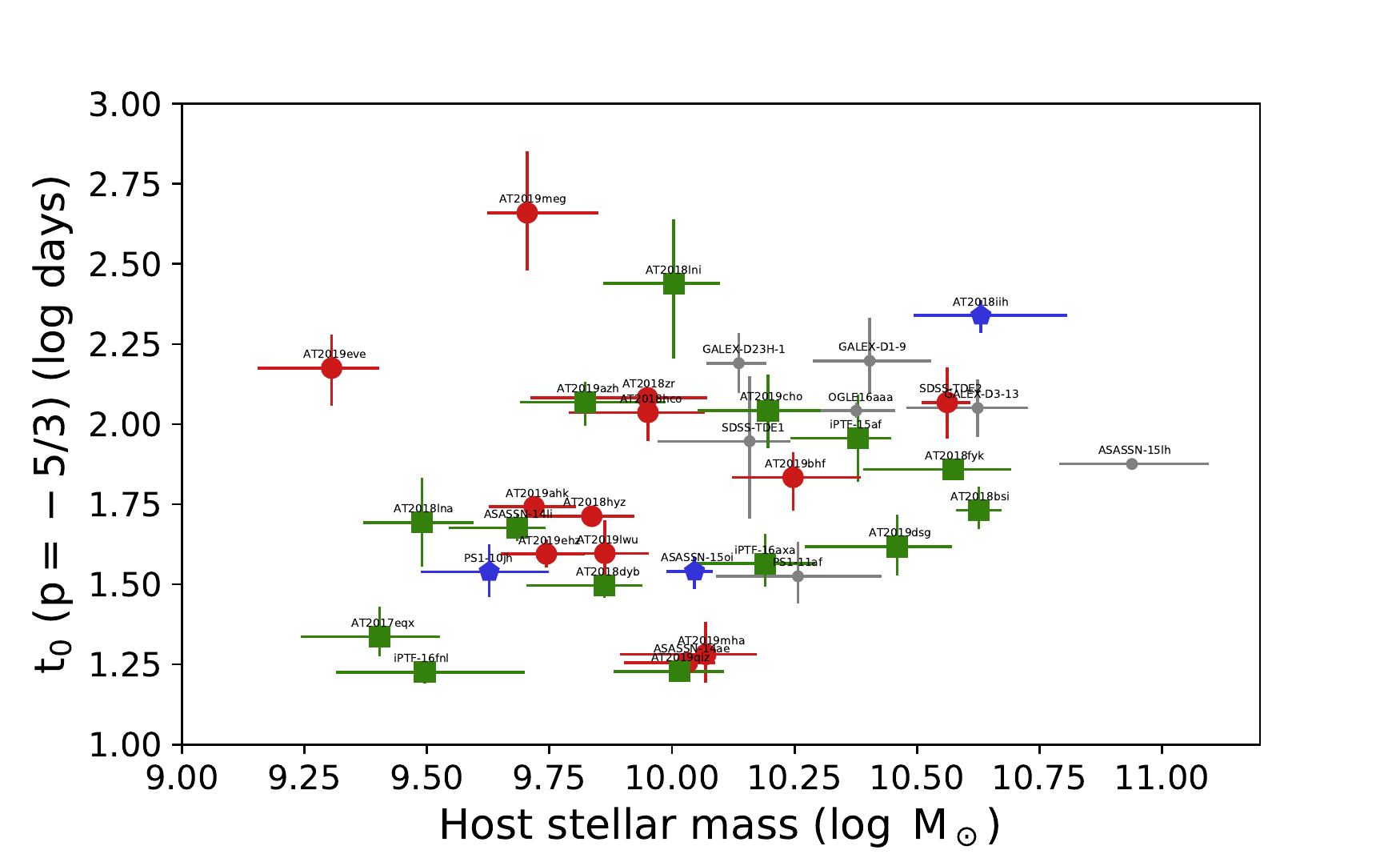}{0.33 \textwidth}{} 
\fig{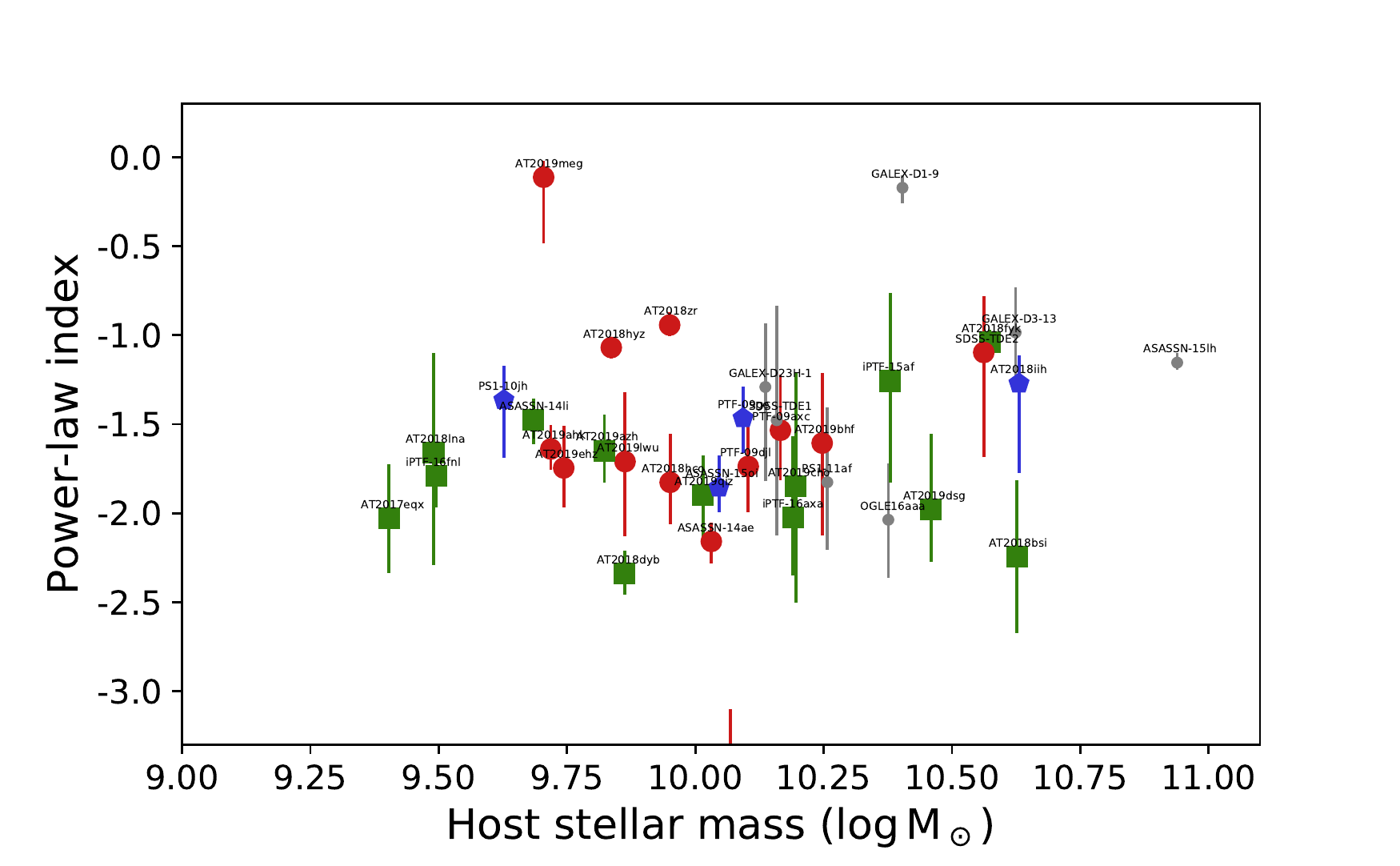}{0.33 \textwidth}{} \\[-15pt]}

\caption{TDE light curve properties versus the total stellar mass obtained from the host photometry. The marker symbol and color indicate the TDE spectroscopic class (the legend is printed in the first panel of the last row). We find a significant correlation between host mass and and three TDE properties: the blackbody luminosity, the monochromatic decay timescale ($\tau$, Eq.~\ref{eq:exp}), and the normalization of the power-law decay of the blackbody light curve ($t_0$, Eq.~\ref{eq:pl}). }\label{fig:vsmass}
\end{figure*}

\subsection{Correlations between TDE light curve properties}
In the previous two sections we presented differences between the TDE spectroscopic classes. We now focus on correlations between TDE light curve properties, using all 39 TDEs in our sample. When considering the correlation between a pair of parameters, we remove sources with an uncertainty larger than 0.3~dex. The results of a Kendall's tau test are listed in Table~\ref{tab:Kendall}. This test only considers the rank of pairs of data points. If we instead use a Pearson's test, which assumes the data is follows a normal distribution, we typically find lower (i.e., more significant) $p$-values. 

When considering the correlations between a given set of parameters, we need to keep in mind that our dataset as a whole shows a large degree of correlation. If all parameters would be uncorrelated, the 45 $p$-values of the correlation test in Table~\ref{tab:Kendall} should follow a uniform distribution between 0 and 1 and for a given limit on the significance $p<p_{\rm test}$ we should find $p_{\rm test}\times 45$ pairs. Instead, we find that $33/45=0.73$ pairs have $p<0.5$, $20/45=0.44$ have $p<0.1$, and $14/45=0.31$ have $p<0.05$. This means that, similar to what we found in the previous section, spurious correlations are unlikely to be important. However, the large degree of correlation makes it harder to find the causal relation between the parameters. 

We find a significant correlation between the blackbody radius and black body temperature ($p<10^{-8}$). The two properties follow the relation expected for a single blackbody spectrum $L_{\rm bb} \propto R^2T^{4}$, with $L_{\rm bb}\approx 10^{44}\,{\rm erg}\,{\rm s}^{-1}$. This correlation simply confirms that our TDEs are well-described by a blackbody spectrum, the scatter around the median luminosity is only 0.3~dex. Since most sources are selected based on optical observations and the bolometric luminosity is largely determined by the temperature estimated from UV follow-up observations, the relative small scatter cannot be entirely explained by Malmquist bias in our flux-limited sample. As expected, we also find a positive correlation ($p=0.03$) between the blackbody temperature and the blackbody luminosity. 

As shown in Fig.~\ref{fig:rise}, the rise time of the flare appears to be correlated with the bolometric luminosity and anti-correlated with the blackbody radius. If we consider the ratio of the bolometric luminosity to the blackbody radius, $L_{\rm bb}/R$, we find a significant correlation with rise time ($p=0.01$). In section~\ref{sec:diff}, we find this could be explained by a longer diffusion time at higher densities.   

Notably, we find no correlation between the rise timescale and the exponential decay timescale ($\tau$ in Eq.~\ref{eq:exp}) or the fallback timescale ($t_0$ with $p=-5/3$ in Eq.~\ref{eq:pl}). 

\subsection{Correlations with host galaxy properties}\label{sec:hostcorr}
A correlation of TDE light curve features with host galaxy properties is anticipated because the black hole mass and the density of the disrupted star should influence the TDE light curve. Black hole mass estimates from the stellar velocity dispersion are not (yet) available for most TDEs in our sample and we therefore use the total host galaxy mass obtained from the host photometry (\S \ref{sec:host}) as a proxy for black hole mass.  

In Fig.~\ref{fig:vsmass} we show a number of TDE properties as a function of total galaxy mass and in Table~\ref{tab:Kendall} we list the significance. We find a significant correlation between stellar mass and blackbody luminosity, $p=0.04$ for a Kendall's tau test. Using this statistic, the strongest correlations are found for the monochromatic decay timescale measured during the first 100~days ($\tau$, $p=0.03$) and the normalization of a power-law fit to the decay of the bolometric light curve ($t_0$, $p=0.01$). The host galaxy mass also appears to be correlated with the fallback timescale (i.e., $t_0$ obtained when $p=-5/3$), but the scatter is larger the correlation is weaker ($p=0.11$). 

We also looked at the principal components of the host galaxy population synthesis parameters (see \S \ref{sec:host}) and found further support that the host galaxy stellar population properties contain information about the TDE properties. Using all 39 TDEs, we find a significant correlation ($p=0.02$) between the blackbody radius and PC4 (which is dominated by the dust optical depth and starformation rate e-folding time). We note that PC4 shows no correlation with redshift or host galaxy mass. We also find an equally strong correlation between the second PCA component, which is  dominated by the mass and metalicity parameter, and the TDE rise time. This is interesting because the correlation between rise time and stellar mass alone is not very strong ($p=0.26$).

\section{Discussion}\label{sec:discussion}
\begin{figure*}

\gridline{	\fig{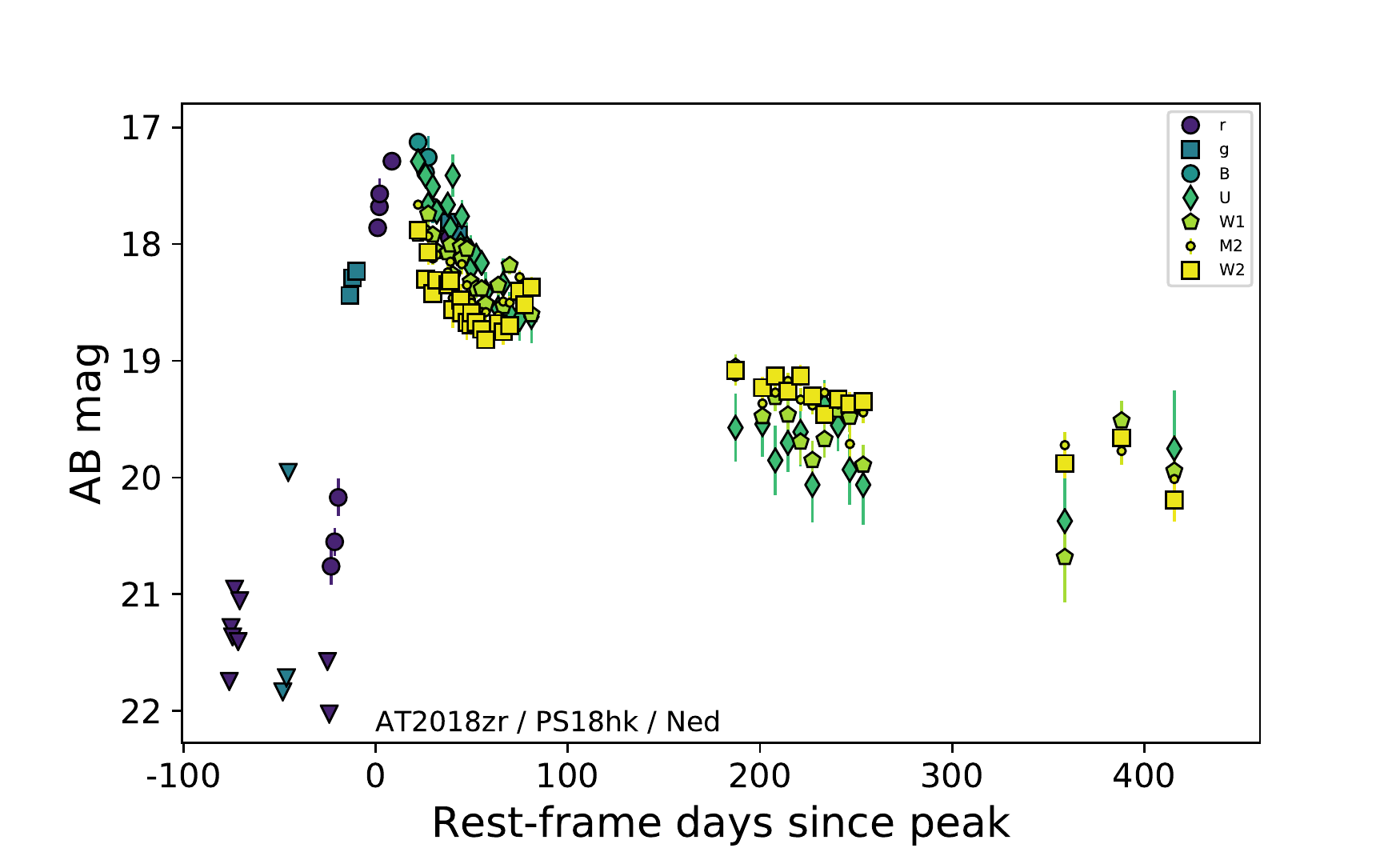}{0.4 \textwidth}{} 
			\fig{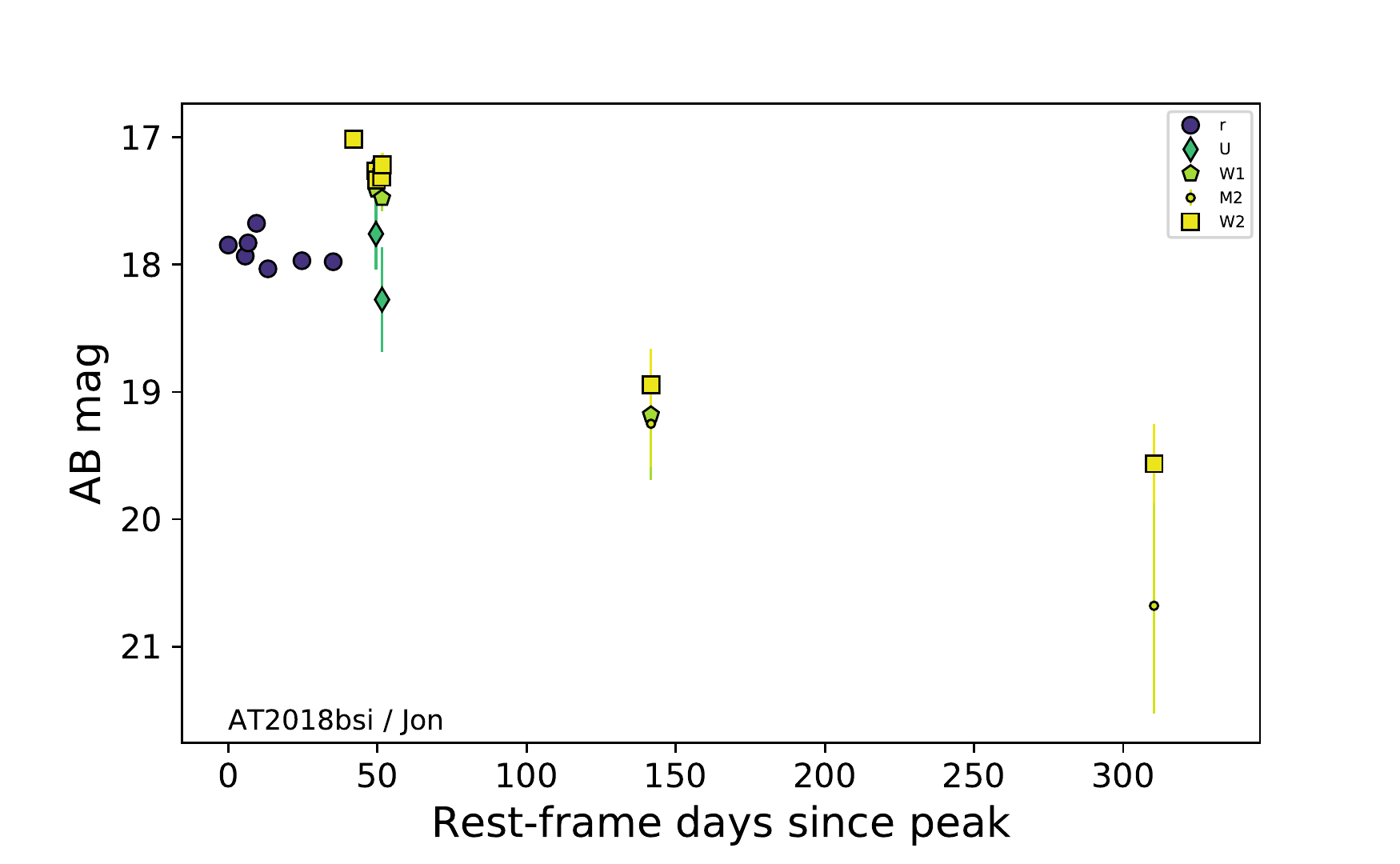}{0.4 \textwidth}{} 
			\\[-38pt]}
			
\gridline{  \fig{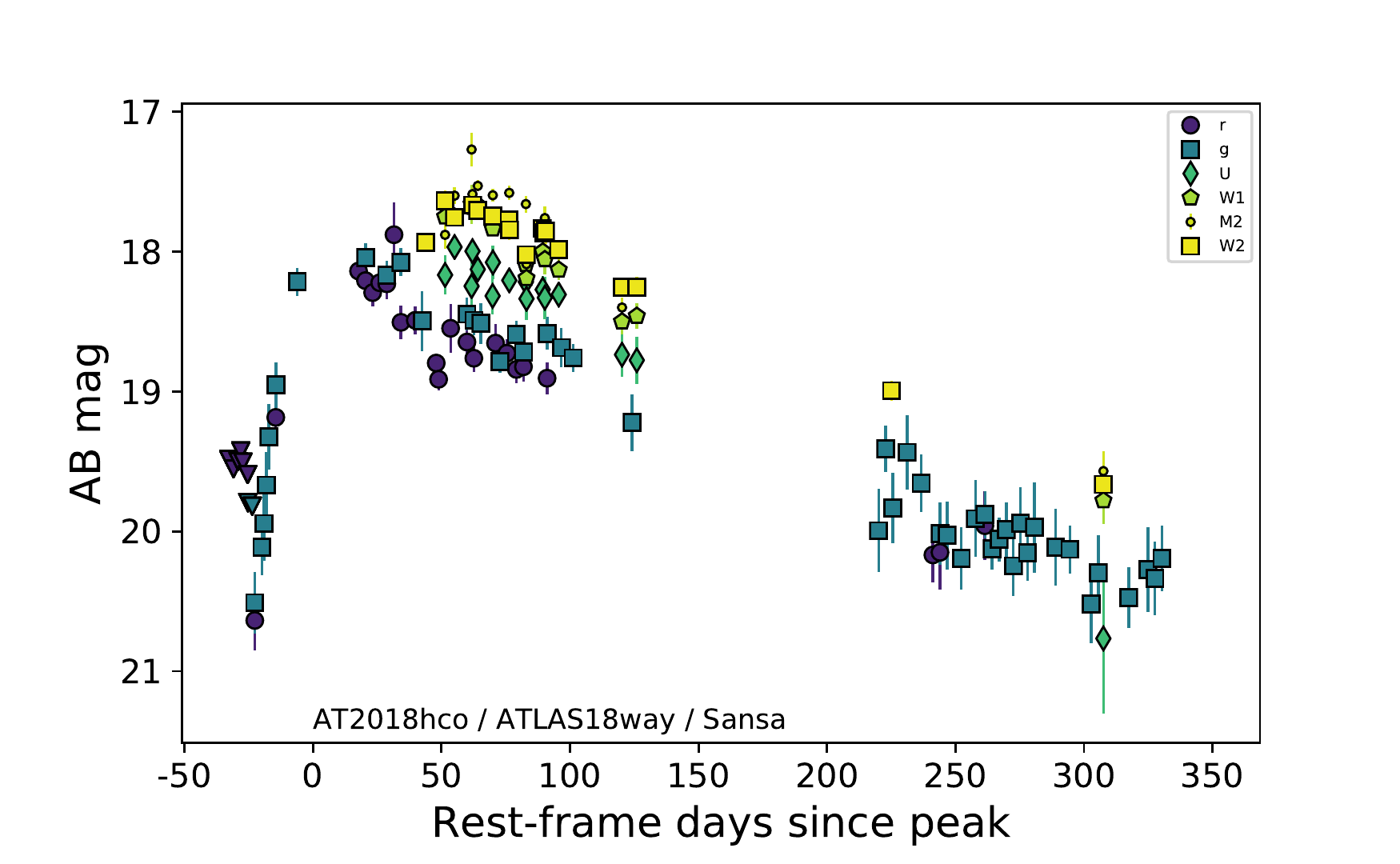}{0.4 \textwidth}{}
            \fig{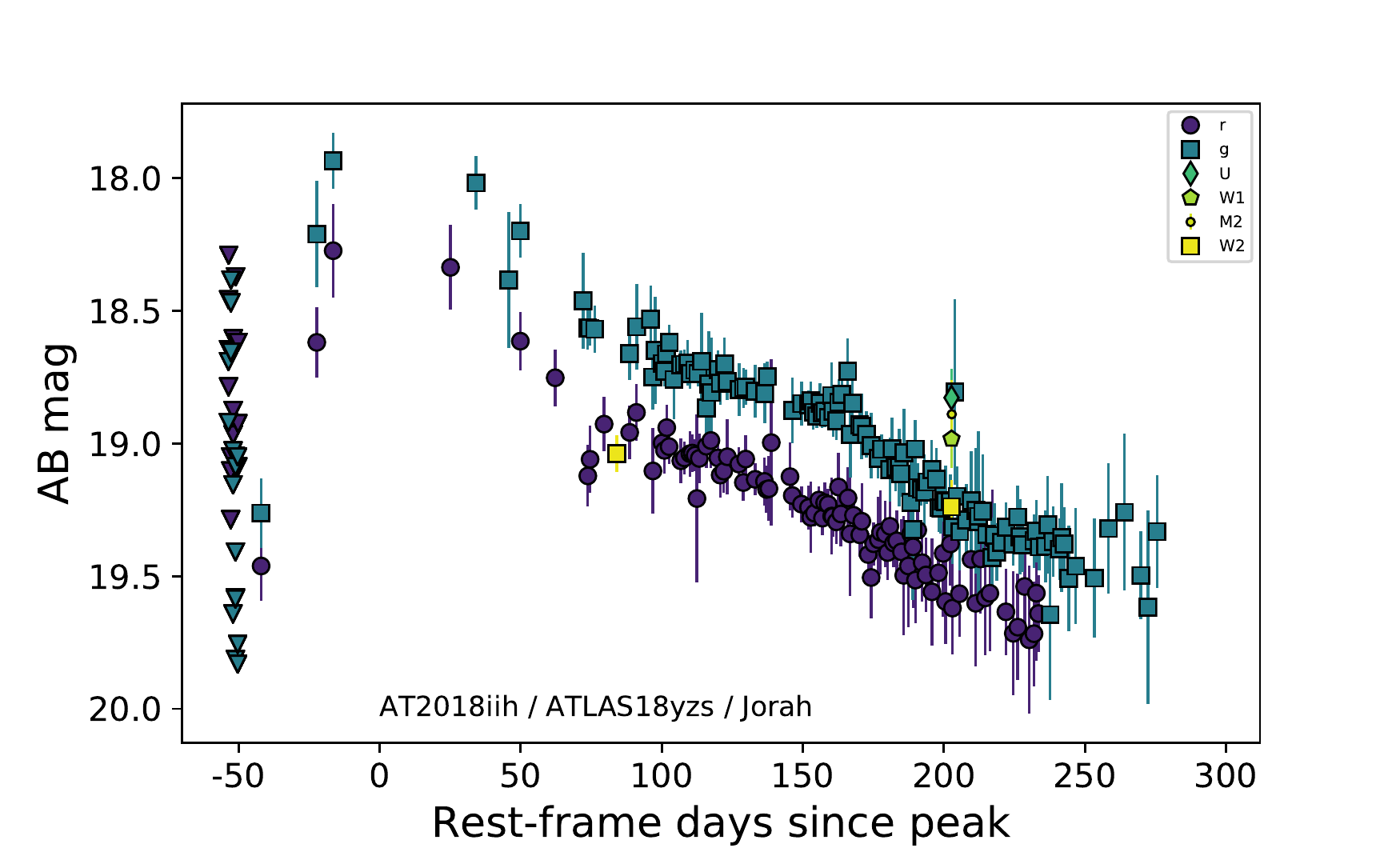}{0.4 \textwidth}{} 
            \\[-38pt]}
            
\gridline{  \fig{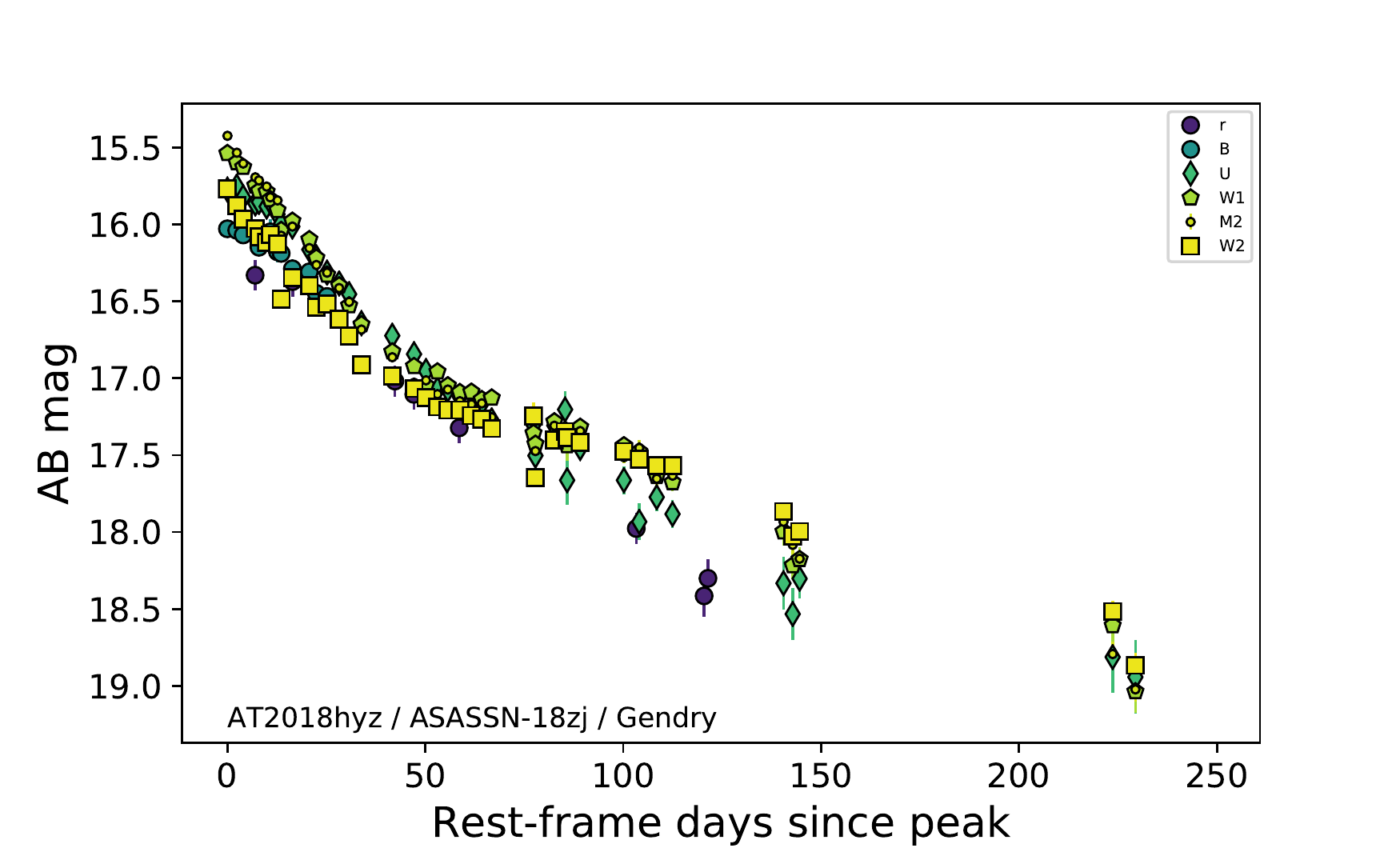}{0.4 \textwidth}{} 
            \fig{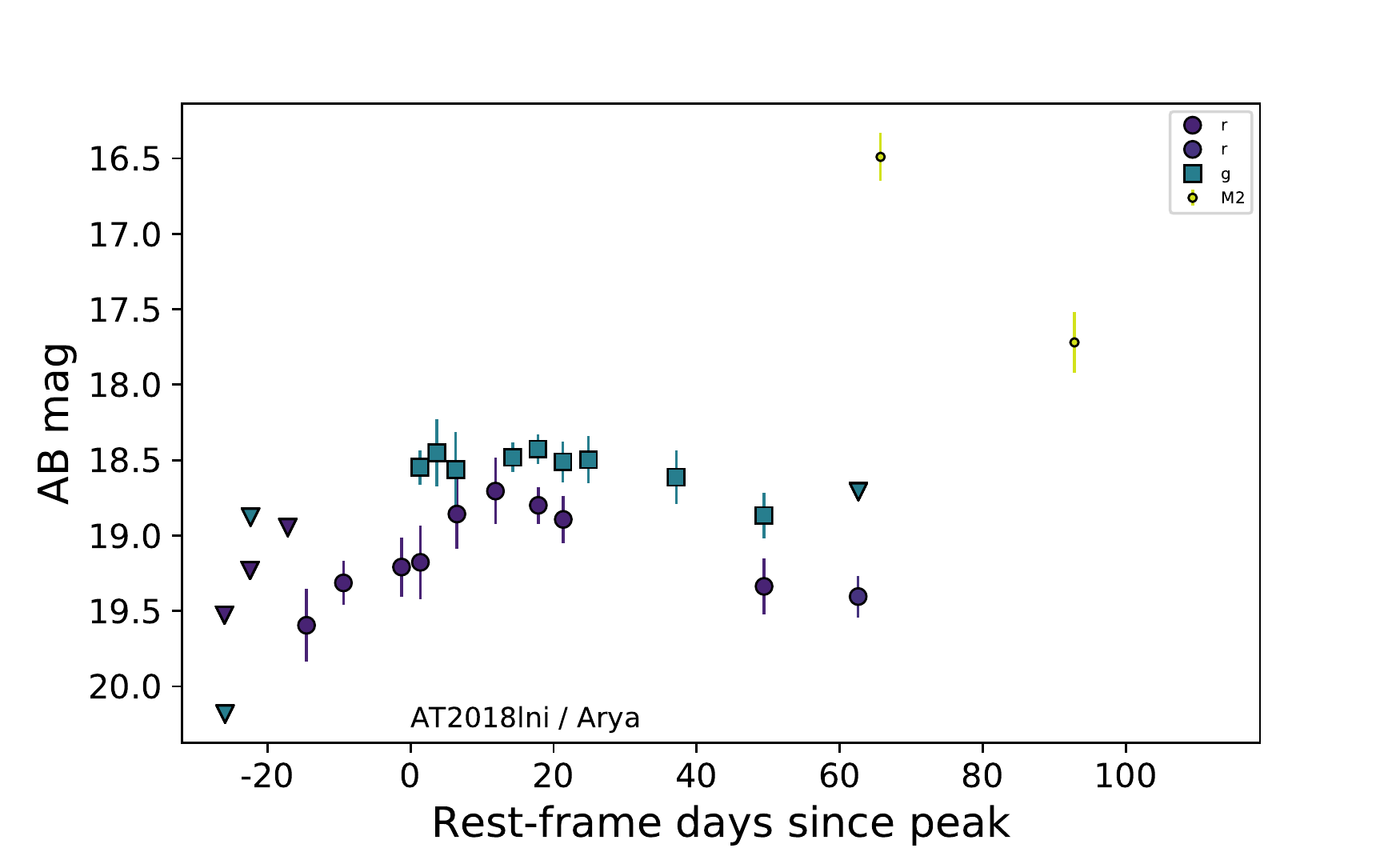}{0.4 \textwidth}{}
            \\[-38pt]}          

\gridline{  \fig{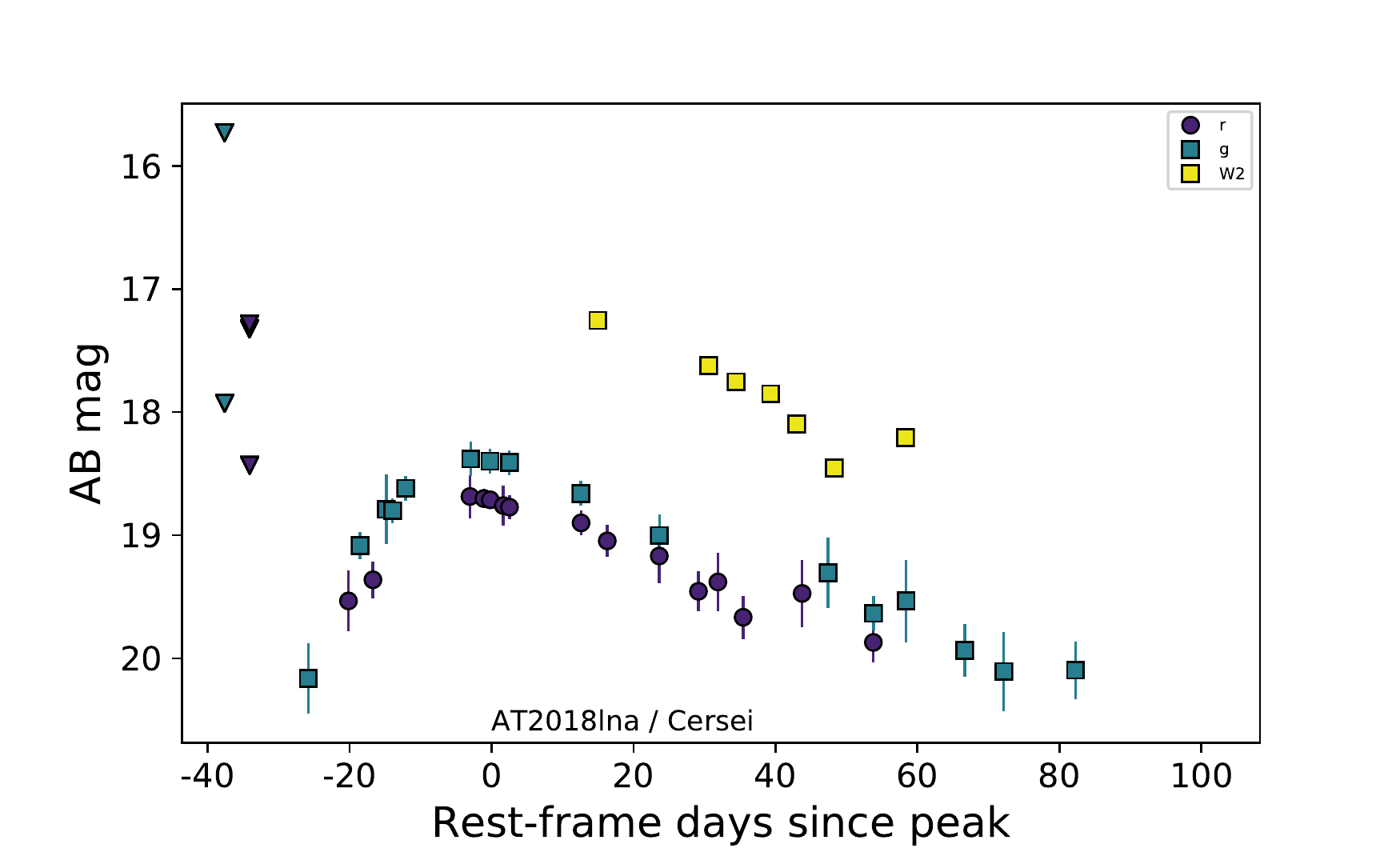}{0.4 \textwidth}{}
            \fig{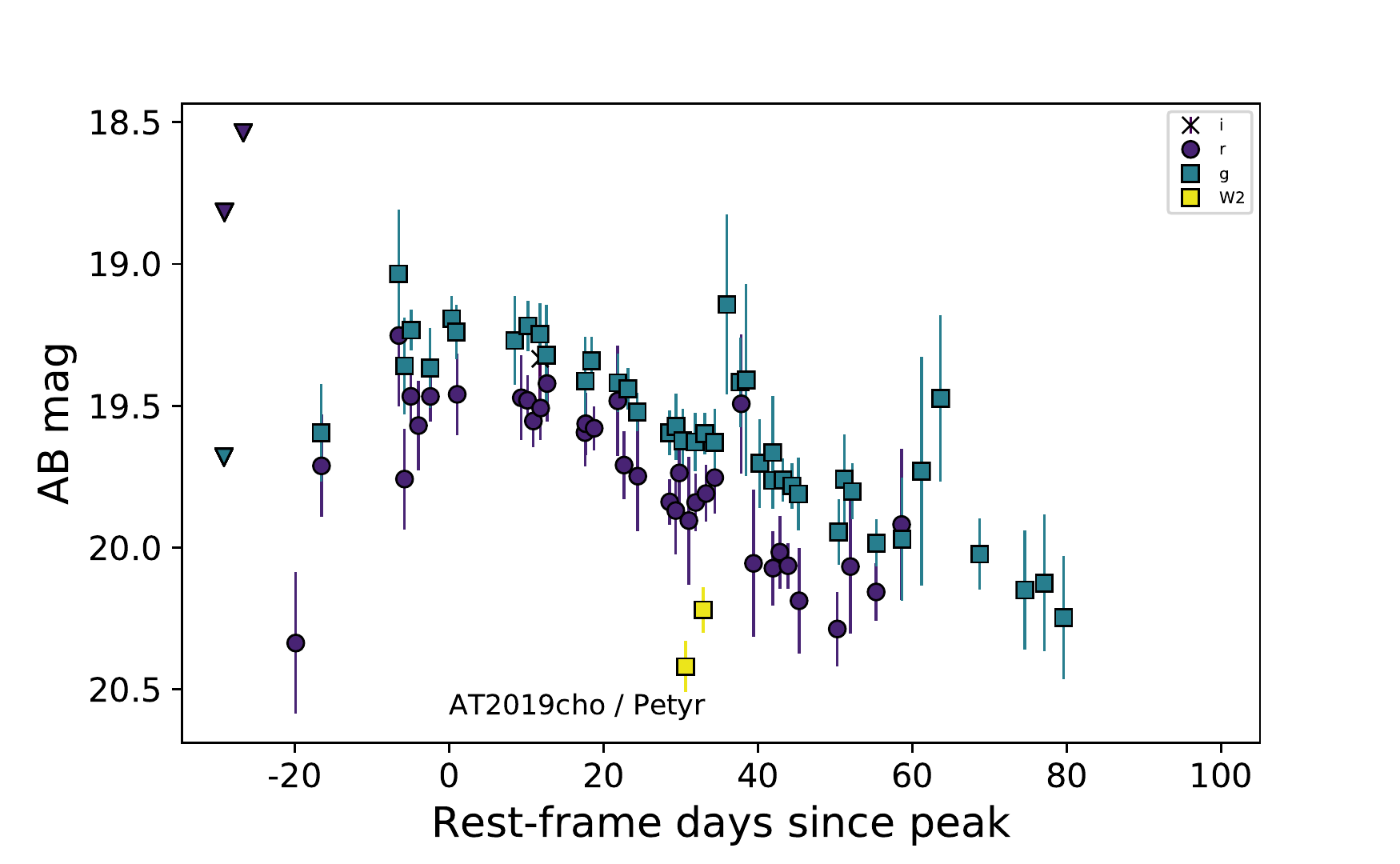}{0.4 \textwidth}{}
             \\[-38pt]}

\gridline{  \fig{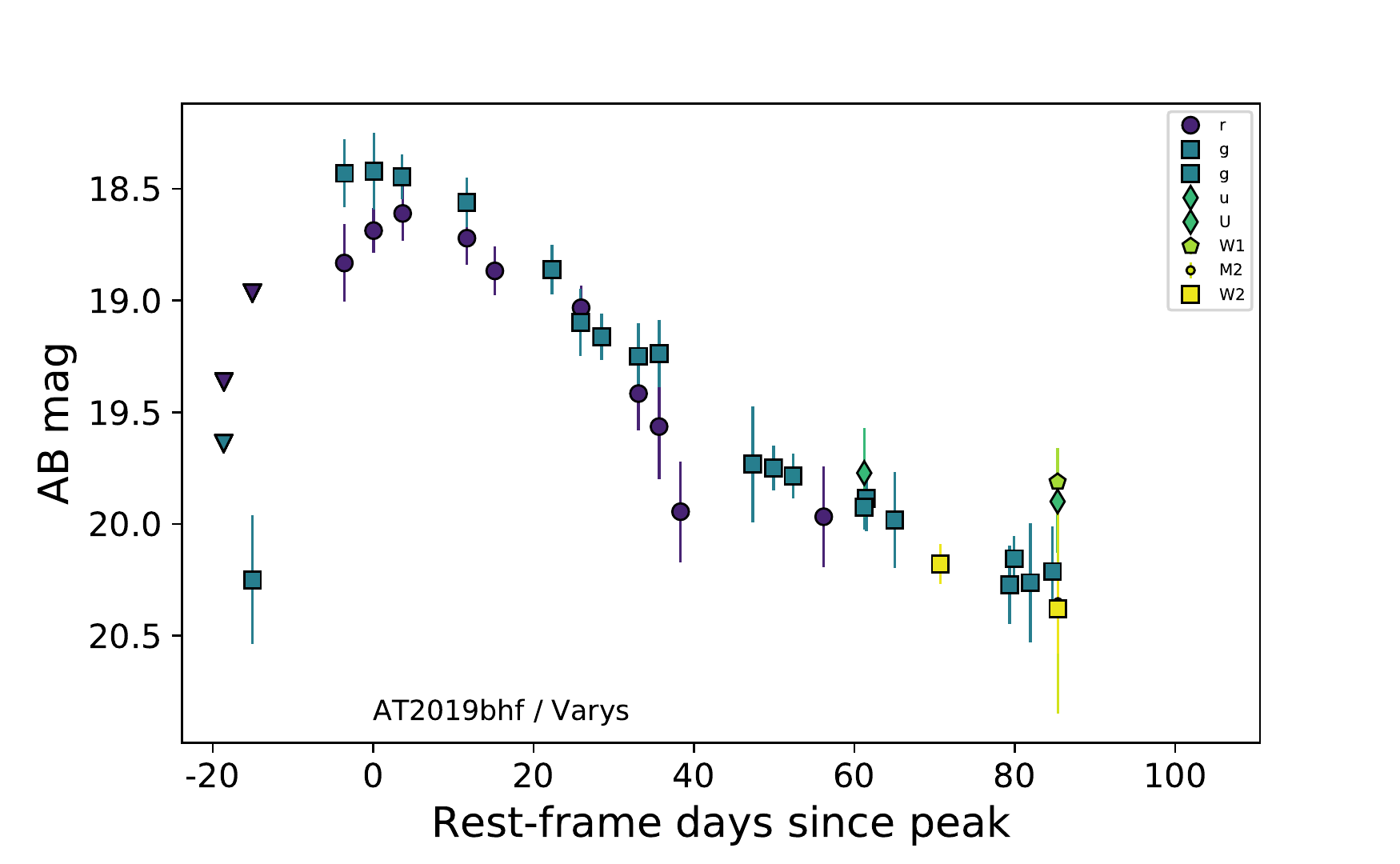}{0.4 \textwidth}{} 
            \fig{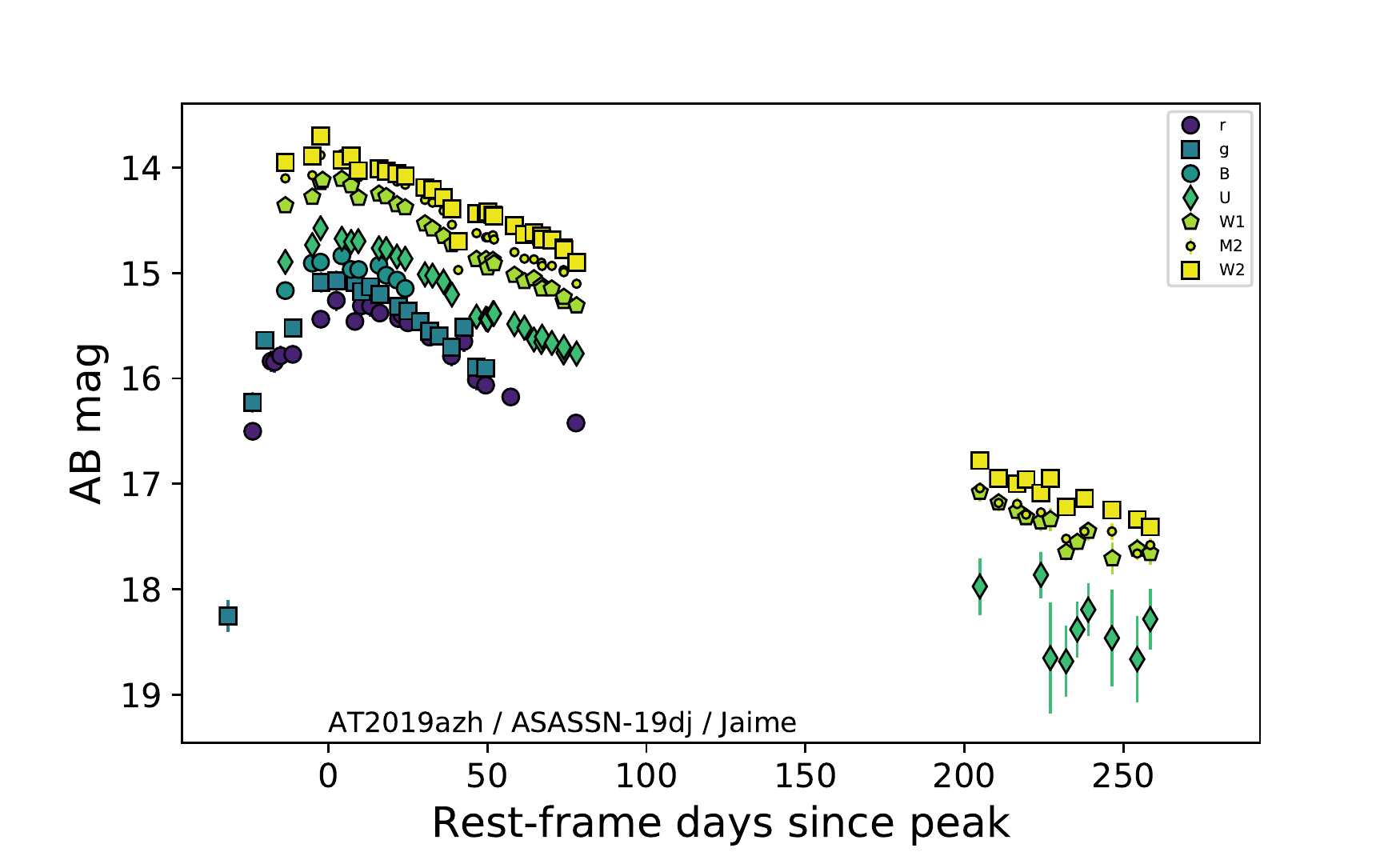}{0.4 \textwidth}{}
            \\[-10pt]}   

\caption{Opitcal/UV light curves based on ZTF, {\it Swift}/UVOT, SEDM, and LT photometry. Arrows indicate 5$\sigma$ upper limits, based on the ZTF alert photometry.}\label{fig:lcs1}
\end{figure*}

The main challenge for a TDE emission model is to turn the fallback rate of the stellar debris, which can be calculated or simulated with reasonable accuracy  \citep[e.g.,][]{Lodato09,Guillochon13}, into an electromagnetic output.  
%Rees88,EvansKochanek89
As shown in Fig.~\ref{fig:lcs1}, when TDEs are observed for longer than $\sim 100$~days a power-law decay is required to explain the observed light curves. The median power-law index of the 39 TDEs in our sample is $p=-1.6$ which is close to the value expected for the full disruption of a star, $p=-5/3$. This result has been noticed in earlier, smaller samples of TDEs \citep[e.g.,][]{Gezari09,Piran15,Hung17} and is an important motivation to construct TDE emission models that couple the (post-peak) bolometric luminosity to the fallback rate   \citep{Guillochon14,Piran15,Krolik16,Mockler18,Bonnerot19}. 

In this work we find a correlation between the decay time scale and total galaxy stellar mass, which is consistent with previously detected correlations between decay time and black hole mass  \citep{Blagorodnova17,Wevers17}. This supports the idea that the fallback timescale can be measured from the post-peak TDE light curve. Indeed, \citet{Mockler18} find that the light curves of an earlier sample of TDEs (most without pre-peak detections) are consistent with ``prompt" emission (i.e. the light curves that directly trace the theoretical fallback rate).

Our new sample contains 21 spectroscopic TDEs with well-measured rise-times, providing a new regime to test models for the emission mechanism. We find no correlation between the rise timescale of the light curve and the decay timescale nor any significant correlation of the rise time with total galaxy mass.  In the next section we discuss the lack of correlation between rise time and galaxy stellar mass (acting as a proxy for black hole mass) is in fact expected for two separate theoretical scenarios of optical emission from TDE. 

\subsection{Photon Advection and Diffusion}\label{sec:diff}
In the model by \citet{Metzger16}, the optical radiation will be advected through an outflowing wind until it reaches the trapping radius ($R_{\rm tr}$), the location at which the radiative diffusion time through the remaining debris is shorter than the outflow expansion time. It is useful to introduce the trapping time $t_{\rm tr}$, which is the time photons are losing a significant amount of energy from being trapped in the wind and adiabatically transferring energy to the outflow. For low mass black holes, $M_{\rm BH} \lesssim 7 \times 10^{6} M_\odot$,  \citet{Metzger16} find that $t_{\rm tr} > t_{\rm fb}$, and thus adiabatic losses suppress and delay the peak of the TDE light curve. In this case, the predicted correlation between the peak luminosity  $L_{\rm pk}$ and $M_{\rm BH}$ is extremely weak, $L_{\rm pk} \propto M_{\rm BH}^{0.06}$. This results from a cancellation of effects: for larger $M_{\rm BH}$, the longer $t_{\fm fb}$ causes the accretion rate powering the outflow to be lower, but the photons are also less trapped in the outflow and so retain more of their energy. This weak correlation between $L_{\rm pk}$ and $M_{\rm BH}$ also manifests as a weak correlation between $t_{\rm pk}$ and $M_{\rm BH}$, again for these lower black hole masses. At higher black hole mass, the relations $L_{\rm pk} \propto M_{\rm BH}^{-1/2}$ and $t_{\rm pk} \propto M_{\rm BH}^{1/2}$, as expected from the mass fallback relations, should reappear.

Alternatively, in the description of \citet{Piran15}, there is no outflow, and the size of the UV/optical emitting region is tied to the apocenter of the most bound stellar debris. For sufficiently low mass black holes, the diffusion time $t_{\rm diff}$ for photons to escape the shock-heated debris will provide a timescale that must be convolved with the shock heating rate set by $t_{\rm fb}$, to produce the final light curve. For sufficiently low black hole mass, $t_{\rm diff} $ may become long enough that the black hole mass dependence inherent in $t_{\rm fb}$ may be washed out by the diffusion time, which is itself more strongly correlated with the mass and structure of the disrupted star than with the black hole mass. For higher mass black holes, as was the case for the \citet{Metzger16}, these radiative transfer effects should diminish in importance, we again expect the mass fallback relations to dictate the shape of the light curve. 

To conclude, the lack of significant correlations between light curve rise time and host galaxy stellar mass in our sample of TDEs could be explained by photon advection or diffusion, since these will interfere with seeing an unmitigated signal from the mass fallback rate. 

The diffusion timescale for electron scattering scales as $t_{\rm diff} \propto \rho R^2$ \citep[e.g.,][]{Metzger17}. For a spherical distribution of mass within the photosphere radius $R$ we find $t_{\rm diff} \propto M/R$. If the blackbody luminosity ($L_{\rm bb}$) is proportional to this mass, we obtain  $t_{\rm diff} \propto L_{\rm bb}/R$. This scaling of the diffusion time and the ratio of blackbody luminosity and radius could explain the observed correlation between $L_{\rm bb}/R$ and the rise timescale (Fig.~\ref{fig:rise}). 

\begin{figure}
\centering
\includegraphics[trim=3mm 2mm 8mm 8mm, clip, width=0.48 \textwidth]{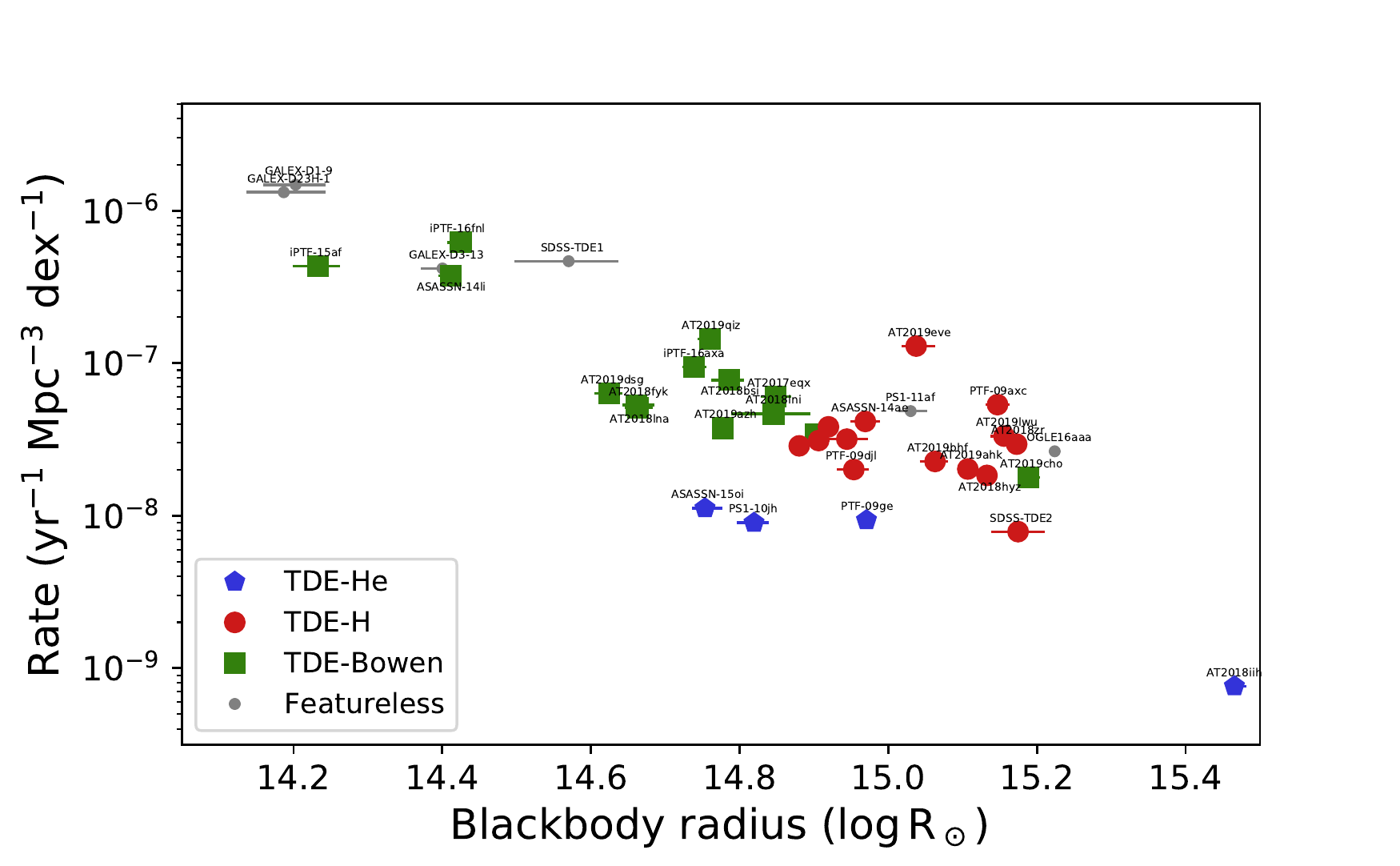}
\caption{The approximate event rate as a function of blackbody radius at peak. For each TDE, we plot the rate corresponding to its $g$-band luminosity, using the empirical luminosity function of \citet{vanVelzen18}. For the spectroscopic TDEs we scale the rate by the fraction of events in each class. We see a strong decrease of the event rate with blackbody radius (roughly scaling as $dN/dR\propto R^{-3}$). If the photosphere is proportional to the mass of the disrupted star (\S \ref{sec:ratemass}), this trend can be explained by the slope of the initial mass function of stars (i.e., low mass stars, and thus small radii, are more common).  }\label{fig:rate}
\end{figure}

\subsection{Event Rate and Stellar Mass}\label{sec:ratemass}
An important property of the TDE-Bowen class is their low optical luminosity. Since they are detected in equal numbers as the H-only class, this low luminosity implies a higher intrinsic rate. To estimate the magnitude of this effect we use the empirical $g$-band luminosity function of TDEs \citep{vanVelzen18} to assign a rate to each TDE based on its observed $g$-band luminosity. To make an approximate correction for the low number of He-only TDE we scale the rate from the luminosity function using the relative number of TDEs in each spectroscopic class. In Fig.~\ref{fig:rate} we show the result as a function of blackbody radius; we find a steep dependence on radius, $dN/dR\propto R^{-3}$. 

\begin{figure}
\includegraphics[trim=0mm 4mm 0mm 20mm, clip, width=0.48\textwidth]{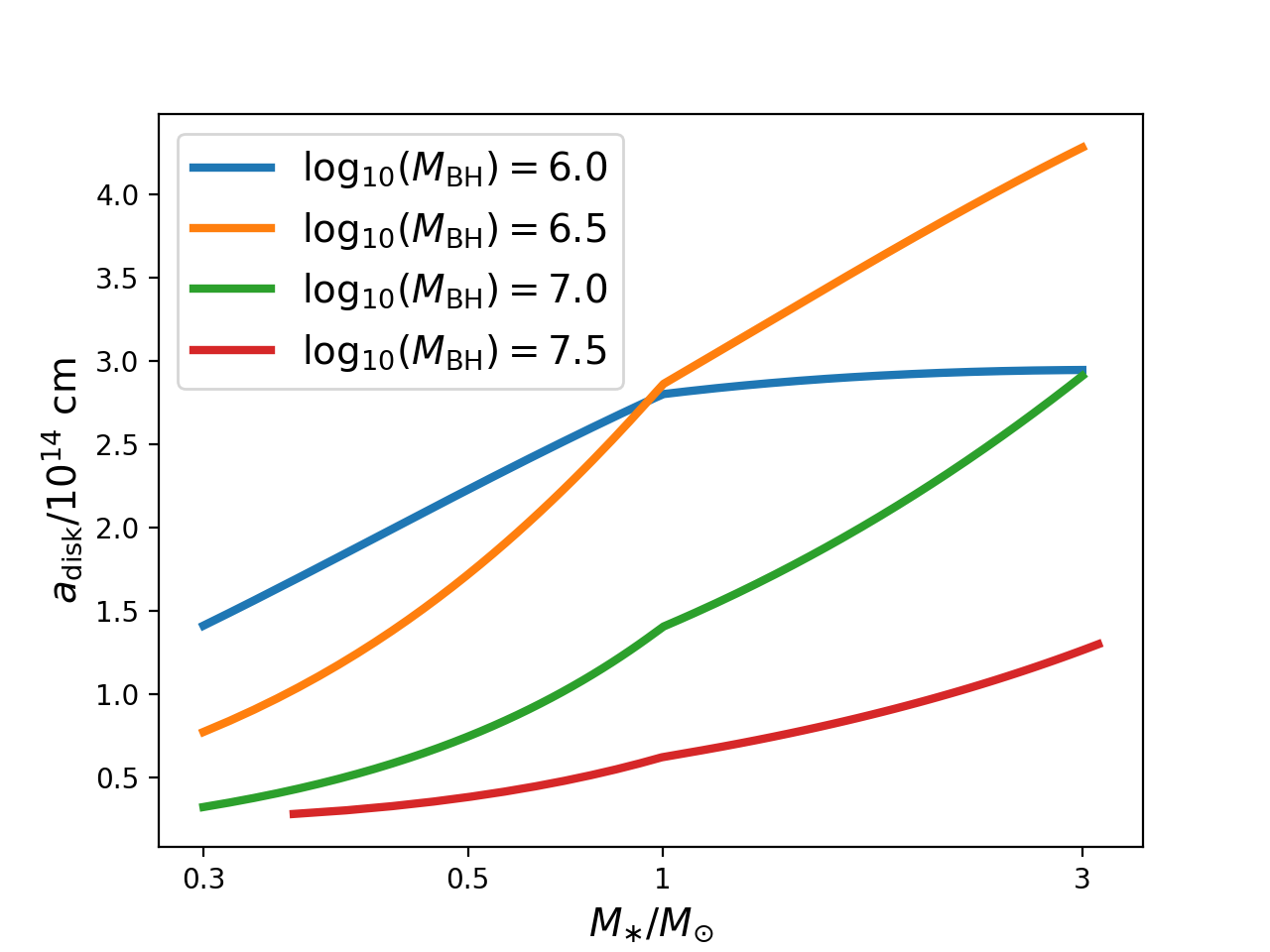}
\caption{The estimated size of the disk following dissipation at the stream intersection shock, computed following \cite{Dai15}. We show this radius as a function of stellar mass and black hole mass, for stellar orbits with a pericenter equal to the tidal radius ($\beta=1$, which are expected to be most common). We see that the disk size decreases with stellar mass. Since low-mass stars are more numerous and thus get disrupted more often, this connection between stellar mass and disk radius could explain the observed decrease of the disruption rate as a function of blackbody radius (Fig.~\ref{fig:rate}). }\label{fig:dai}
\end{figure}

If the optical/UV blackbody radius would be proportional to the mass of the star, we obtain a potential explanation for steep decline of the event rate with blackbody radius (Fig.~\ref{fig:rate}). Such a scaling is in fact expected if the photosphere is proportional to the self-intersection radius of stream. 
In Fig.~\ref{fig:dai} we show the radius of an accretion disk created from energy dissipated at the  self-intersection radius as a function of black hole mass and stellar mass, obtained using the formalism\footnote{\citet{Dai15} provide formulas for the location of the stellar debris intersection and estimates for the resulting rate of energy dissipation at the intersection shock, by using an impulse approximation for general relativistic precession, to leading post-Newtonian order. In this way they derive an approximate formula for the semi-major axis of the elliptical disk that forms immediately following the stream intersection, which depends on the mass of the star, the mass of the black hole, and the dimensionless impact parameter $\beta$.} of \citet{Dai15} and the mass-radius relations for high/low mass main-sequence stars from \citet{Kippenhahn90}. 
If we assume that the blackbody photospheric radius $R$ is proportional to the disk size, from Fig.~\ref{fig:dai} we see that, all else being equal, lower mass stars are associated with smaller values of $R$; at $M_{\rm BH}=10^{6.5}\,M_\odot$, $\log(R) \approx 0.8 \log(M_*)$. Using this result, we find that smaller blackbody radii imply higher blackbody temperatures, consistent with the observed scaling $T\propto R^{-1/2}$ (Fig.~\ref{fig:RT}). On the Rayleigh–Jeans tail we have $L_\nu \propto L_{\rm bb}^{1/4} R^2 T \propto L_{\rm bb}^{1/4} R^{3/2} \propto L_{\rm bb}^{1/4} M_*^{1.2}$. For a typical initial mass function ($dN/dM~\propto M^{-2.3}$), the higher number density of low-mass stars thus provides a simple explanation for the steep decrease of the event rate with blackbody radius ($dN/dR~\propto R^{-3}$, Fig.~\ref{fig:rate}).

The greatest distinction between the TDE-H and TDE-Bowen classes, is that the latter is characterized by smaller radii (e.g., Fig.~\ref{fig:rise} or Fig.~\ref{fig:cumu}). The Bowen fluorescence mechanism requires both a high flux of EUV photons and a high gas density. The observed blackbody luminosity is similar for the TDE-Bowen and TDE-H class, which leads us to consider that small radii of the TDE-Bowen class can be explained by the high-density conditions that enable their fluorescent lines. If we again assume that the optical/UV blackbody radius is related to the stream self-intersection radius (Fig.~\ref{fig:dai}) the density within a spherical emission region ($\rho \sim M_* / R^3)$ will decrease with stellar mass.

It thus appears that we can explain both a higher density within the photosphere and a higher rate of the TDE-Bowen class with the disruption of lower mass stars. However this scenario may not explain a third distinct property of Bowen TDEs: their longer rise time. The diffusion timescale for electron scattering scales as $t_{\rm diff} \propto R^2 \rho$ or $t_{\rm diff} \propto M/R$ for a spherical distribution of mass. In most TDE emission models, this leads to a positive scaling of the diffusion time with stellar mass \citep[][]{Piran15,Metzger17}, although we can speculate that a negative scaling of $t_{\rm diff}$ with stellar mass could be obtained if the mass that is relevant for electron scattering is decoupled from the mass of the disrupted star (e.g., due to the Eddington limit). The high densities that are required for the production of Bowen lines could also be obtained by disruptions with a high impact parameter $\beta$. However such plunging orbits are rare (in the empty loss-cone regime $\beta\leq 1$ and for a full loss-cone $dN/d\beta \propto \beta^{-1}$; \citealt{Lightman_Shapiro77}) and therefore not consistent with the observation that Bowen TDEs are common.  

The TDE-He class presents an interesting case to test the idea that the photosphere radius is set by properties of the disrupted star. The He-only class has the longest rise times and highest luminosities, yet relatively large photosphere radii (Fig.~\ref{fig:rise}). 
%While the high luminosity could be explained by higher mass stars, for main-sequence stars with $M> 1 M_\odot$ in the increase in the diffusion time is modest (Eq.~\ref{eq:diff}). 
We either need a high mass star that is relatively dense, or a high-mass star and a high impact parameter. Both of these scenarios would explain why this spectroscopic class is rare. The lack of H emission would then be a signature either of the lack of hydrogen in a dense He star \citep{Gezari12}, or the radiative transfer effects in a dense reprocessing region \citep{Guillochon14,Roth15}. 

\subsection{Correlations with Host Galaxy Stellar Population Parameters}
The observation that the TDE-H and TDE-Bowen class show a significantly different distribution of stellar population parameters (Table~\ref{tab:KS}) could in principle be used to shed light of which stellar properties (e.g., density, composition, or impact parameter) influence the TDE light curve and spectrum. Unfortunately, the interpretation the population parameters PCA is not straightforward. 

The principal component that yields the largest separation of the two TDE classes is mainly driven by the dust optical depth minus the star formation decay e-folding time ($\tau_{\rm sfh}$), such that the TDE-Bowen class have higher value of $E(B-V) - \tau_{\rm sfh}$. Yet the Bowen TDEs also have the highest blackbody temperatures, implying they are not systematically affected by dust in their host galaxies. The PCA can also capture degeneracies between the population synthesis parameters. Based on the typical posterior distributions it appears that $\tau_{\rm sfh}$ is correlated with the age of the stellar population, while the dust content shows a negative correlation with age. At this point we can speculate that the age of the stellar population is the underlying cause for the difference between the TDE spectroscopic classes. But the correct path forward is to improve our inference of the stellar population properties by including additional information, such as the WISE photometry and the absorption line diagnostics \citep[e.g.,][]{French17}. This will be the subject of future work.

\subsection{Surprises from the Optical/X-ray Ratio}

Most models that use reprocessing of photons from close to the black hole to explain the observed optical emission of TDEs predict that at some point the reprocessing layer becomes transparent to X-rays. In the outflow model by \citet{Metzger17}, the inner wind becomes transparent to X-ray radiation once it is fully ionized by emission from the inner accretion disk, which happens at $t_{\rm ion} \approx 0.8 t_{\rm fb} (M_{\rm BH}/10^6M_\odot)^{-0.8} (M_* / M_\odot)^{0.4}$. At this point the reprocessing efficiency decreases, and one would expect an increase of the ratio of the X-ray to optical/UV luminosity ($L_X/L_{\rm bb}$). Alternatively, if our view of the inner accretion disk is unobscured and the optical emission originates from the stream intersection point \citep{Piran15}, an increase of the X-ray importance is evidence for delayed accretion onto the black hole \citep[as seen in the simulations of][]{Shiokawa15}. 

Since the inner accretion disk itself should also produce optical/UV emission \citep{Cannizzo90,strubbe_quataert09,Lodato11}, the optical luminosity is unlikely to completely vanish when the reprocessing layer is fully ionized. Indeed late-time observations of TDEs ($\sim {\rm few}$ years after peak) show a near-constant luminosity that is consistent with an accretion disk \citep{vanVelzen18_FUV,Jonker19,Mummery19}. 

The dramatic brightening of \Jaime in the soft X-rays 7 months after peak is similar to the behavior of TDE ASASSN-15oi \citep{Liu19}, which was interpreted by \citet{Gezari17} as a result of delayed accretion. However, the faint flux in the soft X-rays could also be explained by a suppression from adiabatic losses due to electron scattering \citep[][]{Dai18}. In contrast to the preferential suppression of soft X-rays associated with atomic absorption, these adiabatic losses leave a subtler imprint on the spectral slope of the attenuated X-ray spectrum, potentially consistent with the lack of strong evolution in the X-ray spectra when $L_{\rm bb}/L_X$ decreases. 

One TDE in our sample (\Brienne) shows a remarkable, and hitherto unseen, evolution of $L_{\rm bb}/L_X$. We observe three X-ray flares during the first months of post-peak observation (Fig.~\ref{fig:LbbLX}), increasing its X-ray luminosity by almost two orders of magnitude to $L_X \approx 5 \times 10^{44}$\,erg\,s$^{-1}$ on a timescale of days. The peak X-ray luminosity of the flares is just below the optical/UV blackbody luminosity measured at the same time (Fig.~\ref{fig:LbbLX_all}). If stream collisions are the main power source of early-time optical TDE emission (i.e., accretion is energetically unimportant), the X-ray flares of \Brienne~could in principle be explained by parcels of gas that are deflected toward the black hole from the stream collision site. However, the short timescale of the observed flare implies these discrete parcels would have to be aimed very precisely, which is not expected; the simulation by \citet{Shiokawa15} does show fluctuations in the mass accretion rate, but these occur on a timescale that is longer than the fallback time. We also note that it would be a coincidence that for each of the three flares the X-rays from the gas deflected by shocks reach $L_{\rm bb}/L_X\sim 1$. On the other hand, an equal amount of intrinsic accretion luminosity and observed optical/UV luminosity is a generic feature of a reprocessing layer with a high covering factor. In this scenario, the X-ray flares are not due to an increase of the accretion rate, but a decrease of the optical depth to our line of sight of the compact X-ray emitting region. 

Obtaining a few brief glimpses of the central soft X-ray emission would be possible for a reprocessing region that is moderately patchy. Since the size of this accretion disk is $\sim 100$ times smaller than the optical photosphere, our hypothetical patchy reprocessing layer can have many small ``gaps" that provide a view of the disk while the reprocessing efficiency remains high. Adapting the equation for the orbital timescale of a gas cloud at $r_{\rm orb}$ from \citet{LaMassa15}, we get a crossing timescale of 
\begin{equation}
t_{\rm cross} = 0.22 \left [ \frac{r_{\rm orb}}{\rm lightday} \right ]^{3/2} \left(\frac{M_{\rm BH}}{10^7 M_\odot}\right)^{-1/2} \arcsin \left [\frac{r_{\rm src}}{r_{\rm orb}} \right ] {\rm yr}.
\end{equation}
For a small gap the distance of the optical photosphere of $10^{14.5}$~cm = $0.1$~lightday, we get a crossing time of $2.5$ days, in agreement with the duration of the soft X-ray flares.

%Rapid X-ray flares have also been detected recently, XPEs

%Another prediction of the model for which X-ray emission is suppressed due to electron scattering, is that the line widths should decrease as the optical depth decreases \citep{Roth18}.  

\section{Conclusions}
\begin{figure*}

\gridline{	\fig{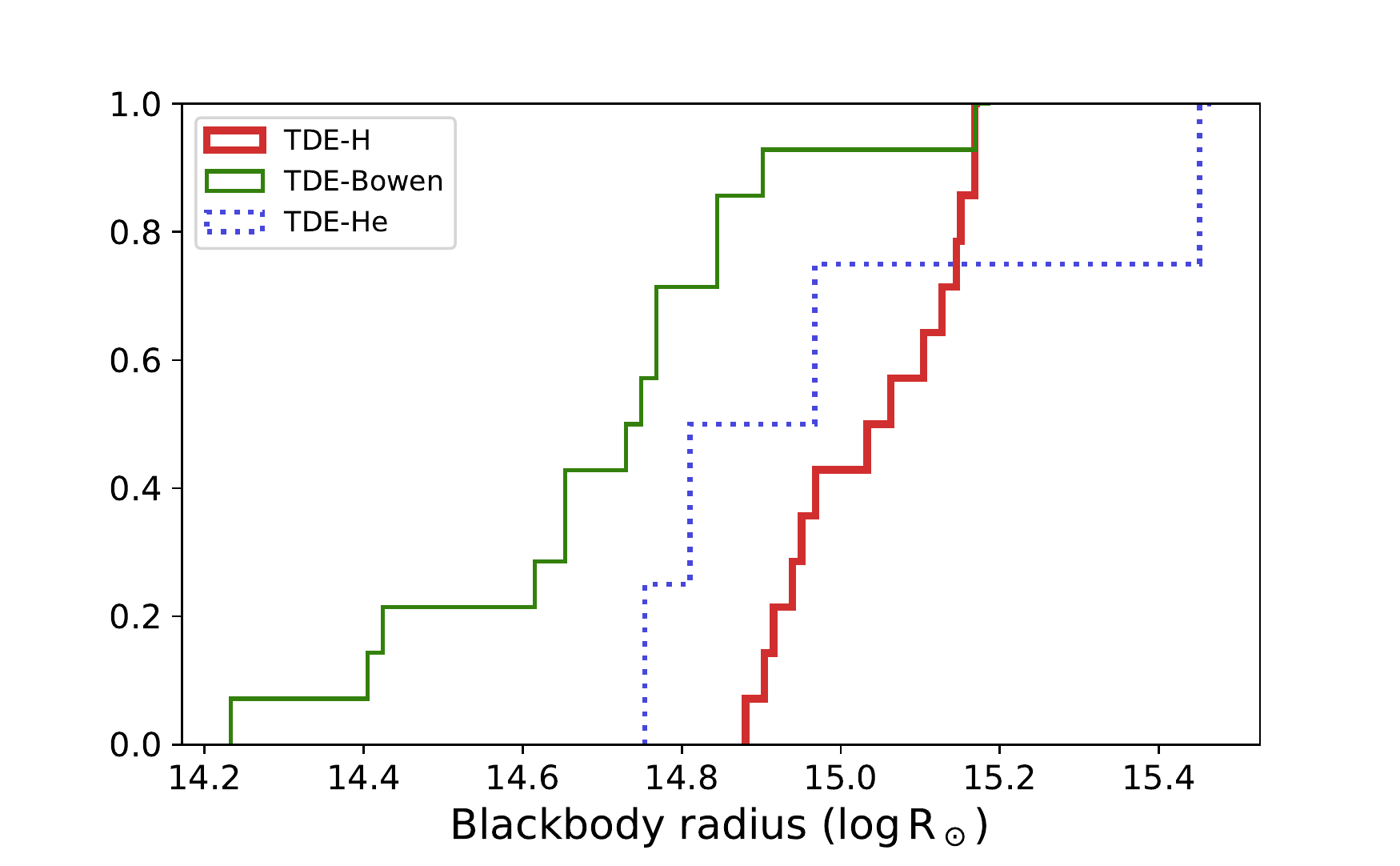}{0.48 \textwidth}{}
            \fig{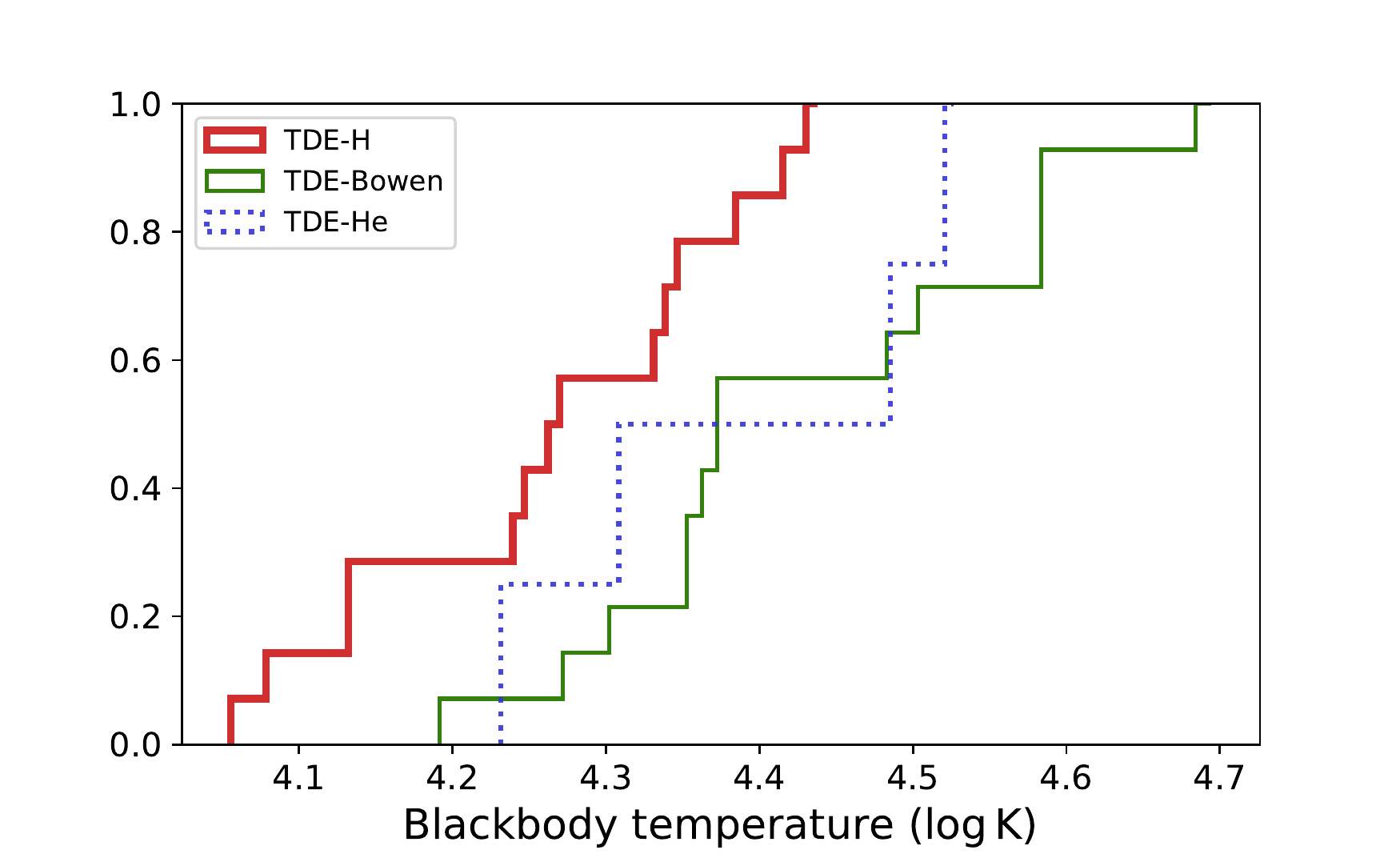}{0.48 \textwidth}{}
			 \\[-30pt]}

\gridline{	\fig{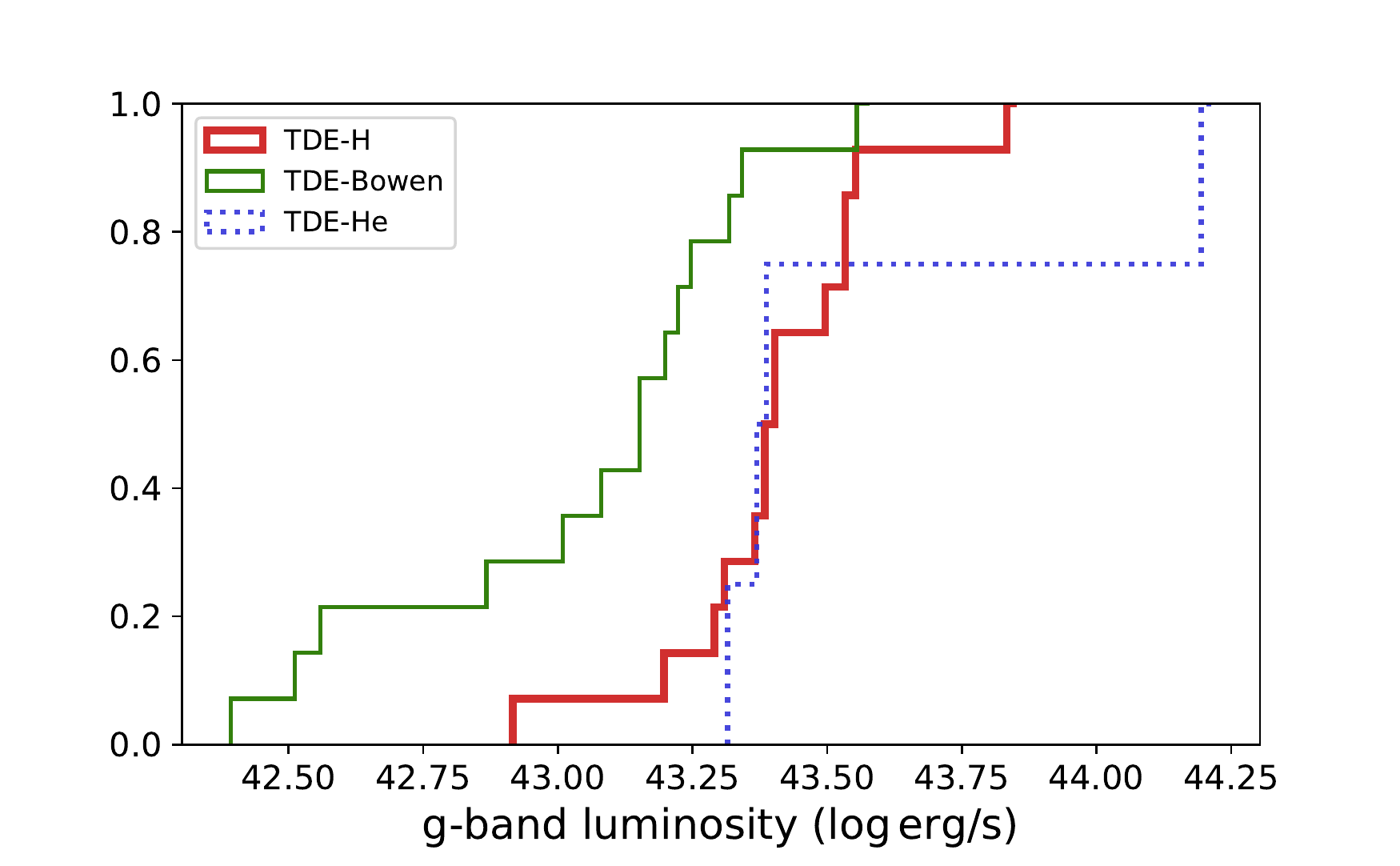}{0.48 \textwidth}{}
            \fig{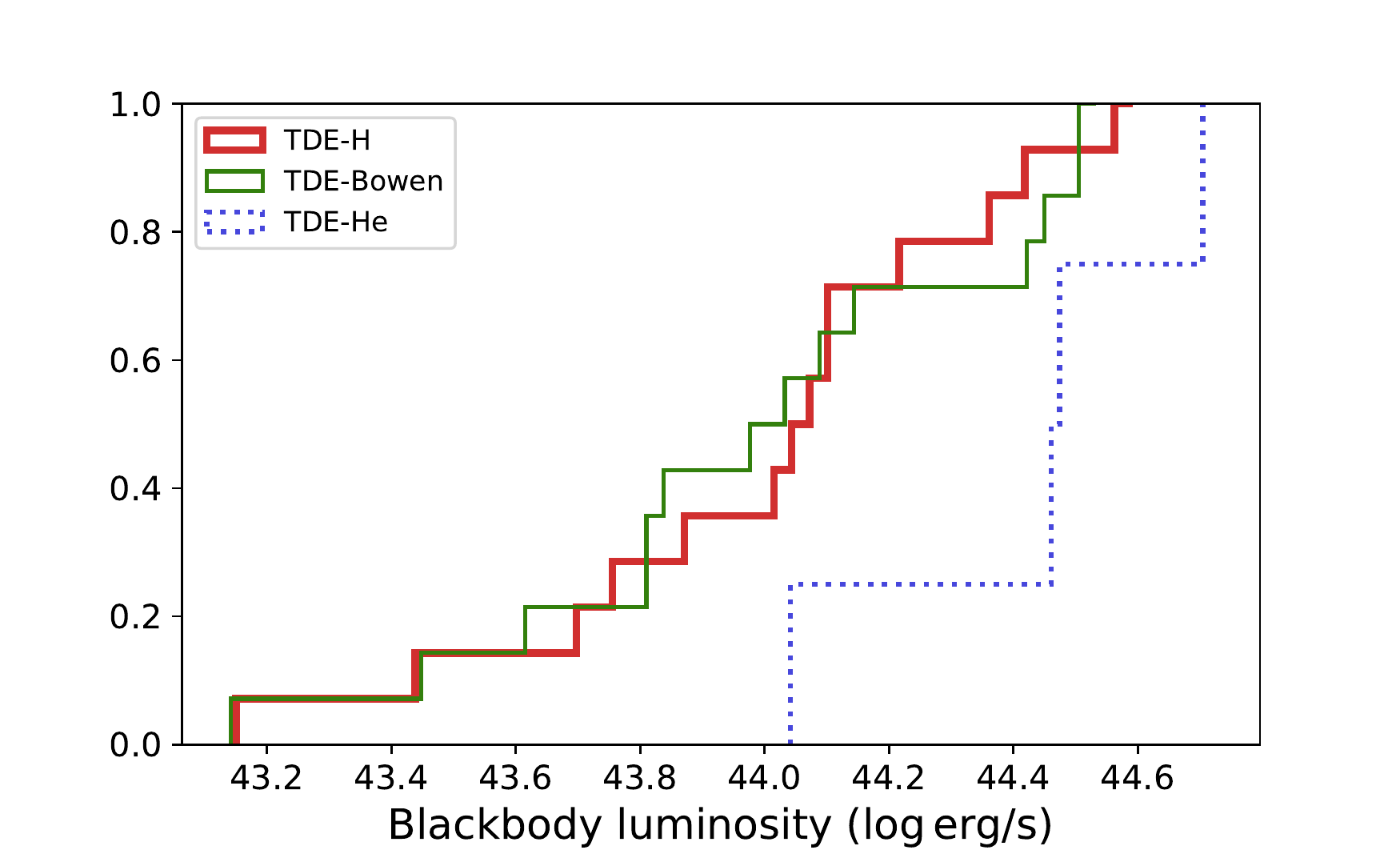}{0.48 \textwidth}{}
			 \\[-30pt]}			
			
\gridline{ \fig{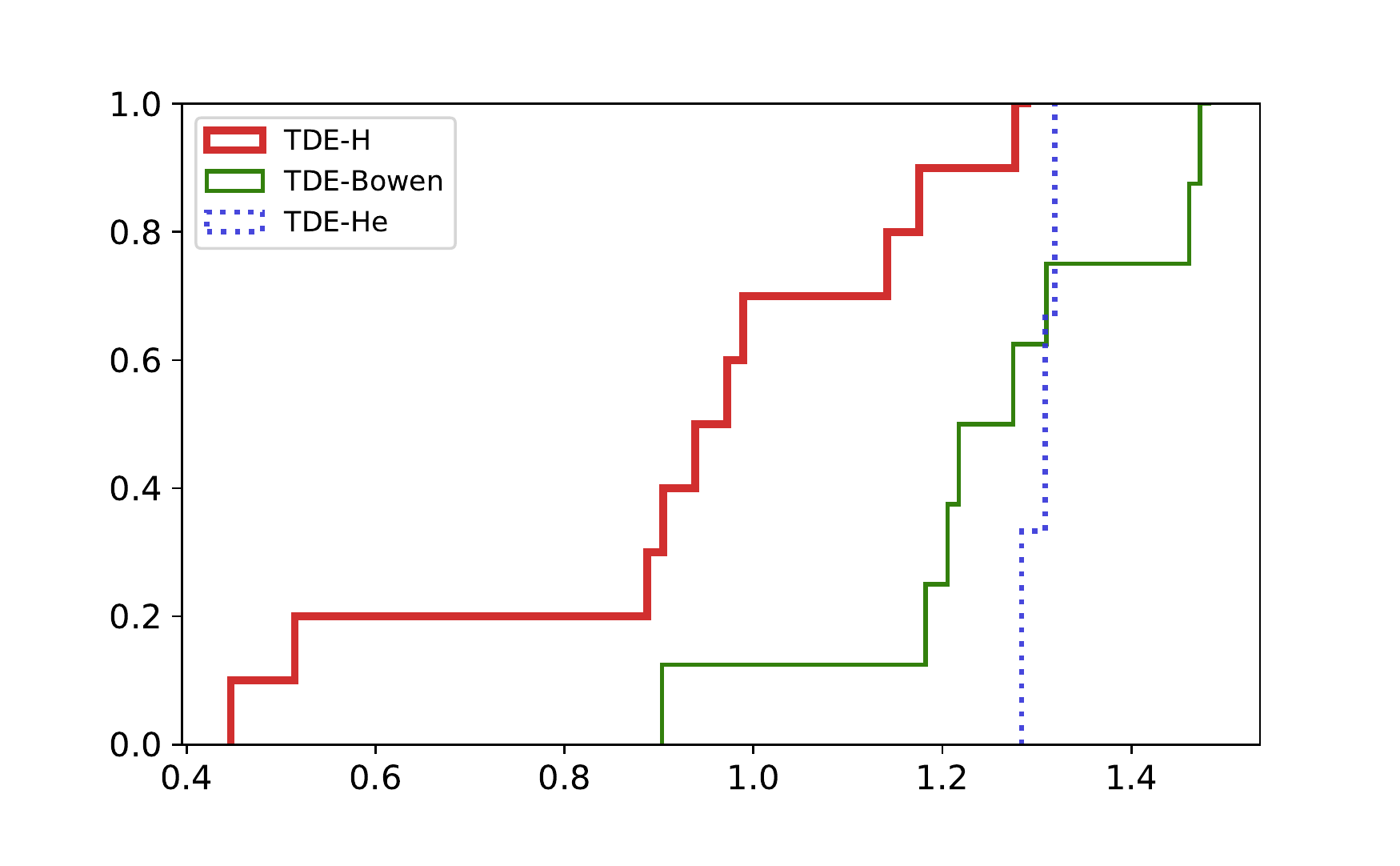}{0.48 \textwidth}{}              
            %\fig{t0_53_flex_cum.pdf}{0.48 \textwidth}{} \\[-30pt]}
            \fig{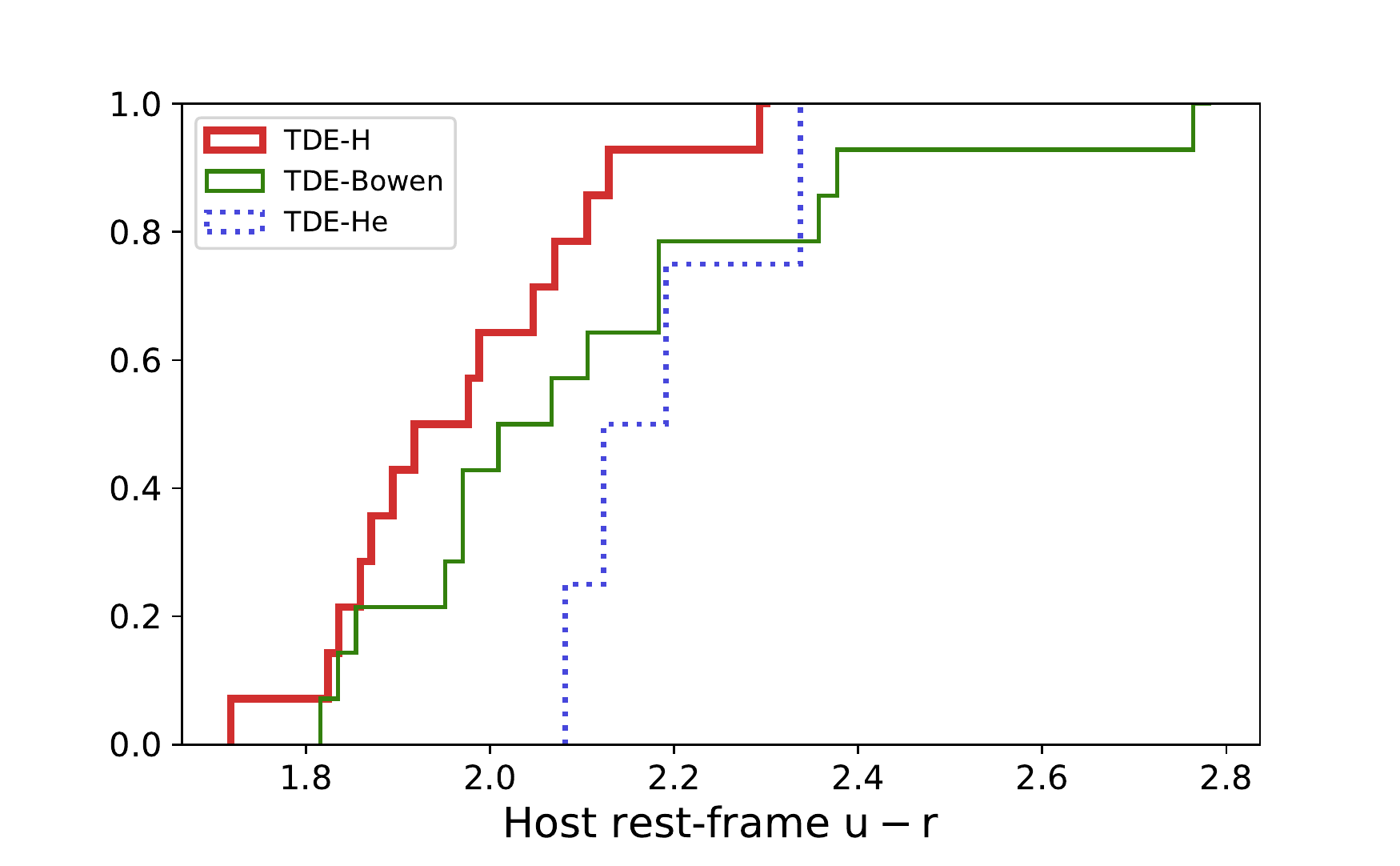}{0.48 \textwidth}{} \\[-30pt]}

\gridline{  \fig{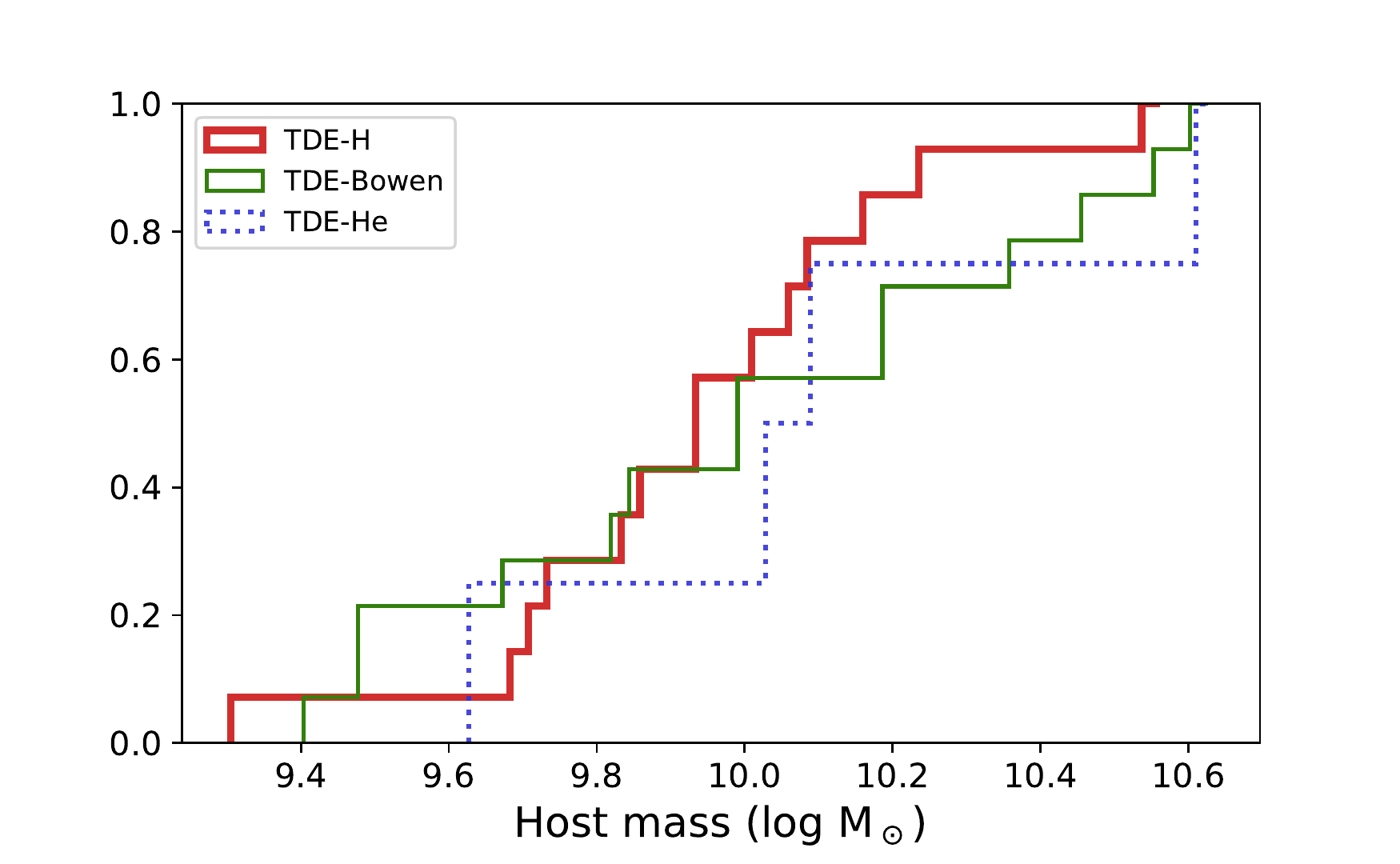}{0.48 \textwidth}{} 
            \fig{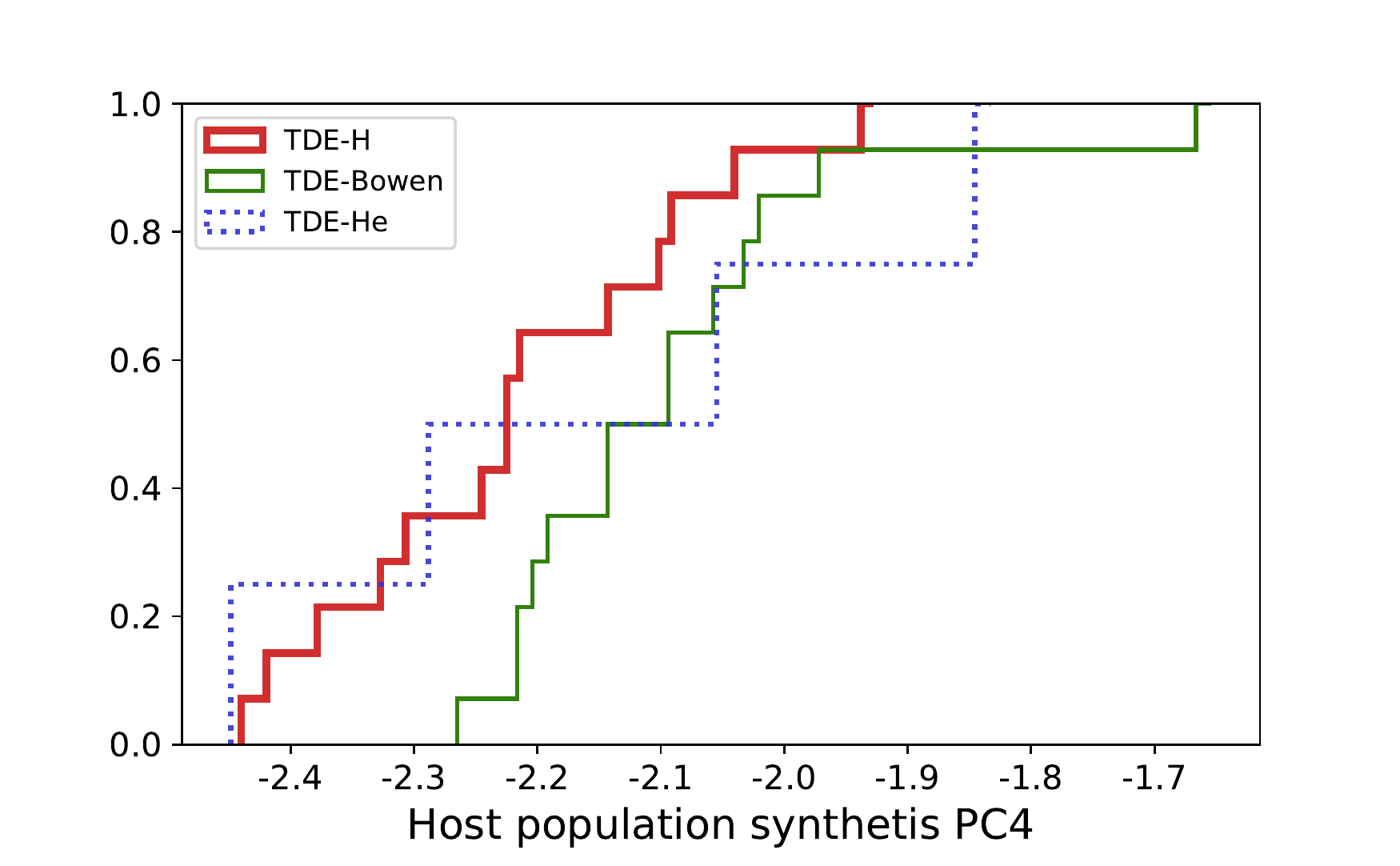}{0.48 \textwidth}{} \\[-10pt]}

\caption{Cumulative distribution of light curve and host galaxy properties for different TDE spectroscopic classes. The TDE-H and TDE-H+He/Bowen spectroscopic class show a significantly different distribution of blackbody radius and temperature, as well a significant difference in their host galaxy stellar population properties encoded in the 4rd principal component. The He-only TDEs appears to separate from the other two groups by their higher luminosity and longer rise time. }\label{fig:cumu}
\end{figure*}

\begin{itemize}
    \item We present 17 TDEs with light curves from ZTF, selected based on the photometric properties of nuclear ZTF transients (Fig.~\ref{fig:photosel}). Galaxies in the green valley are over represented in our ZTF sample by a factor of $\approx 5$ (Fig.~\ref{fig:urhost}). 
 
   % \item Using the ZTF $g$ and $r$-band light curve and {\it Swift}/UVOT photometry we measured the blackbody luminosity, radius, and temperature as function of time (Fig.~\ref{fig:LRT}). 
    
    \item Based on the ZTF and {\it Swift}/UVOT photometry we find that most of the TDEs in our sample show an increase of the temperature with time (Fig.~\ref{fig:LRT}, Table~\ref{tab:lcfit}). 
 
  \item After including spectroscopic TDE from the literature we obtain 32 sources that we classify into three classes: TDE-H, TDE-He, and TDE-Bowen (\S \ref{sec:specclass}. %We also include 7 more TDEs without spectroscopic observations to obtain a total sample size of 39. 
  
    \item We find significant differences between the photometric properties of the TDEs in each spectroscopic class (Table~\ref{tab:KS}). Most notably, the TDE-Bowen class has lower radii and longer rise times (Figs.~\ref{fig:RT} \& \ref{fig:cumu}). 
    
    \item Below an optical/UV photosphere radius of $10^{14.9}$~cm, {\it all} TDEs show either Bowen fluorescence lines or only Helium in their optical spectra. 
  
    \item We find statistically significant differences in the host galaxy population synthesis properties (dust, SFH, metallicity) for the TDE-H and TDE-Bowen class. Using the entire sample of photometric+spectroscopic TDEs, we also detect correlations between linear combinations of stellar population properties and the TDE blackbody radius and rise timescale (\S \ref{sec:hostcorr}).  
    
    \item We find a correlation between host galaxy total mass and the decay timescale of the light curve (Fig.~\ref{fig:vsmass} and Table~\ref{tab:Kendall}), suggesting that shape of the post-peak light curve is related to the fallback timescales and thus contains information about the mass of the black hole that disrupted the star.
    
    \item We identified a significant correlation between the rise timescale and $L_{\rm bb}/R$ (Fig.~\ref{fig:rise}). The rise time is not correlated with host galaxy mass nor with the decay timescale. These results can be explained by photon diffusion, which delays the time to maximum light (\S \ref{sec:diff}). 
   
    \item Four sources are detected in {\it Swift}/XRT observations. In one case (\Brienne) we observed three rapid X-ray flares (Figs.~\ref{fig:LbbLX} \& \ref{fig:LbbLX_all}). The peak luminosity of the X-ray flares approaches the optical/UV blackbody luminosity measured at the time of the. These similar luminosities at different wavelengths can be explained if the optical light is due to reprocessing of accretion power in a region with a high covering factor and some patches that allow the central engine to be briefly visible. 
\end{itemize}

\acknowledgments
{\bf Acknowledgments}  ---
% We thank the referee for the useful comments. 
Based on observations obtained with the Samuel Oschin Telescope 48-inch and the 60-inch Telescope at the Palomar Observatory as part of the Zwicky Transient Facility project. ZTF is supported by the National Science Foundation under Grant No. AST-1440341 and a collaboration including Caltech, IPAC, the Weizmann Institute for Science, the Oskar Klein Center at Stockholm University, the University of Maryland, the University of Washington, Deutsches Elektronen-Synchrotron and Humboldt University, Los Alamos National Laboratories, the TANGO Consortium of Taiwan, the University of Wisconsin at Milwaukee, and Lawrence Berkeley National Laboratories. Operations are conducted by COO, IPAC, and UW. SED Machine is based upon work supported by the National Science Foundation under Grant No. 1106171. 

These results made use of the Discovery Channel Telescope at Lowell Observatory. Lowell is a private, non-profit institution dedicated to astrophysical research and public appreciation of astronomy and operates the DCT in partnership with Boston University, the University of Maryland, the University of Toledo, Northern Arizona University and Yale University. The upgrade of the DeVeny optical spectrograph has been funded by a generous grant from John and Ginger Giovale.
The W. M. Keck Observatory is operated as a scientific partnership among the California Institute of Technology, the University of California, and NASA; the Observatory was made possible by the generous financial support of the W. M. Keck Foundation. 
The Liverpool Telescope is operated on the island of La Palma by Liverpool John Moores University in the Spanish Observatorio del Roque de los Muchachos of the Instituto de Astrofisica de Canarias with financial support from the UK Science and Technology Facilities Council.
Research at Lick Observatory is partially supported by a generous gift from Google.

We thank the Swift team, including the  Observation  Duty  Scientists, and the science planners for promptly  approving  and executing our Swift observations.  We also acknowledge the use of public data from the Swift data archive.
This work made use of data supplied by the UK Swift Science Data Centre at the University of Leicester.
We thank J. Brown and M. Siebert for help with Keck observations. 

S. van Velzen is supported by the James Arhtur Fellowship. S. Gezari is supported in part by NSF CAREER grant 1454816 and NSF AAG grant 1616566. N.R. acknowledges the support of a Joint Space-Science Institute prize postdoctoral fellowship. This project has received funding from the European Research Council (ERC) under the European Union's Horizon 2020 research and innovation programme (grant agreement No. 759194 - USNAC). A.Y.Q.H. is supported by a National Science Foundation Graduate Research Fellowship under Grant No.\,DGE‐1144469. This work was supported by the GROWTH project funded by the National Science Foundation under PIRE Grant No.\,1545949. C.~Fremling gratefully acknowledges support of his research by the Heising-Simons Foundation (\#2018-0907). The UCSC transient team is supported in part by NSF grant AST-1518052, NASA/{\it Swift} grant 80NSSC19K1386, the Gordon \& Betty Moore Foundation, the Heising-Simons Foundation, and by a fellowship from the David and Lucile Packard Foundation to R.J.F.

\software{
Ampel \citep{Nordin19},
Astropy \citep{Astropy-Collaboration18}, 
catsHTM \citep{Soumagnac18},
emcee \citep{Foreman-Mackey13},
extcats (\url{github.com/MatteoGiomi/extcats}),
gPhoton \citep{Million16},
HEAsoft \citep{Arnaud96},
FSPS \citep{Conroy09,Conroy10,Foreman-Mackey14},
Prospector \citep{Johnson17}.
}

%\clearpage
% if you have references to add that are not in general_desk.bib, put them in others.bib
\bibstyle{aasjournals}
\bibliography{general_desk,others}

%----------
%
%\appendix
%
%----------

\begin{figure*}

\gridline{  \fig{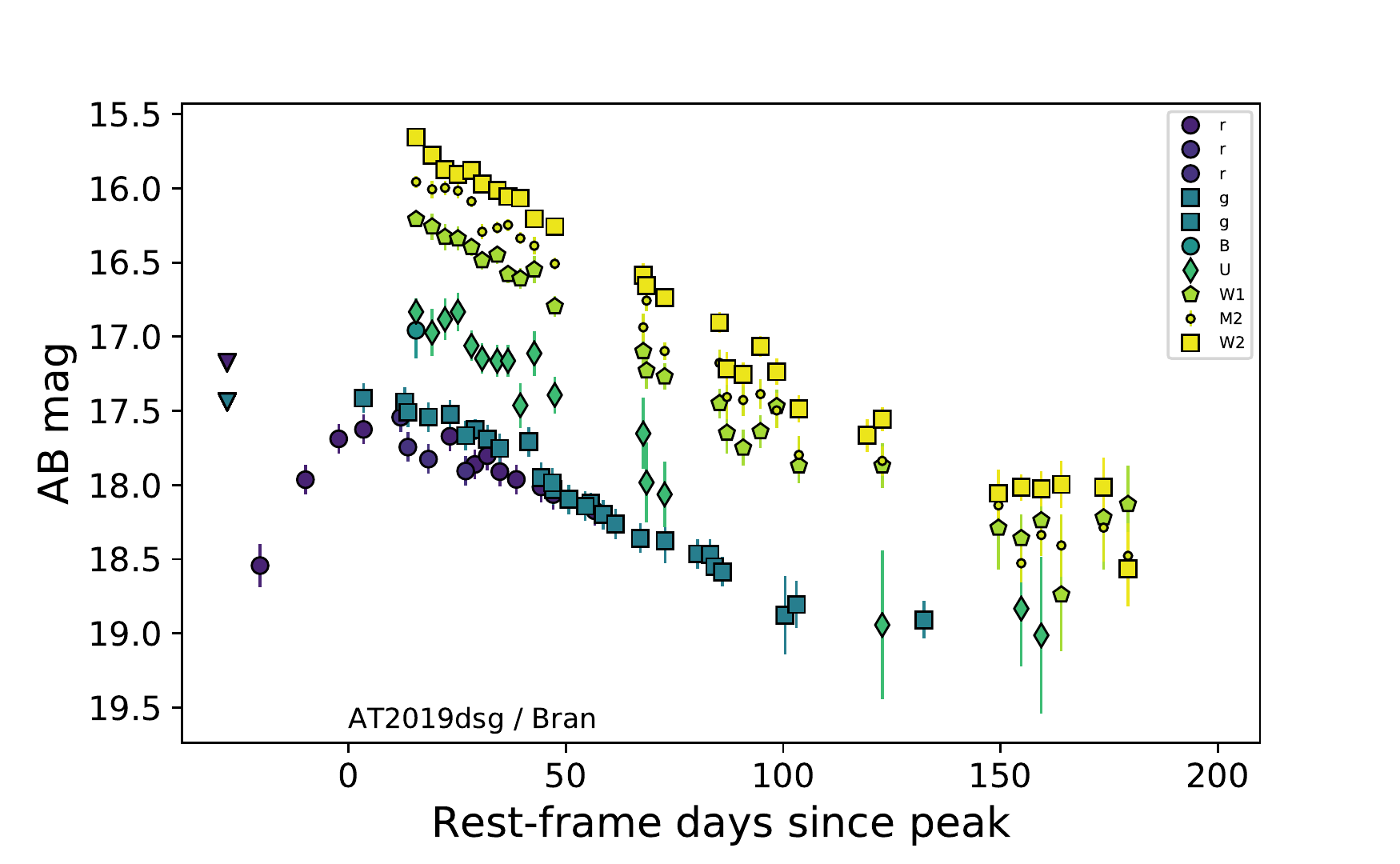}{0.4 \textwidth}{} 
            \fig{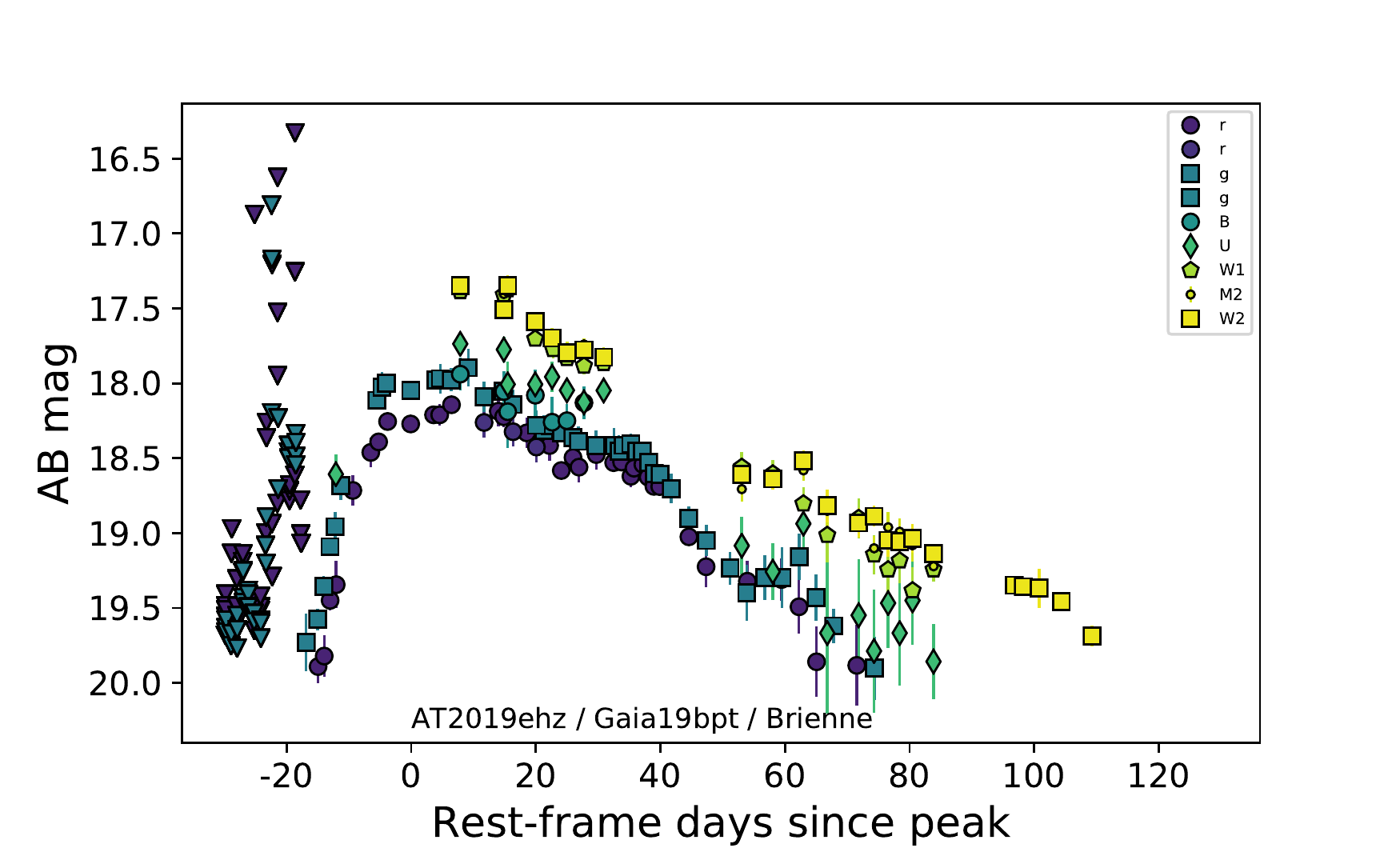}{0.4 \textwidth}{} 
            \\[-38pt]}    

\gridline{ \fig{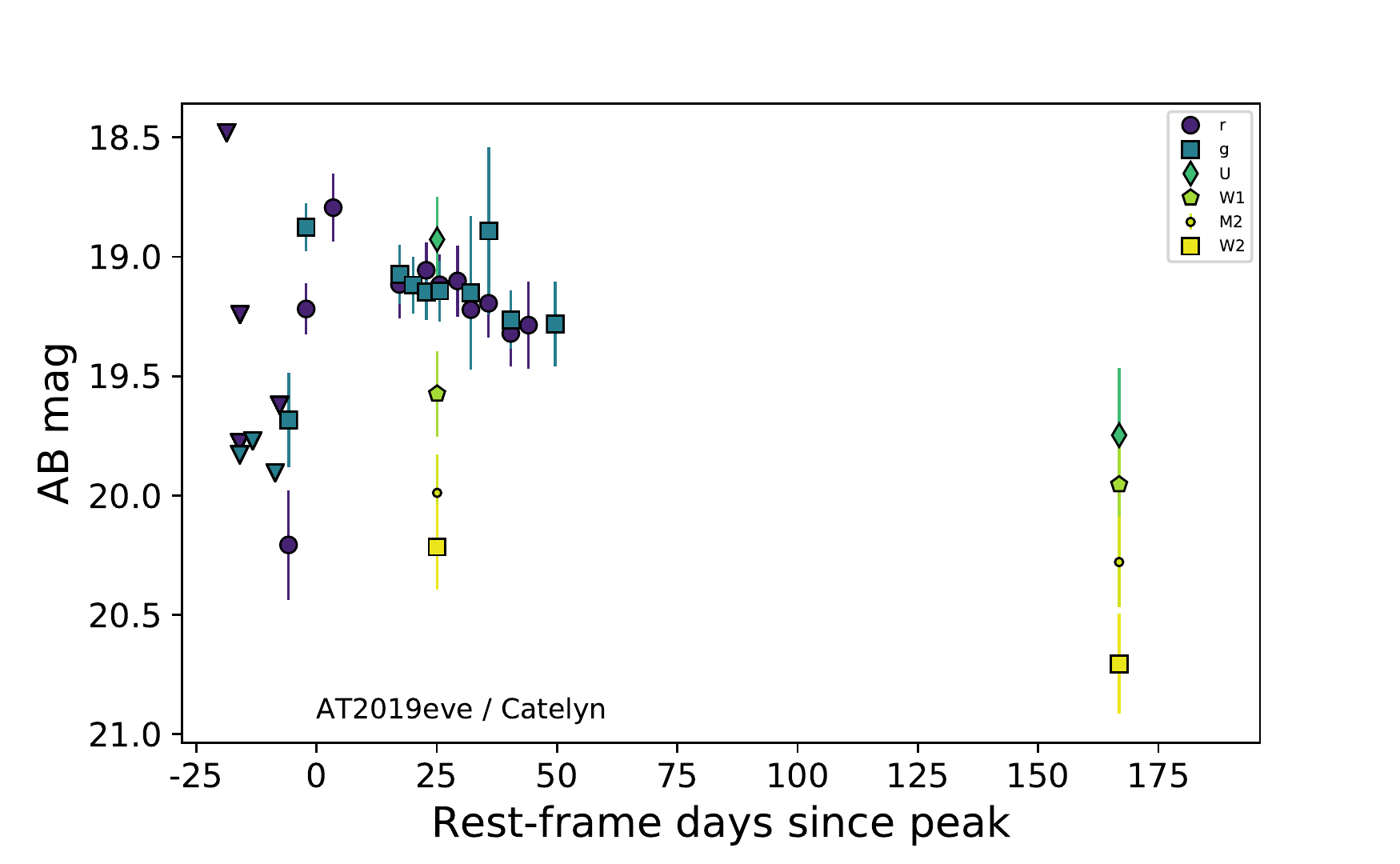}{0.4 \textwidth}{}
            \fig{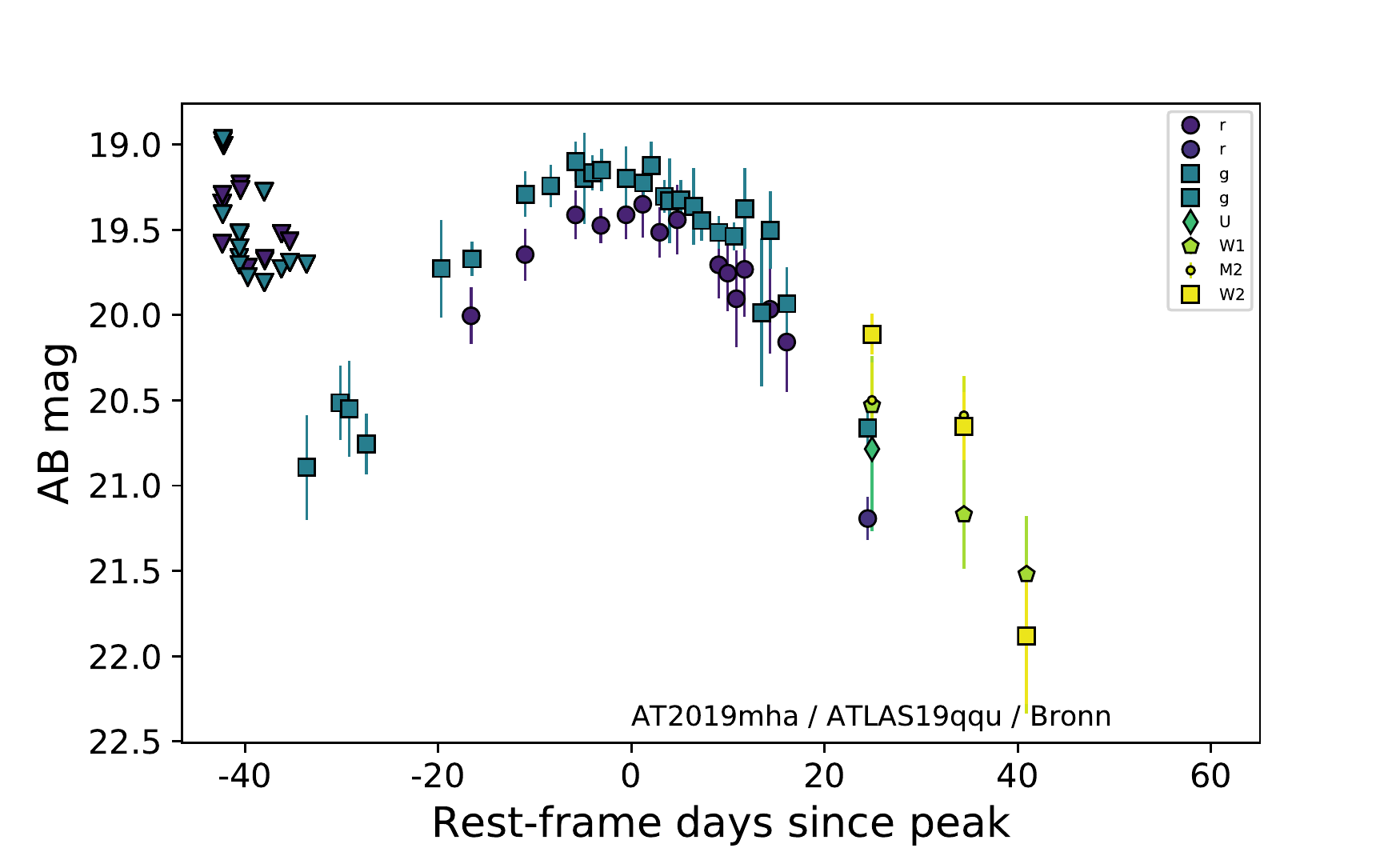}{0.4 \textwidth}{}
           \\[-38pt]}    

\gridline{  \fig{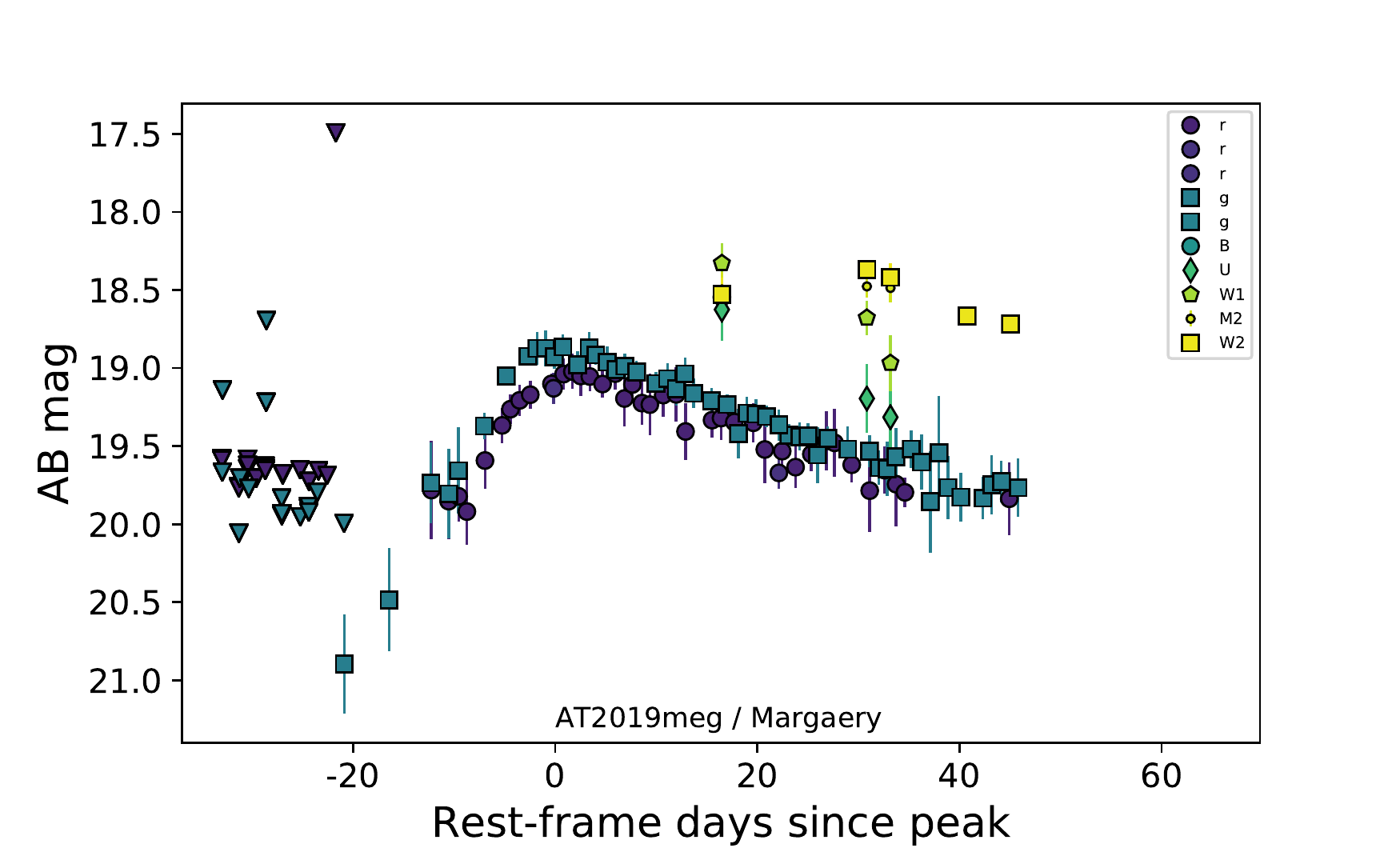}{0.4 \textwidth}{}
            \fig{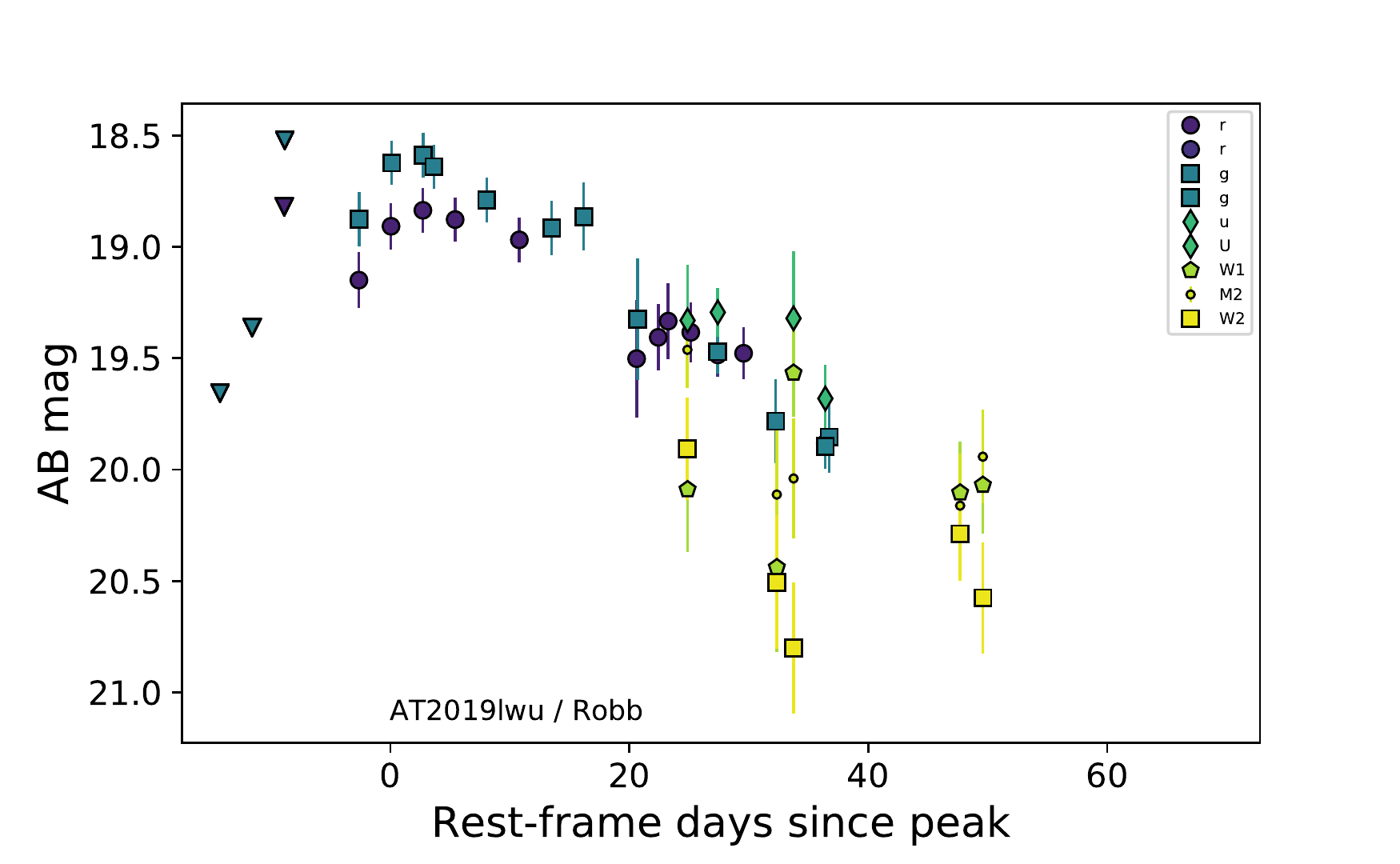}{0.4 \textwidth}{} 
             \\[-38pt]}
            
\gridline{  \fig{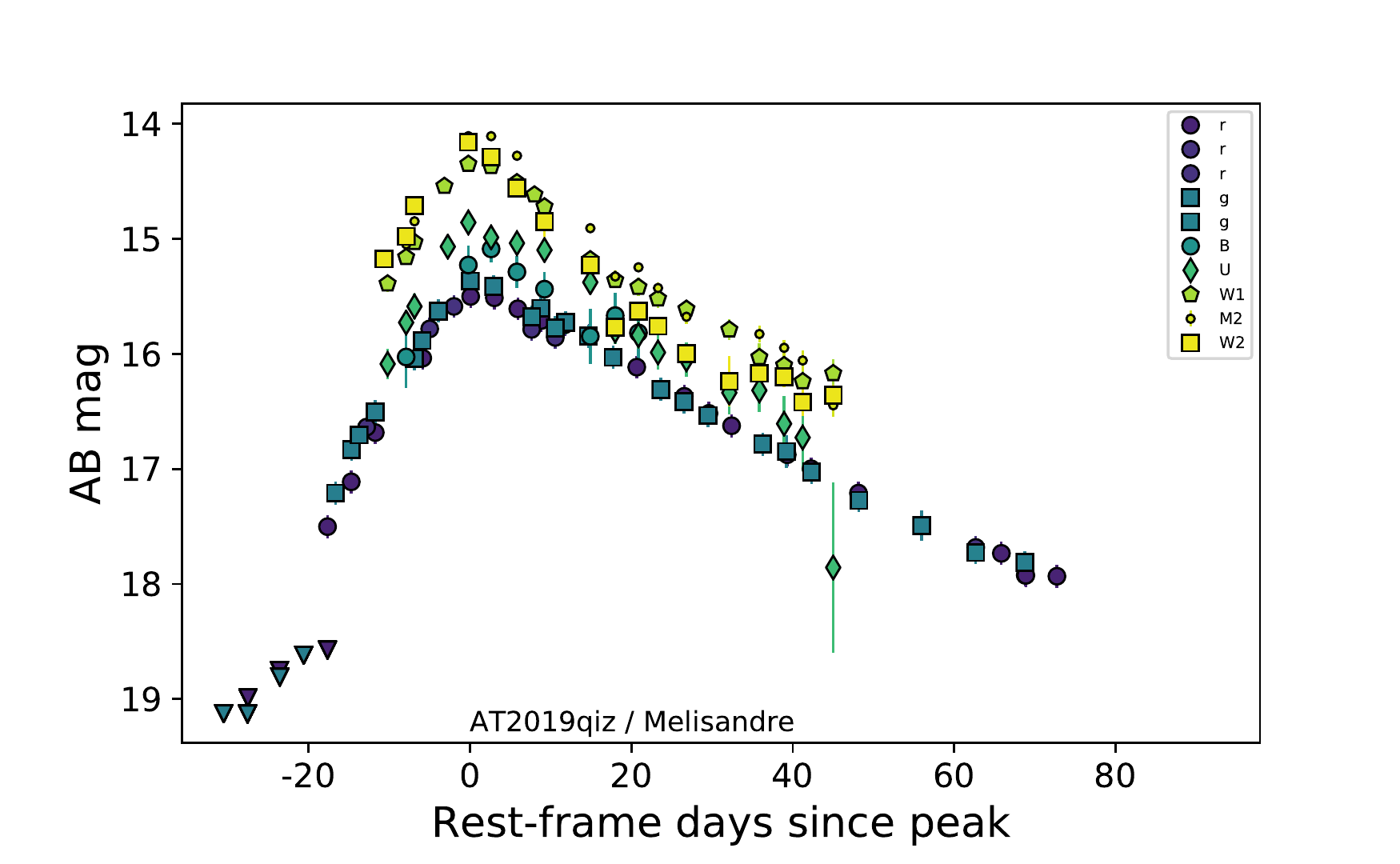}{0.4 \textwidth}{}   
             }   
\caption{Optical/UV light curves (continued).}\label{fig:lcs2}
\end{figure*}

%\fig{lc_DavosSeaworth.pdf}{0.4 \textwidth}{}

\begin{figure*}
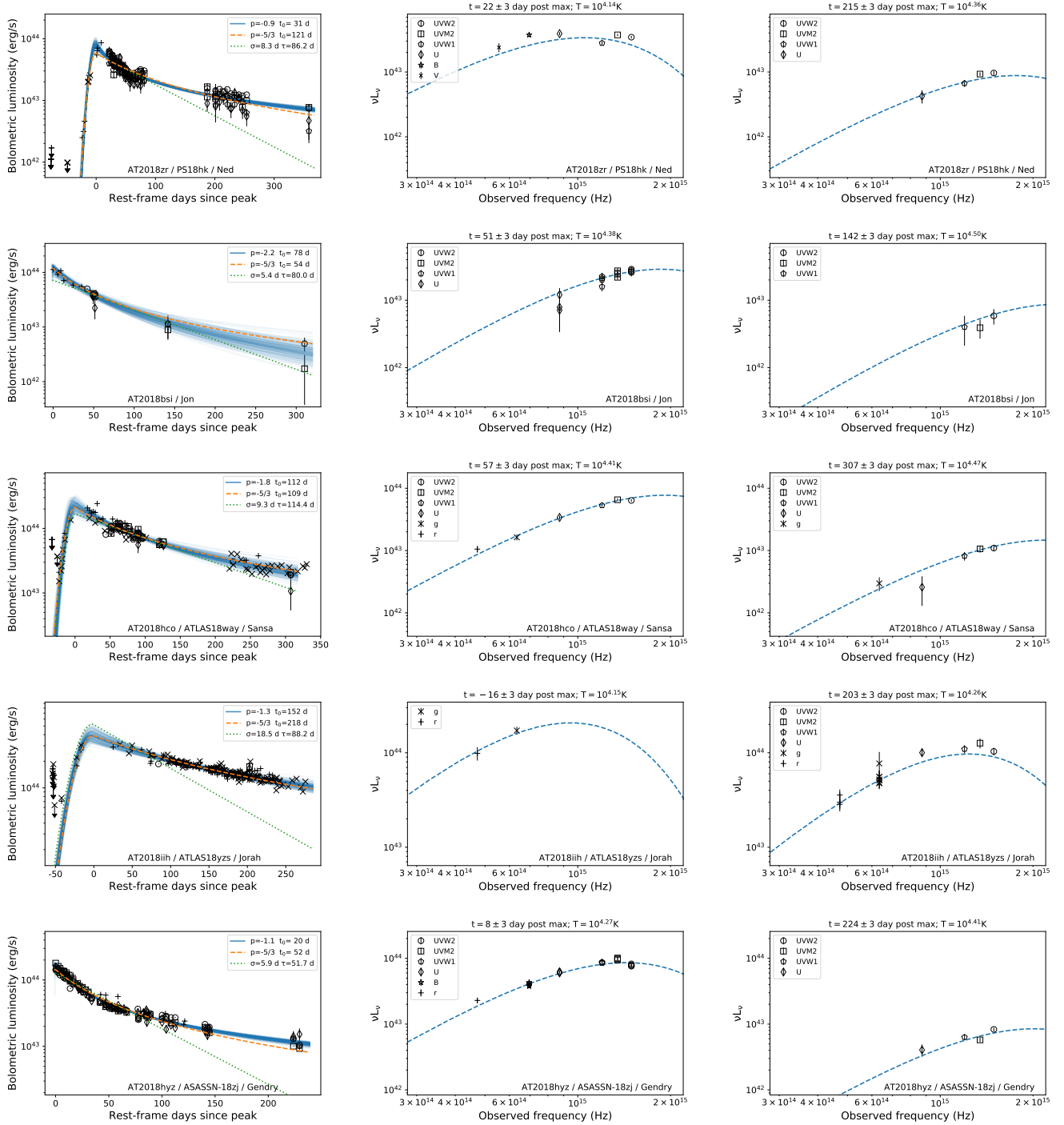


\gridline{	\fig{NedStark/NedStark_lcfit_lin_t365.pdf}{0.33 \textwidth}{} 
			\fig{NedStark/NedStark_sedfit_t19_t365.pdf}{0.33 \textwidth}{} 
			\fig{NedStark/NedStark_sedfit_t214_t365.pdf}{0.33 \textwidth}{} \\[-23pt] }

\gridline{	\fig{JonSnow/JonSnow_lcfit_lin_t365.pdf}{0.33 \textwidth}{} 
			\fig{JonSnow/JonSnow_sedfit_t49_t365.pdf}{0.33 \textwidth}{} 
			\fig{JonSnow/JonSnow_sedfit_t139_t365.pdf}{0.33 \textwidth}{} \\[-23pt] }

\gridline{	\fig{SansaStark/SansaStark_lcfit_lin_t365.pdf}{0.33 \textwidth}{} 
			\fig{SansaStark/SansaStark_sedfit_t55_t365.pdf}{0.33 \textwidth}{} 
			\fig{SansaStark/SansaStark_sedfit_t304_t365.pdf}{0.33 \textwidth}{} \\[-23pt] }

\gridline{	\fig{JorahMormont/JorahMormont_lcfit_lin_t365.pdf}{0.33 \textwidth}{} 
			\fig{JorahMormont/JorahMormont_sedfit_t-17_t365.pdf}{0.33 \textwidth}{} 
			\fig{JorahMormont/JorahMormont_sedfit_t202_t365.pdf}{0.33 \textwidth}{} \\[-23pt] }

\gridline{	\fig{GendryBaratheon/GendryBaratheon_lcfit_lin_t365.pdf}{0.33 \textwidth}{} 
			\fig{GendryBaratheon/GendryBaratheon_sedfit_t4_t365.pdf}{0.33 \textwidth}{} 
			\fig{GendryBaratheon/GendryBaratheon_sedfit_t220_t365.pdf}{0.33 \textwidth}{} \\[-23pt] 
			}

\caption{Blackbody light curves and two example SEDs. For each TDE we show the bolometric light curve, as obtained by fitting Eq.~\ref{eq:pl} to the multi-band photometry (solid blue and dashed orange curves, for the latter the power-law index is fixed at $p=-5/3$). The results for an exponential decay are also shown (Eq.~\ref{eq:exp}, green dotted line). The dispersion in the  power-law model (blue lines) is visualized by drawing samples from the posterior distribution of light curves. For each source we also show two SEDs, one close to peak and a second one at later times.  }
\end{figure*}

\begin{figure*}
\gridline{	\fig{AryaStark/AryaStark_lcfit_lin_t365.pdf}{0.33 \textwidth}{} 
			\fig{AryaStark/AryaStark_sedfit_t-2_t365.pdf}{0.33 \textwidth}{} 
			\fig{AryaStark/AryaStark_sedfit_t46_t365.pdf}{0.33 \textwidth}{} \\[-23pt] }

\gridline{	\fig{CerseiLannister/CerseiLannister_lcfit_lin_t365.pdf}{0.33 \textwidth}{} 
			\fig{CerseiLannister/CerseiLannister_sedfit_t-18_t365.pdf}{0.33 \textwidth}{} 
			\fig{CerseiLannister/CerseiLannister_sedfit_t54_t365.pdf}{0.33 \textwidth}{} \\[-23pt] }

\gridline{	\fig{PetyrBalish/PetyrBalish_lcfit_lin_t365.pdf}{0.33 \textwidth}{} 
			\fig{PetyrBalish/PetyrBalish_sedfit_t25_t365.pdf}{0.33 \textwidth}{} 
			\fig{PetyrBalish/PetyrBalish_sedfit_t55_t365.pdf}{0.33 \textwidth}{} \\[-23pt] }

\gridline{	\fig{Varys/Varys_lcfit_lin_t365.pdf}{0.33 \textwidth}{} 
			\fig{Varys/Varys_sedfit_t-2_t365.pdf}{0.33 \textwidth}{} 
			\fig{Varys/Varys_sedfit_t82_t365.pdf}{0.33 \textwidth}{} \\[-23pt] }

\gridline{	\fig{JaimeLannister/JaimeLannister_lcfit_lin_t365.pdf}{0.33 \textwidth}{} 
			\fig{JaimeLannister/JaimeLannister_sedfit_t-5_t365.pdf}{0.33 \textwidth}{} 
			\fig{JaimeLannister/JaimeLannister_sedfit_t73_t365.pdf}{0.33 \textwidth}{} \\[-23pt] }

\gridline{	\fig{BranStark/BranStark_lcfit_lin_t365.pdf}{0.33 \textwidth}{} 
			\fig{BranStark/BranStark_sedfit_t10_t365.pdf}{0.33 \textwidth}{} 
			\fig{BranStark/BranStark_sedfit_t151_t365.pdf}{0.33 \textwidth}{} \\[-23pt] }

\caption{Blackbody light curves and two example SEDs (continued).}
\end{figure*}

\begin{figure*}

\gridline{	\fig{Brienne/Brienne_lcfit_lin_t365.pdf}{0.33 \textwidth}{} 
			\fig{Brienne/Brienne_sedfit_t-14_t365.pdf}{0.33 \textwidth}{} 
			\fig{Brienne/Brienne_sedfit_t82_t365.pdf}{0.33 \textwidth}{} \\[-23pt] }

\gridline{	\fig{CatelynStark/CatelynStark_lcfit_lin_t365.pdf}{0.33 \textwidth}{} 
			\fig{CatelynStark/CatelynStark_sedfit_t19_t365.pdf}{0.33 \textwidth}{}
			\fig{CatelynStark/CatelynStark_sedfit_t163_t365}{0.33 \textwidth}{} \\[-23pt] }

\gridline{	\fig{Bronn/Bronn_lcfit_lin_t365.pdf}{0.33 \textwidth}{} 
			\fig{Bronn/Bronn_sedfit_t-2_t365.pdf}{0.33 \textwidth}{} 
			\fig{Bronn/Bronn_sedfit_t19_t365.pdf}{0.33 \textwidth}{} \\[-23pt] }

\gridline{	\fig{MargaeryTyrell/MargaeryTyrell_lcfit_lin_t365.pdf}{0.33 \textwidth}{} 
			\fig{MargaeryTyrell/MargaeryTyrell_sedfit_t-5_t365.pdf}{0.33 \textwidth}{} 
			\fig{MargaeryTyrell/MargaeryTyrell_sedfit_t28_t365}{0.33 \textwidth}{} \\[-23pt] }

\gridline{	\fig{RobbStark/RobbStark_lcfit_lin_t365.pdf}{0.33 \textwidth}{} 
			\fig{RobbStark/RobbStark_sedfit_t-5_t365.pdf}{0.33 \textwidth}{} 
			\fig{RobbStark/RobbStark_sedfit_t31_t365.pdf}{0.33 \textwidth}{} \\[-23pt] }

\gridline{	\fig{Melisandre/Melisandre_lcfit_lin_t365.pdf}{0.33 \textwidth}{} 
			\fig{Melisandre/Melisandre_sedfit_t-5_t365.pdf}{0.33 \textwidth}{} 
			\fig{Melisandre/Melisandre_sedfit_t40_t365.pdf}{0.33 \textwidth}{} \\[-23pt] }

\caption{Blackbody light curves and two example SEDs (continued).}
\end{figure*}

\begin{table*}
\centering
\caption{Host properties}\label{tab:host}
\begin{tabular}{l c c c c c c}
\hline
 name                 & mass         & $^{0.0}u-r$         & dust             &  age & $\tau_{\rm sfh}$ & $Z/Z_\odot$ \\ 
 	                 & log\,$M_\odot$         &          &     $E(B-V)$        &  Gyr & Gyr & log \\ 
\hline
 \Ned                 &  $9.95_{0.24}^{0.12}$ &  $2.30_{0.06}^{0.08}$ &  $0.42_{0.32}^{0.40}$ &  $4.89_{3.21}^{3.26}$ &  $0.20_{0.08}^{0.17}$ &  $-0.48_{0.59}^{0.44}$\\
 \Jon                 &  $10.63_{0.05}^{0.05}$ &  $2.12_{0.04}^{0.04}$ &  $0.77_{0.27}^{0.16}$ &  $2.95_{0.67}^{0.73}$ &  $0.75_{0.18}^{0.17}$ &  $-0.12_{0.24}^{0.16}$\\
 \Sansa               &  $9.95_{0.16}^{0.12}$ &  $1.86_{0.05}^{0.07}$ &  $0.21_{0.15}^{0.23}$ &  $5.79_{2.97}^{4.14}$ &  $0.30_{0.16}^{0.42}$ &  $-1.56_{0.31}^{0.53}$\\
 \Jorah               &  $10.63_{0.14}^{0.18}$ &  $2.34_{0.07}^{0.07}$ &  $0.52_{0.37}^{0.31}$ &  $3.82_{1.93}^{3.38}$ &  $0.22_{0.10}^{0.29}$ &  $-0.31_{0.51}^{0.38}$\\
 \Gendry              &  $9.84_{0.14}^{0.09}$ &  $1.90_{0.04}^{0.04}$ &  $0.27_{0.14}^{0.16}$ &  $4.74_{1.40}^{2.98}$ &  $0.23_{0.09}^{0.23}$ &  $-1.41_{0.37}^{0.44}$\\
 \Arya                &  $10.00_{0.14}^{0.09}$ &  $1.98_{0.07}^{0.08}$ &  $0.32_{0.20}^{0.28}$ &  $6.52_{3.36}^{3.62}$ &  $0.26_{0.11}^{0.31}$ &  $-1.48_{0.39}^{0.53}$\\
 \Cersei              &  $9.49_{0.12}^{0.11}$ &  $1.98_{0.09}^{0.07}$ &  $0.23_{0.17}^{0.21}$ &  $6.40_{2.50}^{3.61}$ &  $0.28_{0.14}^{0.37}$ &  $-1.25_{0.53}^{0.44}$\\
 \Petyr               &  $10.20_{0.14}^{0.11}$ &  $2.08_{0.09}^{0.08}$ &  $0.25_{0.18}^{0.25}$ &  $6.78_{2.87}^{3.29}$ &  $0.28_{0.15}^{0.36}$ &  $-0.97_{0.51}^{0.39}$\\
 \Varys               &  $10.25_{0.12}^{0.14}$ &  $2.08_{0.07}^{0.06}$ &  $0.72_{0.46}^{0.20}$ &  $3.45_{1.21}^{1.66}$ &  $0.49_{0.26}^{0.32}$ &  $-1.05_{0.67}^{0.72}$\\
 \Jaime               &  $9.82_{0.13}^{0.16}$ &  $1.82_{0.04}^{0.04}$ &  $0.38_{0.16}^{0.14}$ &  $2.33_{1.10}^{1.26}$ &  $0.20_{0.07}^{0.20}$ &  $-1.16_{0.53}^{0.71}$\\
 \Bran                &  $10.46_{0.19}^{0.11}$ &  $2.19_{0.09}^{0.07}$ &  $0.44_{0.29}^{0.29}$ &  $6.05_{2.85}^{4.44}$ &  $0.29_{0.15}^{0.37}$ &  $-0.99_{0.71}^{0.52}$\\
 \Brienne             &  $9.74_{0.09}^{0.08}$ &  $2.06_{0.05}^{0.05}$ &  $0.09_{0.07}^{0.21}$ &  $6.83_{2.07}^{2.23}$ &  $0.25_{0.12}^{0.27}$ &  $-0.79_{0.23}^{0.19}$\\
 \Catelyn             &  $9.31_{0.15}^{0.10}$ &  $1.82_{0.07}^{0.06}$ &  $0.24_{0.19}^{0.38}$ &  $5.31_{2.47}^{3.27}$ &  $0.40_{0.23}^{0.35}$ &  $-1.68_{0.24}^{0.59}$\\
 \Bronn               &  $10.07_{0.18}^{0.10}$ &  $1.99_{0.05}^{0.07}$ &  $0.49_{0.28}^{0.24}$ &  $3.95_{1.96}^{2.78}$ &  $0.23_{0.10}^{0.20}$ &  $-1.24_{0.44}^{0.55}$\\
 \Margaery            &  $9.70_{0.08}^{0.15}$ &  $1.99_{0.06}^{0.06}$ &  $0.24_{0.17}^{0.26}$ &  $4.02_{1.18}^{2.74}$ &  $0.63_{0.42}^{0.24}$ &  $-0.62_{0.55}^{0.64}$\\
 \Robb                &  $9.86_{0.13}^{0.09}$ &  $1.92_{0.06}^{0.04}$ &  $0.13_{0.10}^{0.16}$ &  $5.69_{2.03}^{2.98}$ &  $0.24_{0.11}^{0.22}$ &  $-1.16_{0.50}^{0.40}$\\
 \Melisandre          &  $10.01_{0.13}^{0.09}$ &  $2.01_{0.09}^{0.07}$ &  $0.27_{0.20}^{0.29}$ &  $6.40_{2.60}^{3.30}$ &  $0.33_{0.18}^{0.34}$ &  $-1.19_{0.54}^{0.40}$\\

\hline
\end{tabular}
\end{table*}

\begin{table*}
\centering
\caption{Light curve shape parameters.}\label{tab:lcfit}
\begin{tabular}{l c c c c c c c c c c}
\hline
name                 & $L_g$         & $L_{\rm bb}$   & $T$               & $dT/dt$ & $t_{\rm peak/max}$ & $\sigma$          & $\tau$            & $p$               & $t_0$           & $t_0|_{p=-5/3}$   \\
 & $\log$ erg/s & $\log$ erg/s & $\log$ K     & $10^{2}$ K/day & MJD            & $\log$ day   & $\log$ day   &                   & $\log$ day   & $\log$ day     \\
 \hline
\ \Ned                 &  $43.41_{0.02}^{0.02}$ &  $43.76_{0.02}^{0.02}$ &  $4.14_{0.01}^{0.01}$ &  $0.49_{0.06}^{0.06}$ &  $58180.1_{1.0}^{1.1}$ &  $1.0_{0.04}^{0.04}$ &  $1.93_{0.03}^{0.03}$ &  $-0.9_{0.1}^{0.1}$ &  $1.49_{0.14}^{0.10}$ &  $2.08_{0.02}^{0.02}$\\
 \Jon                 &  $42.69_{0.07}^{0.08}$ &  $43.86_{0.10}^{0.20}$ &  $4.53_{0.08}^{0.09}$ &  $0.82_{0.97}^{0.82}$ &  $58217.1             $ & --            &  $1.72_{0.06}^{0.05}$ &  $-1.8_{1.3}^{0.8}$ &  $1.84_{0.46}^{0.45}$ &  $1.86_{0.14}^{0.16}$\\
 \Sansa               &  $43.39_{0.02}^{0.02}$ &  $44.22_{0.04}^{0.04}$ &  $4.39_{0.01}^{0.01}$ &  $-0.00_{0.09}^{0.11}$ &  $58403.4_{2.0}^{2.1}$ &  $1.0_{0.05}^{0.04}$ &  $2.06_{0.04}^{0.05}$ &  $-1.8_{0.2}^{0.3}$ &  $2.05_{0.14}^{0.12}$ &  $2.04_{0.09}^{0.08}$\\
 \Jorah               &  $44.21_{0.03}^{0.03}$ &  $44.72_{0.03}^{0.03}$ &  $4.23_{0.01}^{0.01}$ &  $0.20_{0.05}^{0.04}$ &  $58449.7_{0.4}^{0.8}$ &  $1.3_{0.02}^{0.02}$ &  $1.95_{0.03}^{0.04}$ &  $-1.3_{0.5}^{0.2}$ &  $2.18_{0.09}^{0.18}$ &  $2.34_{0.05}^{0.05}$\\
 \Gendry              &  $43.57_{0.01}^{0.01}$ &  $44.10_{0.01}^{0.01}$ &  $4.25_{0.01}^{0.01}$ &  $0.20_{0.05}^{0.06}$ &  $58428.0             $ & --            &  $1.71_{0.01}^{0.01}$ &  $-1.1_{0.1}^{0.0}$ &  $1.30_{0.05}^{0.06}$ &  $1.71_{0.01}^{0.01}$\\
 \Arya                &  $43.26_{0.03}^{0.03}$ &  $44.03_{0.11}^{0.19}$ &  $4.38_{0.05}^{0.07}$ &  $0.32_{0.40}^{0.38}$ &  $58461.9_{6.1}^{5.2}$ &  $0.9_{0.43}^{0.28}$ &  $2.30_{0.16}^{0.22}$ &  $-1.3_{1.6}^{0.9}$ &  $2.47_{0.48}^{0.36}$ &  $2.44_{0.23}^{0.20}$\\
 \Cersei              &  $43.23_{0.02}^{0.02}$ &  $44.53_{0.05}^{0.07}$ &  $4.59_{0.03}^{0.03}$ &  $1.06_{1.03}^{0.66}$ &  $58508.6_{2.1}^{2.2}$ &  $1.2_{0.06}^{0.06}$ &  $1.65_{0.03}^{0.03}$ &  $-1.7_{0.6}^{0.6}$ &  $1.74_{0.29}^{0.26}$ &  $1.69_{0.14}^{0.14}$\\
 \Petyr               &  $43.58_{0.01}^{0.02}$ &  $44.00_{0.02}^{0.02}$ &  $4.19_{0.01}^{0.01}$ &  $0.66_{0.42}^{0.28}$ &  $58547.5_{10.1}^{2.1}$ &  $1.2_{0.34}^{0.07}$ &  $1.88_{0.04}^{0.05}$ &  $-1.8_{0.7}^{0.6}$ &  $2.15_{0.30}^{0.32}$ &  $2.04_{0.12}^{0.11}$\\
 \Varys               &  $43.50_{0.02}^{0.06}$ &  $44.05_{0.05}^{0.04}$ &  $4.27_{0.02}^{0.02}$ &  $0.67_{0.27}^{0.20}$ &  $58542.1_{8.1}^{2.5}$ &  $0.9_{0.13}^{0.08}$ &  $1.63_{0.03}^{0.03}$ &  $-1.6_{0.5}^{0.4}$ &  $1.79_{0.23}^{0.18}$ &  $1.83_{0.10}^{0.08}$\\
 \Jaime               &  $43.33_{0.01}^{0.01}$ &  $44.44_{0.02}^{0.02}$ &  $4.51_{0.01}^{0.01}$ &  $0.88_{0.17}^{0.18}$ &  $58558.5_{2.1}^{1.7}$ &  $1.3_{0.05}^{0.05}$ &  $1.85_{0.01}^{0.02}$ &  $-2.2_{0.3}^{0.4}$ &  $2.07_{0.11}^{0.10}$ &  $1.89_{0.07}^{0.05}$\\
 \Bran                &  $43.16_{0.03}^{0.03}$ &  $44.46_{0.05}^{0.05}$ &  $4.59_{0.02}^{0.02}$ &  $0.24_{0.44}^{0.55}$ &  $58603.1_{4.1}^{3.7}$ &  $1.2_{0.10}^{0.08}$ &  $1.76_{0.01}^{0.01}$ &  $-2.0_{0.3}^{0.4}$ &  $1.79_{0.18}^{0.14}$ &  $1.62_{0.09}^{0.10}$\\
 \Brienne             &  $43.32_{0.01}^{0.01}$ &  $44.04_{0.02}^{0.02}$ &  $4.34_{0.01}^{0.01}$ &  $-0.24_{0.13}^{0.14}$ &  $58612.7_{0.7}^{0.6}$ &  $0.9_{0.02}^{0.02}$ &  $1.67_{0.01}^{0.01}$ &  $-1.7_{0.2}^{0.2}$ &  $1.62_{0.12}^{0.10}$ &  $1.59_{0.04}^{0.05}$\\
 \Catelyn             &  $42.92_{0.03}^{0.03}$ &  $43.15_{0.03}^{0.03}$ &  $4.06_{0.01}^{0.01}$ &  $0.07_{0.08}^{0.09}$ &  $58613.0_{0.4}^{0.8}$ &  $0.4_{0.04}^{0.06}$ &  $2.24_{0.17}^{0.22}$ &  $-0.7_{1.2}^{0.2}$ &  $1.63_{0.16}^{0.56}$ &  $2.18_{0.12}^{0.10}$\\
 \Bronn               &  $43.38_{0.01}^{0.01}$ &  $44.12_{0.05}^{0.06}$ &  $4.35_{0.03}^{0.03}$ &  $0.90_{1.20}^{0.73}$ &  $58704.7_{0.7}^{0.7}$ &  $1.2_{0.02}^{0.02}$ &  $1.20_{0.02}^{0.02}$ &  $-3.8_{0.7}^{0.7}$ &  $1.62_{0.14}^{0.10}$ &  $1.28_{0.09}^{0.10}$\\
 \Margaery            &  $43.42_{0.01}^{0.01}$ &  $44.36_{0.03}^{0.04}$ &  $4.44_{0.01}^{0.01}$ &  $1.95_{0.09}^{0.04}$ &  $58696.7_{0.6}^{0.6}$ &  $0.9_{0.03}^{0.03}$ &  $1.70_{0.02}^{0.02}$ &  $-0.1_{0.4}^{0.1}$ &  $2.18_{0.69}^{0.55}$ &  $2.66_{0.18}^{0.19}$\\
 \Robb                &  $43.37_{0.02}^{0.02}$ &  $43.70_{0.03}^{0.03}$ &  $4.14_{0.01}^{0.01}$ &  $0.60_{0.29}^{0.23}$ &  $58691.0_{1.0}^{1.4}$ &  $0.5_{0.29}^{0.14}$ &  $1.56_{0.03}^{0.03}$ &  $-1.7_{0.4}^{0.4}$ &  $1.61_{0.17}^{0.17}$ &  $1.60_{0.11}^{0.10}$\\
 \Melisandre          &  $42.88_{0.01}^{0.01}$ &  $43.46_{0.02}^{0.02}$ &  $4.27_{0.01}^{0.01}$ &  $-0.85_{0.22}^{0.20}$ &  $58763.4_{0.5}^{0.5}$ &  $0.9_{0.01}^{0.01}$ &  $1.38_{0.01}^{0.01}$ &  $-1.9_{0.3}^{0.2}$ &  $1.34_{0.09}^{0.09}$ &  $1.23_{0.03}^{0.03}$\\

\hline
\end{tabular}
\end{table*}

\begin{table*}
\centering
\caption{KS $p$-value comparing TDEs separated in three spectral classes.}
\label{tab:KS}
\begin{tabular}{l l l l}
\hline
                               & H-only vs. H+He/Bowen   & H-only vs. He-only      & H+He/Bowen vs. He-only \\
\hline
Blackbody radius               & $p=0.00002$ (14 vs. 14) & $p=0.34314$ (14 vs. 4) & $p=0.34314$ (14 vs. 4) \\ 
Blackbody temperature          & $p=0.01878$ (14 vs. 14) & $p=0.34314$ (14 vs. 4) & $p=0.91634$ (14 vs. 4) \\ 
g-band luminosity              & $p=0.00490$ (14 vs. 14) & $p=0.81176$ (14 vs. 4) & $p=0.02288$ (14 vs. 4) \\ 
Blackbody luminosity           & $p=0.92052$ (14 vs. 14) & $p=0.07190$ (14 vs. 4) & $p=0.14118$ (14 vs. 4) \\ 
Rise e-folding time            & $p=0.03910$ (11 vs. 9) & $p=0.02198$ (11 vs. 3) & $p=0.09091$ (9 vs. 3) \\ 
Decay e-folding time           & $p=0.76724$ (12 vs. 14) & $p=0.99121$ (12 vs. 4) & $p=0.74967$ (14 vs. 4) \\ 
Fallback time                  & $p=0.76724$ (12 vs. 14) & $p=0.52527$ (12 vs. 3) & $p=0.73529$ (14 vs. 3) \\ 
Power-law index                & $p=0.29992$ (13 vs. 13) & $p=0.48403$ (13 vs. 4) & $p=0.17311$ (13 vs. 4) \\ 
Redshift                       & $p=0.34332$ (14 vs. 14) & $p=0.74967$ (14 vs. 4) & $p=0.54118$ (14 vs. 4) \\ 
Host mass                      & $p=0.63548$ (14 vs. 14) & $p=0.61176$ (14 vs. 4) & $p=0.81176$ (14 vs. 4) \\ 
Host rest-frame $u-r$          & $p=0.63548$ (14 vs. 14) & $p=0.02288$ (14 vs. 4) & $p=0.34314$ (14 vs. 4) \\ 
Time since peak SFR            & $p=0.15493$ (14 vs. 14) & $p=0.54118$ (14 vs. 4) & $p=0.74967$ (14 vs. 4) \\ 
SFH $\tau$                     & $p=0.34332$ (14 vs. 14) & $p=0.74967$ (14 vs. 4) & $p=0.40719$ (14 vs. 4) \\ 
Host metalicity                & $p=0.99959$ (14 vs. 14) & $p=0.18954$ (14 vs. 4) & $p=0.18954$ (14 vs. 4) \\ 
Host dust E(B-V)               & $p=0.15493$ (14 vs. 14) & $p=0.40719$ (14 vs. 4) & $p=0.97353$ (14 vs. 4) \\ 
Host population synthetis PC4  & $p=0.01878$ (14 vs. 14) & $p=0.74967$ (14 vs. 4) & $p=0.34314$ (14 vs. 4) \\ 
\hline

\end{tabular}
\end{table*}

\begin{table*}
\footnotesize
\centering
\caption{Kendall's tau $p$-value comparing photometric and host galaxy properties.}
\label{tab:Kendall}
\begin{tabular}{l c c c c c c c c c c}
\hline
&  $R$       &   $T$       &   $L_{\rm bb}$ &   $L/R$     &   rise      &   decay     &   $z$       &   mass      &   $u$-$r$   &   PC4       \\ 
\hline
$R$              &             &   $10^{-8.9}$ (39)&   0.475 (39)&   $10^{-3.3}$ (39)&   0.508 (27)&   0.663 (36)&   0.119 (39)&   0.446 (39)&   0.586 (39)&   0.018 (39)\\
$T$              &   $10^{-8.9}$ (39)&             &   0.029 (39)&   $10^{-9.7}$ (39)&   0.045 (27)&   0.892 (36)&   0.856 (39)&   0.726 (39)&   0.304 (39)&   0.150 (39)\\
$L_{\rm bb}$     &   0.475 (39)&   0.029 (39)&             &   $10^{-5.8}$ (39)&   0.016 (27)&   0.567 (36)&   0.049 (39)&   0.041 (39)&   0.119 (39)&   0.157 (39)\\
$L/R$            &   $10^{-3.3}$ (39)&   $10^{-9.7}$ (39)&   $10^{-5.8}$ (39)&             &   0.008 (27)&   0.288 (36)&   0.071 (39)&   0.068 (39)&   0.075 (39)&   0.875 (39)\\
rise             &   0.508 (27)&   0.045 (27)&   0.016 (27)&   0.008 (27)&             &   0.941 (24)&   0.901 (27)&   0.263 (27)&   0.080 (27)&   0.184 (27)\\
decay            &   0.663 (36)&   0.892 (36)&   0.567 (36)&   0.288 (36)&   0.941 (24)&             &   0.053 (36)&   0.027 (36)&   0.496 (36)&   0.913 (36)\\
$z$              &   0.119 (39)&   0.856 (39)&   0.049 (39)&   0.071 (39)&   0.901 (27)&   0.053 (36)&             &   0.003 (39)&   0.570 (39)&   0.124 (39)\\
mass             &   0.446 (39)&   0.726 (39)&   0.041 (39)&   0.068 (39)&   0.263 (27)&   0.027 (36)&   0.003 (39)&             &   0.002 (39)&   0.097 (39)\\
$u$-$r$          &   0.586 (39)&   0.304 (39)&   0.119 (39)&   0.075 (39)&   0.080 (27)&   0.496 (36)&   0.570 (39)&   0.002 (39)&             &   0.315 (39)\\
PC4              &   0.018 (39)&   0.150 (39)&   0.157 (39)&   0.875 (39)&   0.184 (27)&   0.913 (36)&   0.124 (39)&   0.097 (39)&   0.315 (39)&             \\
\hline

\hline
%\tablecomments{The $p$-value for the null-hypothesis that the two properties that statistically independent. For the relation between $R$ and $T$, $p<3\times 10^{-5}$. }
\end{tabular}
\end{table*}



\end{document}